\documentclass[pra,amsmath,amssymb,nofootinbib,superscriptaddress,onecolumn,11pt]{revtex4-2}
\usepackage{bm,psfrag,graphicx,setspace}
\usepackage[T1]{fontenc}
\usepackage[left]{lineno}
\usepackage{xcolor}
\usepackage{hyperref}
\usepackage{bbold}
\usepackage{mathtools}
\usepackage{braket}
\usepackage{lineno}

\DeclareRobustCommand{\igmas}{\text{\reflectbox{$\sigma$}}}

\begin{document}
\linespread{1.2}
\selectfont
\title{Conservation of optical chirality in nanoscale light-matter interactions: A study of the Born-Kuhn model system}
\author{Kevin N. Moser}
\thanks{These authors contributed equally to this work.}
\author{Marc R. Bourgeois}
\thanks{These authors contributed equally to this work.}
\author{Elliot K. Beutler}
\author{David J. Masiello}
\email{masiello@uw.edu}
\affiliation{Department of Chemistry, University of Washington, Seattle, USA}
\begin{abstract}
Optical chirality density is a measure of the local handedness of electromagnetic fields. Like energy density, it may be absorbed or scattered through the interaction between light and matter. Here, we utilize the conservation of optical chirality to connect the parity and time-reversal symmetries of the intrinsic excitational eigenmodes of a material to those of their associated electromagnetic eigenfields as dictated by Maxwell's equations. To make this connection explicit, we theoretically examine the Born-Kuhn (BK) system, composed of a pair of plasmonic nanorods of variable separation, as a prototypical material model that is both geometrically chiral in its static structure and truly excitationally chiral in its eigenexcitations and eigenfields. By relaying optical chirality metrics of the BK eigenfields back to their underlying sourcing material degrees of freedom, we derive a unique mechanical chirality measure that is related to, but distinct from, other pseudoscalar metrics recently discussed in the literature. Beyond analysis in the absence of sources, we further derive optical chiral extinction, scattering, and absorption cross sections under external drive and discuss their rigorous connection to more common circular dichroism measurements as well as their limitations in comparison to eigenfield chirality metrics. Lastly, we investigate the conversion of achiral linearly polarized light into chiral elliptically polarized light through interaction with the BK system, illustrating the conservation of optical chirality in the interaction between light and matter through an analytically tractable example.
\end{abstract}


\maketitle
\section{Introduction}

Parity violating chiral structures are ubiquitous in nature. At subatomic length scales the fundamental electroweak interaction is both parity and charge parity violating, coupling only to left handed fermions and right handed antifermions. This intrinsic asymmetry extends beyond particle physics into the realm of chemistry and biology, where molecular handedness undergirds key biochemical functions, most notably in the homochiral features of left handed amino acids, and right handed sugar enantiomers found in nucleic acids and glycoproteins \cite{blackmond2010origin,frank_spontaneous_1953}. Additionally, enantioselective synthesis employed in pharmaceutical chemistry and drug development is crucial to effective and non-harmful medicinal therapies \cite{liu2015catalytic,xu2022enantiomer}. At the nanoscale, chiral structures serve as platforms for chiral sensing \cite{Warning2021,Almousa2024,mun_electromagnetic_2020,Lee2020} and single-molecule detection \cite{Dionne2020,Biteen2024,MaWei2013,zhang2024quantum}. Chiral features persist also at the macroscale in the structure of chiral crystals \cite{Ueda2023, ishito2023truly}, the growth of snail shells \cite{schilthuizen_convoluted_2005}, and in the formation of spiral galaxies \cite{yu2020probing} as well as in countless other examples.


While the presence or absence of improper rotations in the point group of an object, including mirror and inversion symmetries, suffices to classify the chirality of static material structures, parity violation alone is incapable of describing chirality arising in dynamic systems \cite{Barron_1986}. Although an extensive body of literature is now dedicated to quantifying the degree to which static structures are chiral \cite{Zabrodsky1995,Pinsky2008,Dryzun2011,RevModPhys.71.1745}, it was not until the work of Barron \cite{Barron_1986} and the definitions of \textit{true} and \textit{false} chirality that dynamical chirality could be classified on the basis of parity and time-reversal symmetries. Using this expanded notion, Tang and Cohen \cite{tang2010optical} built on the work of Lipkin \cite{Lipkin_1964} to introduce a measure of {\it optical chirality} for electromagnetic fields that evolve in space and time. Optical chirality quantifies the underlying handedness of light and its asymmetric interaction with chiral matter. Yet, despite the fact that electromagnetic fields are sourced by underlying material excitations as dictated by Maxwell's equations, the connection between optical chirality metrics and those assessing the dynamical chirality of collective low-energy material excitations, e.g., phonon \cite{choi2022chiral, Ueda2023,ishito2023truly,Abraham2024}, plasmon \cite{fan2012chiral,hentschel2017chiral,zhang2019unraveling,KuzykAnton2012Dsoc}, and polariton \cite{cerdan2023chiral,baranov2023toward,ostovar2015through} excitations, remains unestablished. Also unclear is the extent to which traditional observables of optical activity such as circular dischroism \cite{fan2012chiral} are faithful reporters of material excitational chirality since such observables require probes that may also be dynamically chiral, e.g., circularly polarized light, thus convolving the chirality of the probe with that of the material in the measured signal.




To examine these open questions, in this paper we utilize the Born-Kuhn (BK) structure as a simple material model system that has been employed extensively in the literature to investigate chiroptical phenomena \cite{BK,Kuhn1930,Born1918,Lourenco-MartinsHugo2021Opai,BourgeoisMarcR.2022PEEG}. Formed from a pair of rods that are angularly and vertically offset, the BK model system can be physically realized in many different contexts depending on the laws dictating the coupling between each rod, e.g., the transition dipoles within a molecular chromophore dimer, a pair of dielectric or metallic nanophotonic cavities, a duet of acoustic guitarists, or two gravitating massive bodies. Here, we specialize to the case of two plasmonic nanorods interacting through the fully-retarded electromagnetic field. Through a quasinormal mode analysis of the BK system as its geometry is continuously deformed from one handedness to the other, we demonstrate that the optical chirality measures established in Ref. \cite{tang2010optical} can be further leveraged to characterize the excitational chirality of the underlying sourcing eigencurrent distributions and their associated eigenfields across all interaction regimes spanning from the near- to far-field. For small inter-rod separation distances, we explicitly connect the radiated optical chirality of the BK eigenfields to a unique time-averaged pseudoscalar quantity characterizing the chirality of the underlying material excitations. The identified pseudoscalar metric is similar to, but distinct from, other pseudoscalar metrics used in the literature to assess the excitational chirality of molecular systems \cite{Abraham2024,RevModPhys.71.1745,Pinsky2008,Dryzun2011,Zabrodsky1995,sidorova2021quantitative}.


Beyond analysis of the intrinsic excitational chirality of the BK eigenmodes and their associated eigenfields, we also study the behavior of the BK system under plane wave optical drive. In analogy to the extinction, absorption, and scattering of optical power governed by Poynting's theorem, we invoke the conservation theorem for optical chirality to derive closed form expressions for the optical chirality extinction, absorption, and scattering cross sections of the BK system. Through analytical investigation we show how the optical chirality extinction cross section is related to the more familiar circular dichroism measurement, while neither observable captures the full bisignate behavior of the radiated eigenfield chirality with increasing inter-rod separation spanning into the intermediate- and far-fields. Lastly, leveraging the analytical tractability of the BK model dynamics, we provide an illustrative example demonstrating the conservation of optical chirality in light-matter interactions to investigate the conversion of incident linearly polarized achiral light into a rotated and elliptically polarized chiral scattered field through interaction with the BK system. The results are interpreted through the optical theorem for optical chirality together with the Stokes parameters describing the polarization state of the scattered field. Together, this study of the analytically tractable BK model elucidates the flow of optical chirality in nanoscale light-matter interactions and adds to the ongoing development of metrics quantifying collective material excitations, both of which support the continued pursuit of optoelectronic and photonic devices exploiting chiral degrees of freedom. 

The organization of this paper is as follows: In Section \ref{BK_section} we introduce the BK structure and characterize its intrinsic eigenexcitations and their parity and time-reversal symmetries. In Section \ref{SEC_EigSca_OptChi}, we apply the optical chirality metric to the scattered BK eigenfields in the absence of drive and, in Section \ref{SEC_chi_metric_connect}, rigorously connect the radiated eigenfield chirality to a unique pseudoscalar metric characterizing the excitational chirality of the field's underlying material eigenexcitations. In Section \ref{BKdrive}, we introduce a driving optical field to the BK system and derive analytically exact expressions for the BK optical power and circular dichroism spectra (Section \ref{CD_section}) as well as the cross sections for chiral extinction, absorption, and scattering based on the conservation theorem for optical chirality (Section \ref{chiralCS}). Lastly, in Section \ref{QWP} we invoke a generalization of the optical theorem to examine the conservation of optical chirality through an example whereby the BK system converts an achiral incident light field into a chiral scattered field, in analogy to an optical wave plate. A Conclusion (Section \ref{conc}) and Appendices follow. Gaussian units are used throughout.

\section{The Born-Kuhn Structure and its Intrinsic Excitations}
\label{BK_section}

\begin{figure*}
  \begin{center}
\includegraphics[width=1\textwidth]{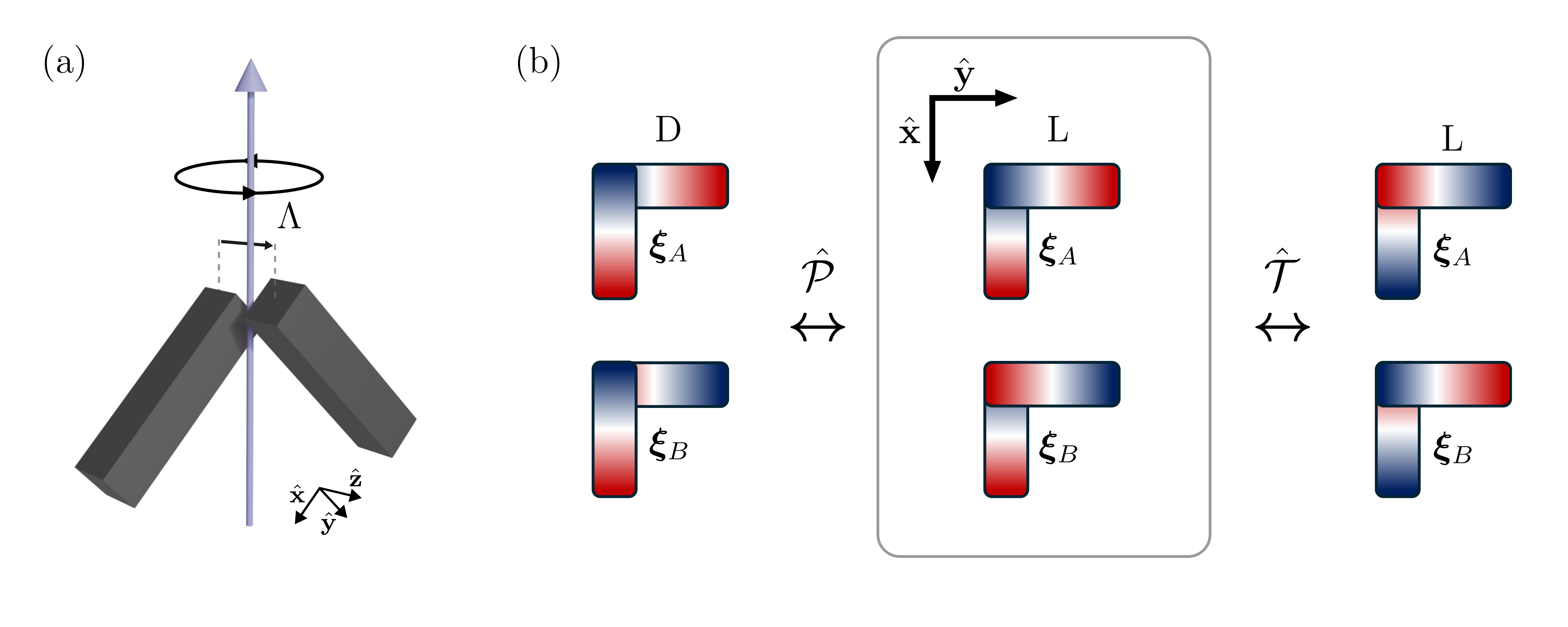}
  \end{center}
  \caption{The chiral Born-Kuhn (BK) structure. (a) Scheme of the BK structure composed of two metallic nanorods separated by distance $\Lambda$ along $\hat{\mathbf{z}}$ ($\Lambda < 0 $ shown). (b) The in-phase ($\bm{\xi}_B$) and out-of-phase ($\bm{\xi}_A$) plasmonic eigenexcitations of the L enantiomer are depicted in the boxed middle panel along with the transformed modes resulting from parity ($\hat{\cal{P}}$) and time-reversal ($\hat{\cal{T}}$) operations. In the case of parity, an additional proper rotation $(\hat{\cal{R}}$) is applied to align the transformed structure with its original orientation. Colors indicate the induced polarizations of the plasmonic dipole on each rod.}
  \label{fig1}
\end{figure*}
The Born-Kuhn (BK) structure depicted in Fig. \ref{fig1} represents a prototypical material system that exhibits chiral symmetry in both its static geometry and in its excitations \cite{Kuhn1930,Born1918}. The BK structure is composed of two rods that are angularly offset in the $xy$ plane and displaced along the $z$ axis by the distance $\Lambda = z_1 - z_2$. Its static geometry belongs to the $C_2$ point group containing the identity and rotations by $\pi$ around one axis, indicated by the vertical arrow in Fig. \ref{fig1}(a). The lack of inversion symmetry, or any other improper rotation, renders the BK structure geometrically chiral, with two distinct enantiomers labeled by L ($\Lambda < 0$) and D ($\Lambda > 0$).

Here, we examine the case of two identical metallic nanorods, each hosting a single dipolar surface plasmon mode ${\bf d}_i=\hat{\bf{n}}_id_i$ ($i=1,2$) oriented along the nanorod's long axis of length $a$ with all other plasmonic modes detuned away from the spectral region of investigation. The two plasmonic modes interact via the fully retarded electric dipole field and are governed by the complex-frequency, transcendental quasinormal mode equation \cite{Trichoidal_Dichroism}
\begin{equation}
 \begin{bmatrix}
    \alpha_1^{-1}(\omega) & -g(\omega) \\
    -g(\omega) & \alpha_2^{-1}(\omega)
\end{bmatrix}
\bm{\xi} = \mathbf{0},
\label{amat}
\end{equation}
where $\alpha_1(\omega) = \alpha_2(\omega) = ({e^2}/{m})/{\left(\omega_d^2-i\gamma(\omega)\omega-\omega^2\right)}$ are the polarizabilities of the two nanorods of frequency $\omega_1=\omega_2\equiv\omega_d$, effective mass $m_1=m_2\equiv m$ \cite{cherqui2014combined, cherqui2016characterizing, Smith2019}, elementary charge $-e$, and interaction energy $-d_1g(\omega)d_2$, with $g(\omega)= \hat{\bf{n}}_1 \cdot \tensor{\bf{G}}(\omega) \cdot \hat{\bf{n}}_2=\hat{\bf{n}}_2 \cdot \tensor{\bf{G}} \cdot \hat{\bf{n}}_1$. Here $\tensor{\bf G}(\omega)$ is the ${\bf x}={\bf r}_1$ and ${\bf x}'={\bf r}_2$ component of the vacuum dipole relay tensor $\tensor{\bf{G}}({\bf x},{\bf x}',\omega)=[k^2+\nabla\nabla']e^{ik|{\bf x}-{\bf x}'|}/|{\bf x}-{\bf x}'|,$ where $k=\omega/c$ is the free space wavenumber at frequency $\omega$, $c$ is the speed of light in vacuum, and $|{\bf r}_1-{\bf r}_2|$ is the separation distance between the dipoles located at positions ${\bf r}_1=(a/2,0,\Lambda/2)$ and ${\bf r}_2=(0,a/2,-\Lambda/2).$ The dipole decay rate $\gamma(\omega)=\gamma_{\textrm{nr}}+\gamma_{\textrm{rad}}(\omega)$ is composed of both nonradiative $\gamma_{\textrm{nr}}$ and radiative $\gamma_{\textrm{rad}}(\omega)=(2e^2/3mc^3)\omega^2$ components, where the former is the Drude electron-ion scattering rate accounting for Ohmic losses in the bulk metal and the latter is the radiative dipole decay rate accompanying the effect of radiation reaction \cite{devkota_making_2019,YangYanhe2022Rtpr}.

In the quasistatic and lossless limits where $c\to\infty$ and $\gamma_{\textrm{nr}}\to0$, Eq. \eqref{amat} simplifies considerably and admits the following plasmonic eigenmodes 
\begin{equation}
    \bm{\xi}_{A/B} = \begin{bmatrix} d_1^{A/B} \\
    d_2^{A/B}
    \end{bmatrix}
    =\frac{1}{\sqrt{2}}\begin{bmatrix} 1 \\
    \pm 1
    \end{bmatrix} d,
    \label{eigenmodes}
\end{equation}
and associated real-valued eigenfrequencies
\begin{equation}
    \omega_{A/B} = \sqrt{\omega_d\pm\frac{e^2}{m}g},
    \label{eigenmodes2}
\end{equation}
where $A$ $(+)$ and $B$ $(-)$ label the irreducible representations $\Gamma \in \{A,B\}$ of the $C_2$ point group, and frequency-independent $g>0$. Eq. \eqref{eigenmodes} represents the vectorized form of $\bm{\xi}_\Gamma={\bf d}_{1}^\Gamma +{\bf d}_2^\Gamma=\hat{\bf n}_1d_1^\Gamma +\hat{\bf n}_2d_2^\Gamma,$ where $d_{i}^\Gamma$ are the mode components projected along each nanorod oriented in the $\hat{\bf n}_1=\hat{\bf x}$ and $\hat{\bf n}_2=\hat{\bf y}$ directions as depicted in Fig. \ref{fig1}(a). Physically, $\bm{\xi}_\Gamma$ are the in-phase ($-$) and out-of-phase ($+$) plasmonic eigenmodes of magnitude $d$ with fixed energetic ordering. Outside of the quasistatic and lossless limits, however, $g(\omega)$, $\alpha_{1}(\omega),$ and $\alpha_{2}(\omega)$ are generally complex-valued, frequency-dependent functions, and the fully-retarded quasinormal mode solutions of Eq. \eqref{amat} must be determined numerically \cite{montoni2018tunable, bourgeois2022optical}. Outside of Eqs. \eqref{eigenmodes} and \eqref{eigenmodes2}, which are presented only to provide a simple physical understanding of the intrinsic BK model excitations, all subsequent discussion involves the fully-retarded BK quasinormal mode solutions. For simplicity these quasinormal modes and their associated quasinormal mode electromagnetic fields will be referred to as eigenmodes and eigenfields.

\begin{figure}
\centering
\includegraphics[width=0.5\textwidth]{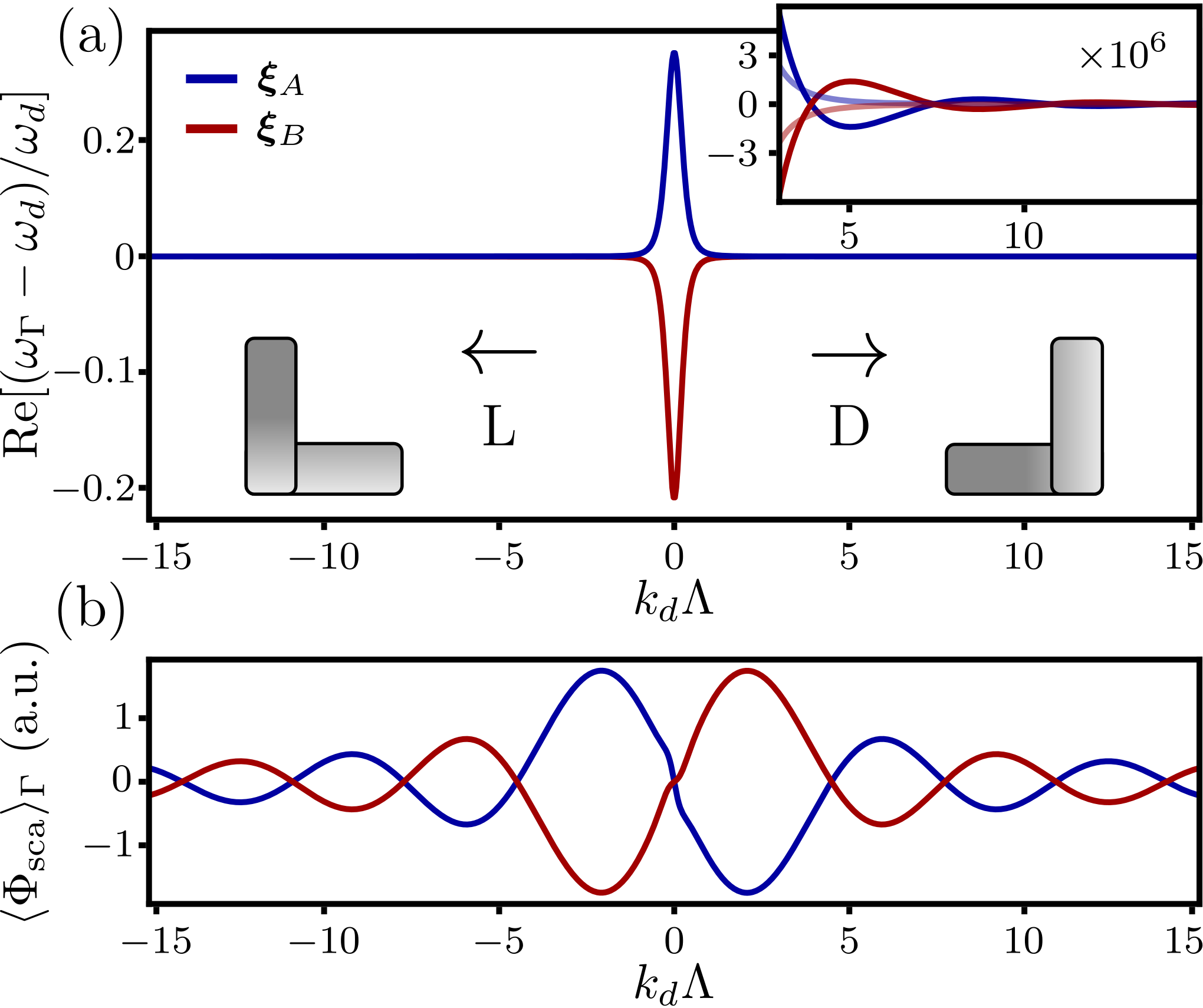}
\caption{Eigenfrequencies and radiated eigenfield chirality of the BK structure. (a) BK quasinormal mode eigenfrequencies $\omega_\Gamma$ from Eq. \eqref{eigenmodes2} and (b) radiated eigenfield chirality $\langle \Phi_\text{sca} \rangle_\Gamma$ evaluated at $\omega_\Gamma$ versus $k_d\Lambda$ spanning from near- to far-field regimes. The inset in panel (a) shows a zoom-in comparing the exact QNM frequency oscillation (solid) and quasistatic eigenfrequency decay (transparent) with $k_d\Lambda$ in the intermediate- to far-field. (b) Evaluated at the QNM frequencies $\omega_\Gamma$, $\langle \Phi_\text{sca} \rangle_\Gamma$ also oscillates with $k_d\Lambda$, signaling fluctuations in sign of the QNM optical chirality with increasing interdipole separation within the intermediate- to far-field.}
\label{fig3}
\end{figure}
Fig. \ref{fig3}(a) displays the dependence of the BK eigenmode frequencies $\omega_\Gamma$ upon interdipole separation $k_d\Lambda$ in units of the plasmon resonance wavelength $\lambda_d=2\pi/k_d=2\pi c/\omega_d$, spanning from the near- to the far-field of both L and D enantiomers \footnote{All calculations and figures use the following set of BK material parameters typical of plasmonic noble metals: $\hbar\omega_d=3$ eV, $\hbar\gamma_{\textrm{nr}}=0.07$ eV, and $m=2\times10^{-33}$ g.}. The $\omega_\Gamma$ are determined numerically by searching for the complex-valued roots of the transcendental function $f(\omega)=\alpha_1^{-1}(\omega)\alpha_2^{-1}(\omega)-g^2(\omega)$ stemming from the determinant of the dynamical matrix in Eq. \eqref{amat}. At values of $|k_d\Lambda|\ll1,$ the $\omega_\Gamma$ are maximally split, fall to $\omega_d$ in the intermediate-field, and then oscillate around $\omega_d$ in the far-field whenever the interdipole separation distance $\Lambda$ approximately equals an integer number of resonant wavelengths $\lambda_d$ \cite{Novotny}. Such oscillatory behavior of $\omega_\Gamma$ in the far-field is to be expected as a pair of radiating dipoles may causally exchange energy at all separation distances, resulting in in-phase and out-of-phase hybrid modes that alternate in energy ordering as a function of separation (see inset) \cite{Novotny}. For comparison, Fig. \ref{fig3}(a) also plots the quasistatic eigenmode frequencies (as transparent curves) which decay in $|k_d\Lambda|$ but do not oscillate.


For the L enantiomer with $k_d\Lambda\ll1$, the eigenmodes $\bm{\xi}_\Gamma$, represented by their dipole polarizations are illustrated in the middle panel of Fig. \ref{fig1}(b). Application of the parity ($\hat{\cal{P}}$) operation interconverts both L eigenmodes to their D-enantiomer counterparts up to a proper rotation ($\hat{\cal R}$) by inverting the location and reversing the orientation of the plasmonic dipole moment within each nanorod. Application of the time-reversal ($\hat{\cal{T}}$) operation, meanwhile, does not interconvert L and D enantiomers, as only the dipole directions are reversed. Thus, according to the definitions of Barron \cite{Barron_1986}, the BK eigenmodes $\bm{\xi}_\Gamma$ are said to be {\it truly chiral} above and beyond the geometric chirality dictated by the static BK structure. We emphasize that the true chirality of Barron is tied to the excitational chirality of the system's Hamiltonian eigenmodes, which, while related, is distinctly different from the chirality of its underlying geometry.

As any oscillating distribution of charge produces electromagnetic fields according to Maxwell's equations, associated with the $\bm{\xi}_\Gamma$ eigenmodes are the eigenfields ${\bf E}_\Gamma,$ which are also truly excitationally chiral. Specifically,
\begin{equation}
\begin{split}
\hat{\cal P}:{\bf E}^{\textrm{L}}_{\Gamma}({\bf x})&\to\hat{\cal R}{\bf E}^{\textrm{D}}_{\Gamma}({\bf x})\\
\hat{\cal T}:{\bf E}^{\textrm{L}}_{\Gamma}({\bf x})&\to{\bf E}^{\textrm{L}}_{\Gamma}({\bf x})e^{i\phi},
\end{split}
\end{equation}
where $\phi$ is a constant phase offset. Appendix \ref{SEC_Geo_Exc_Sym} illustrates how the true chiralities of ${\bf E}_\Gamma$ are directly inherited from the $\hat{\mathcal{P}}$ and $\hat{\mathcal{T}}$ symmetries of the underlying eigenmodes $\boldsymbol{\xi}_\Gamma$, while the optical chiralities associated with ${\bf E}_\Gamma$ are presented in section \ref{SEC_EigSca_OptChi}. The connection between the optical chirality metric characterizing the electromagnetic field and a momentum pseudoscalar characterizing the chirality of the time-harmonic material currents associated with the BK eigenmodes is established in section \ref{SEC_chi_metric_connect}.

\subsection{Radiation of eigenfield chirality}
\label{SEC_EigSca_OptChi}

Whether in the presence (or absence) of external driving, the excitations (eigenexcitations) of a material will be accompanied by their associated electromagnetic fields (eigenfields) as dictated by Maxwell's equations. In 2010, Tang and Cohen identified a parity-odd, time-even pseudoscalar function of the electromagnetic fields that satisfies Barron's definition of true chirality and is a reporter of the optical chirality of the electromagnetic field \cite{tang2010optical}.  This quantity, called the optical chirality density
\begin{equation}
    C({\bf x}, t) = \frac{1}{8 \pi} \Big[ {\bf E}({\bf x}, t) \cdot  {\nabla} \times {\bf E}({\bf x}, t) + {\bf B}({\bf x}, t) \cdot {\nabla} \times{\bf B}({\bf x}, t)  \Big]
\label{opt_chiral_density_def}
\end{equation}
was initially discovered in 1964 to be directly derivable from Maxwell's equations as one of Lipkin's ten zilches \cite{Lipkin_1964}, each of which is conserved under Maxwell's equations. Like electromagnetic energy, momentum, and angular momentum \cite{vernon_decomposition_2024,bliokh_transverse_2015,allen_optical_2003,barnett_orbital_1994}, conservation of optical chirality density $C({\bf x}, t)$ is governed by the continuity equation
\begin{equation}
    (d/dt){C}  = - {\nabla} \cdot \boldsymbol{\varphi} + \mathcal{J},
    \label{chi_conservation}
\end{equation}
where $\boldsymbol{\varphi}=(c/8\pi)[{\bf E}\times(\nabla\times{\bf B})-{\bf B}\times(\nabla\times{\bf E})]$ is the chirality flux, and $\mathcal{J}=-(1/2)[{\bf E}\cdot\nabla\times{\bf J}+{\bf J}\cdot\nabla\times{\bf E}]$ represents matter degrees of freedom serving as sources and/or sinks of optical chirality. Using this conservation law, Ref. \cite{tang2010optical} showed that the differential absorption rate of circularly polarized light by molecular enantiomers can be expressed as the product of the optical chirality of the excitation field $C$ and a term containing only material parameters encoding the intrinsic chiral responses of the molecule \cite{Barron2009-sp}, highlighting the distinction between the excitational chiralities of light and matter degrees of freedom.

The continuity theorem for optical chirality can be employed to establish the intrinsic connection between the excitational chirality stored in a material system's eigenmodes and the optical chirality stored in its associated eigenfields. In the absence of external driving, volume integration of the time-averaged chiral continuity equation (Eq. \eqref{chi_conservation}) results in 
\begin{equation}
0=-\int d{\bf x}\nabla\cdot\langle\boldsymbol{\varphi}_{\textrm{sca}}\rangle+\int d{\bf x}\langle\mathcal{J}_{\textrm{sca}}\rangle.
\label{TAconteq}
\end{equation}
The first term on the right-hand side represents the rate at which optical chirality, i.e., net helicity \cite{poulikakos2019optical} of plane waves comprising the plane expansion of a given field \cite{Bliokh2011, coles2012chirality}, flows out of the volume $V$ through its bounding surface $S$ \footnote{Here and in the following, $\langle A\rangle$ indicates the time-average of $A$, and all volume and surface integrals are taken with respect to the volume $V$ and its closed surface $S$, respectively.}, and defines the radiated chirality of the $\Gamma\in\{A,B\}$ eigenfields 
\begin{equation}
\begin{aligned}
\langle \Phi_{\text{sca}}\rangle_\Gamma &\equiv \int d{\bf x}\,\nabla\cdot\langle {\bm\varphi}_{\textrm{sca}}\rangle_\Gamma\\
&=-\frac{\omega_\Gamma}{16\pi}\int r^2d\Omega\,\hat{\bf n}\cdot\textrm{Im}[{\bf E}_{\Gamma}\times{\bf E}_{\Gamma}^{*}+{\bf B}_{\Gamma}\times{\bf B}_{\Gamma}^{*}],\\
\end{aligned}
\label{Phi_sca_plus_minus}
\end{equation}
in terms of the time-averaged scattered optical chirality flux $\langle\bm{\varphi}_{\textrm{sca}}\rangle_\Gamma =- (\omega_\Gamma/16\pi)\textrm{Im}[{\bf E}_{\Gamma}\times{\bf E}_{\Gamma}^*+{\bf B}_{\Gamma}\times{\bf B}_{\Gamma}^*]$ and source $\langle{\cal J}_{\textrm{sca}}\rangle_\Gamma=-(1/4)\textrm{Re}[{\bf E}_{\Gamma}\cdot\nabla\times{\bf J}_\Gamma^*+{\bf J}_\Gamma\cdot\nabla\times{\bf E}_{\Gamma}^*]$, where the eigencurrent densities ${\bf J}_\Gamma({\bf x})=\dot{\bf d}_1^\Gamma\delta({\bf x}-{\bf r}_1)+\dot{\bf d}_2^\Gamma\delta({\bf x}-{\bf r}_2)$ oscillate at the characteristic eigenfrequencies $\omega_\Gamma$ and source the eigenfields $\mathbf{E}_{\Gamma}({\bf x})=({i}/{\omega}_\Gamma)\int d{\bf x}'\tensor{\bf{G}}({\bf x},{\bf x}',\omega_\Gamma)\cdot{\bf J}_\Gamma({\bf x}')$ observed at ${\bf x}=\hat{\bf n}r.$

In the case of monochromatic fields, the time-averaged chirality flux $\langle\bm{\varphi}_{\textrm{sca}}\rangle_\Gamma$ is trivially related to the optical spin angular momentum density $\langle \mathbf{s}\rangle_\Gamma = (8 \pi \omega)^{-1} \text{Im}\big\{ \mathbf{E}_\Gamma^* \times \mathbf{E}_\Gamma + \mathbf{B}_\Gamma^* \times \mathbf{B}_\Gamma \big\}$ \cite{barnett2010rotation, vernon_decomposition_2024} by $\langle\bm{\varphi}_{\textrm{sca}}\rangle _\Gamma= (\omega^2_\Gamma /2)\langle \mathbf{s} \rangle_\Gamma$. As a consequence, Eq. \eqref{Phi_sca_plus_minus} can alternatively be written as 
\begin{equation}
\begin{aligned}
\langle \Phi_{\text{sca}}\rangle_\Gamma &=\frac{\omega^2_\Gamma}{2}\int r^2d\Omega\,\hat{\bf n}\cdot\langle \mathbf{s}\rangle_\Gamma.
\end{aligned}
\end{equation} 
The optical spin angular momentum density $\mathbf{s}({\bf x})$ can itself be decomposed as $\mathbf{s}({\bf x}) = \mathbf{s}_c({\bf x}) + \mathbf{s}_P({\bf x})$, where the canonical $\mathbf{s}_c$ and Poynting $\mathbf{s}_P$ contributions are oriented parallel to and perpendicular to the observation direction $\hat{\mathbf{n}}$, respectively, for fields produced by a collection of localized time-harmonic sources with center of mass at the origin \cite{vernon_decomposition_2024}. In such situations, while the Poynting spin angular momentum of the scattered electromagnetic field generates intrinsic radiative torques acting on the sources \cite{Nieto_2015_oam,Nieto-Vesperinas_15}, $\langle\mathbf{s}_P\rangle_\Gamma$ does not contribute to $\langle \Phi_{\textrm{sca}}\rangle_\Gamma$. Instead, $\langle \Phi_{\textrm{sca}}\rangle_\Gamma$ is determined by the angle-integrated flux of canonical optical spin angular momentum $\langle \mathbf{s}_c \rangle_\Gamma$ out of volume $V$, which is connected to the existence of a non-zero time-averaged chiral pressure force on chiral matter \cite{vernon_decomposition_2024}.


In the case of the the BK structure, Eq. \eqref{Phi_sca_plus_minus} can be evaluated analytically using the Fraunhofer far-fields $\mathbf{E}_\Gamma({\bf x})\approx k_\Gamma^2(e^{ik_\Gamma r}/r)[(\hat{\bf n}\times{\bf d}^\Gamma_1 e^{-ik_\Gamma\hat{\bf n}\cdot{\bf r}_1})\times\hat{\bf n}+(\hat{\bf n}\times{\bf d}^\Gamma_2 e^{-ik_\Gamma\hat{\bf n}\cdot{\bf r}_2})\times\hat{\bf n}]$ to yield the pseudoscalar
\begin{equation}
\begin{split}
   \langle \Phi_{\text{sca}} \rangle_\Gamma &= -c k_\Gamma^5 \text{Re}\big[d^\Gamma_1(d^\Gamma_2)^*\big]j_1(k_\Gamma\Lambda)\\
   &\approx c k_\Gamma^5\text{Re}\big[d^\Gamma_1(d^\Gamma_2)^*\big]\times\begin{cases} 
      -(1/3)(k_\Gamma\Lambda) & |k_\Gamma\Lambda|\ll1 \\
      \cos(k_\Gamma\Lambda)/(k_\Gamma\Lambda) & |k_\Gamma\Lambda|\gg1,
   \end{cases}
   \label{phi_scaig}
   \end{split}
\end{equation}
where $k_\Gamma=\omega_\Gamma/c.$ Fig. \ref{fig3}(b) displays the dependence of the radiated eigenfield chirality $\langle \Phi_{\text{sca}} \rangle_\Gamma$ upon reduced interdipole separation $k_d\Lambda$ for both $\boldsymbol{\xi}_\Gamma$ eigenmodes of both L and D enantiomers. For small $|k_d\Lambda|$, $\langle \Phi_{\text{sca}} \rangle_\Gamma$ calculated at $\omega_\Gamma$ behaves monotonically. However, at large $|k_d\Lambda|,$ Fig. \ref{fig3}(b) demonstrates that $\langle \Phi_{\text{sca}} \rangle_\Gamma$ alternates in sign along with the alternating mode frequency ordering shown in Fig. \ref{fig3}(a), meaning that the true excitational chirality intrinsic to the BK eigenmodes and their eigenfields varies with interdipole separation even though the underlying geometric chirality does not. Such fluidity in sign of the intrinsic excitational radiated chirality $\langle \Phi_{\text{sca}} \rangle_\Gamma$ is generic to chiral materials whose components interact through intermediate- and far-fields, and highlights an important distinction from the static chirality dictated by the material's point group symmetry that remains invariant independent of interaction length scales.

\subsection{Optical chirality and momentum pseudoscalars}
\label{SEC_chi_metric_connect}



The previous section described the quantification of excitational chirality stored in the BK eigenfields. This quantification, however, may also be expressed equivalently in terms of the underlying eigenmode matter degrees of freedom themselves. To make this connection explicit, we utilize a vector spherical harmonic decomposition \cite{Jackson2003} to express the eigenfields in terms of set of co-located multipoles; see Appendix \ref{Multipoles}. Specifically, at $\ell=1$ order, the electromagnetic responses of the displaced BK dipoles ${\bf d}^\Gamma_1(t)$ and ${\bf d}^\Gamma_2(t)$ can be represented by the co-located pair of electric and magnetic dipoles ${\bf d}_\Gamma(t)=\int d{\bf x}\,{\bf x}\rho_\Gamma({\bf x},t)$ and ${\bf m}_\Gamma(t)=(1/2c)\int d{\bf x}\,{\bf x}\times{\bf J}_\Gamma({\bf x},t)$, with asymptotic fields ${\bf E}^\Gamma_{\bf d}=k_\Gamma^2(e^{ik_\Gamma r}/r)[(\hat{\bf n}\times{\bf d}_\Gamma)\times\hat{\bf n}]$, ${\bf B}^\Gamma_{\bf d}=k_\Gamma^2(e^{ik_\Gamma r}/r)[\hat{\bf n}\times{\bf d}_\Gamma]$, ${\bf E}^\Gamma_{\bf m}=-k_\Gamma^2(e^{ik_\Gamma r}/r)[\hat{\bf n}\times{\bf m}_\Gamma]$, ${\bf B}^\Gamma_{\bf m}=k_\Gamma^2(e^{ik_\Gamma r}/r)[(\hat{\bf n}\times{\bf m}_\Gamma)\times\hat{\bf n}]$ observed in the far-field. Here, the charge density $\rho_\Gamma({\bf x},t)=-e\delta({\bf x}-{\bf r}^\Gamma_{\bf d}(t))$ and current density ${\bf J}_\Gamma({\bf x},t)=-e\dot{\bf r}_{\bf m}^\Gamma(t)\delta({\bf x}-{\bf r}^\Gamma_{\bf m}(t))$ correspond to the oscillatory plasmonic motion ${\bf r}^\Gamma_{\bf d}(t)$ and ${\bf r}^\Gamma_{\bf m}(t)$ about the BK (and multipole) origin. Thus, at $\ell=1$ order, the time-averaged optical chirality density and radiated eigenfield chirality in the far-field region ($k_\Gamma r\gg1$) are
\begin{equation}
\begin{split}
   \langle C({\bf x},t) \rangle_\Gamma &=\frac{\omega_\Gamma}{8\pi c} \text{Im}\big[{\bf E}_\text{s}^{\Gamma}\cdot{\bf B}_\text{s}^{\Gamma *}\big]\\
   &=\frac{\omega_\Gamma}{8\pi c} \text{Im}\big[({\bf E}_{\bf d}^\Gamma+{\bf E}^\Gamma_{\bf m})\cdot({\bf B}^\Gamma_{\bf d}+{\bf B}^\Gamma_{\bf m})^*\big]\\
&=\frac{1}{4\pi r^2}\Big(\frac{\omega_\Gamma}{c}\Big)^5\text{Im}\big[{\bf d}_\Gamma\cdot{\bf m}_\Gamma^*-({\bf d}_\Gamma\cdot\hat{\bf n})({\bf m}_\Gamma^*\cdot\hat{\bf n})\big],
\label{Cdm}
\end{split}
\end{equation}
and, from Eq. \eqref{TAconteq},
\begin{equation}
\begin{aligned}
\langle \Phi_{\text{sca}}\rangle_\Gamma  &= \int d{\bf x}\, \langle\mathcal{J}^\Gamma_{\textrm{sca}}\rangle\\
&=-\frac{1}{2} \int d\mathbf{x} \, \text{Re} \left[ \mathbf{E}_\Gamma \cdot (\nabla \times \mathbf{J}_\Gamma^*) \right] \\
&=\frac{k_\Gamma^4 \omega_\Gamma}{8 \pi} \int d\Omega \, \text{Im} \big[  (\mathbf{d}_\Gamma \cdot \mathbf{m}_\Gamma^*) - (\mathbf{d}_\Gamma \cdot \hat{\mathbf{n}})(\mathbf{m}_\Gamma^* \cdot \hat{\mathbf{n}}) \big],
\end{aligned}
\label{Phi_sca_plus_minus_J}
\end{equation}
illustrating how the metrics characterizing the optical chirality of the eigenfields can be expressed purely in terms of the sourcing matter degrees of freedom, e.g., the electric $\mathbf{d}_\Gamma$ and magnetic $\mathbf{m}_\Gamma$ dipole moments. Taken together, the optical chirality density and radiated eigenfield chirality are related by
\begin{equation}
\begin{aligned}
\langle \Phi_{\text{sca}}\rangle_\Gamma  
&=\frac{c}{2}\int r^2d\Omega \, \langle C({\bf x},t) \rangle_\Gamma\\
&=\frac{k_\Gamma^4 \omega_\Gamma}{8 \pi} \int d\Omega \hat{\mathbf{n}} \, \cdot \text{Im} \big\{ \hat{\mathbf{n}}\,( \mathbf{d}_\Gamma^\perp \cdot \mathbf{m}^{\perp *}_\Gamma)\big\}.
\end{aligned}
\label{chi_sca_and_density}
\end{equation}
Here, the resolution of the identity $\tensor{\bf I}=\hat{\bf n}\hat{\bf n}+(\hat{\bf n}\times\ )\times\hat{\bf n}$ was used in the last line to decompose the effective dipole moments $\mathbf{d}_\Gamma$ and $\mathbf{m}_\Gamma$ into their transverse components $\mathbf{d}_\Gamma^\perp$ and $\mathbf{m}_\Gamma^\perp$, where $\mathbf{V}_1^\perp\cdot\mathbf{V}_2^{\perp}=(\mathbf{V}_1 \cdot \mathbf{V}_2) - (\mathbf{V}_1 \cdot \hat{\mathbf{n}})(\mathbf{V}_2 \cdot \hat{\mathbf{n}})=(\hat{\bf n}\times\mathbf{V}_1) \cdot ( \hat{\bf n}\times\mathbf{V}_2)$ for arbitrary vectors ${\bf V}_1$ and ${\bf V}_2$.

Invoking the relationships $\mathbf{d}_\Gamma = (-ie/m\omega_\Gamma)\mathbf{p}_\Gamma$ and $\mathbf{m}_\Gamma = (-e/2mc)\mathbf{L}_\Gamma$ to express the effective electric and magnetic dipoles in terms of their associated linear ($\mathbf{p}_\Gamma$) and angular ($\mathbf{L}_\Gamma$) momenta \cite{Jackson2003}, respectively, Eqs. \eqref{Cdm} and \eqref{Phi_sca_plus_minus_J} can be written as 
\begin{equation}
\begin{split}
   \langle C({\bf x},t) \rangle_\Gamma &=\frac{3c}{16\pi m\omega_\Gamma^2 r^2}\Big(\frac{\omega_\Gamma}{c}\Big)^4\gamma_\text{rad}(\omega_\Gamma)\text{Re} \bigg\{ \mathbf{p}^\perp_\Gamma \cdot\mathbf{L}_\Gamma^{\perp *} \bigg\}  
\label{Cdm2}
\end{split}
\end{equation}
and
\begin{equation}
\begin{aligned}
\langle \Phi_{\text{sca}}\rangle_\Gamma  
&=\frac{3 k_\Gamma^2}{32\pi m} \gamma_\text{rad}(\omega_\Gamma) \int d\Omega \,\text{Re} \bigg\{\mathbf{p}^\perp_\Gamma\cdot\mathbf{L}_\Gamma^{\perp *}\bigg\},
\end{aligned}
\label{Phi_sca_plus_minus_3}
\end{equation}
where $\gamma_{\textrm{rad}}(\omega_\Gamma)=({2e^2}/{3m^2c^3})\omega_\Gamma^2$ is the radiation reaction decay rate introduced in Sec. \ref{BK_section}. Critically, Eqs. \eqref{Cdm2} and \eqref{Phi_sca_plus_minus_3} connect metrics of chirality stored in the electromagnetic field back to the excitational chirality stored in the mechanical momentum and angular momentum of the matter degrees of freedom underlying the $\ell=1$ vector spherical harmonic eigenmodes $\mathbf{d}_\Gamma$ and $\mathbf{m}_\Gamma$. We emphasize that Eqs. \eqref{Cdm2} and \eqref{Phi_sca_plus_minus_3} are derived directly from Maxwell's equations, eliminating the need for choices to be made in the relationship between the field and matter chirality metrics, and may be extended to arbitrary multipole order, as shown in Appendix \ref{Multipoles}.

Intriguingly, although a multiplicity of bisignate pseudoscalar and monosignate nonpseudoscalar metrics have been employed in the literature to quantify structural and dynamic chirality in chemical systems \cite{Abraham2024,RevModPhys.71.1745,Pinsky2008,Dryzun2011,Zabrodsky1995, sidorova2021quantitative}, the product $\mathbf{p}^\perp_\Gamma \cdot\mathbf{L}_\Gamma^{\perp *}$ appearing in Eqs. \eqref{Cdm2} and \eqref{Phi_sca_plus_minus_3} closely resembles the axial momentum pseudoscalar (MPS) introduced in Ref. \cite{Abraham2024} in the analysis of molecular vibrational normal mode chirality. The axial MPS chirality metric for a collection of atomic sites indexed by $i$ is $\text{MPS}^\Gamma_z = \sum_i  \langle p^i_zL^i_z \rangle_\Gamma$, where a choice of $z$ axis must be made. The highest molecular symmetry axis is selected as the $\hat{\bf z}$ direction in Ref. \cite{Abraham2024}. By construction, $\sum_\beta \text{MPS}^\Gamma_\beta = \sum_i\langle\mathbf{p}_i \cdot \mathbf{L}_i \rangle_\Gamma = \sum_i \langle \mathbf{p}_i \cdot (\mathbf{x}_i\times\mathbf{p}_i) \rangle_\Gamma = 0$, where $\beta$ indexes Cartesian directions. In contrast, the linear $\mathbf{p}_\Gamma$ and angular $\mathbf{L}_\Gamma$ momenta appearing in Eqs. \eqref{Cdm2} and \eqref{Phi_sca_plus_minus_3} arise from the distinct effective electric $\mathbf{d}_\Gamma$ and magnetic $\mathbf{m}_\Gamma$ dipolar degrees of freedom, enabling $\mathbf{p}^\perp_\Gamma \cdot\mathbf{L}_\Gamma^{\perp *}\neq0$.

\begin{figure}
\centering
\includegraphics[width=0.5\textwidth]{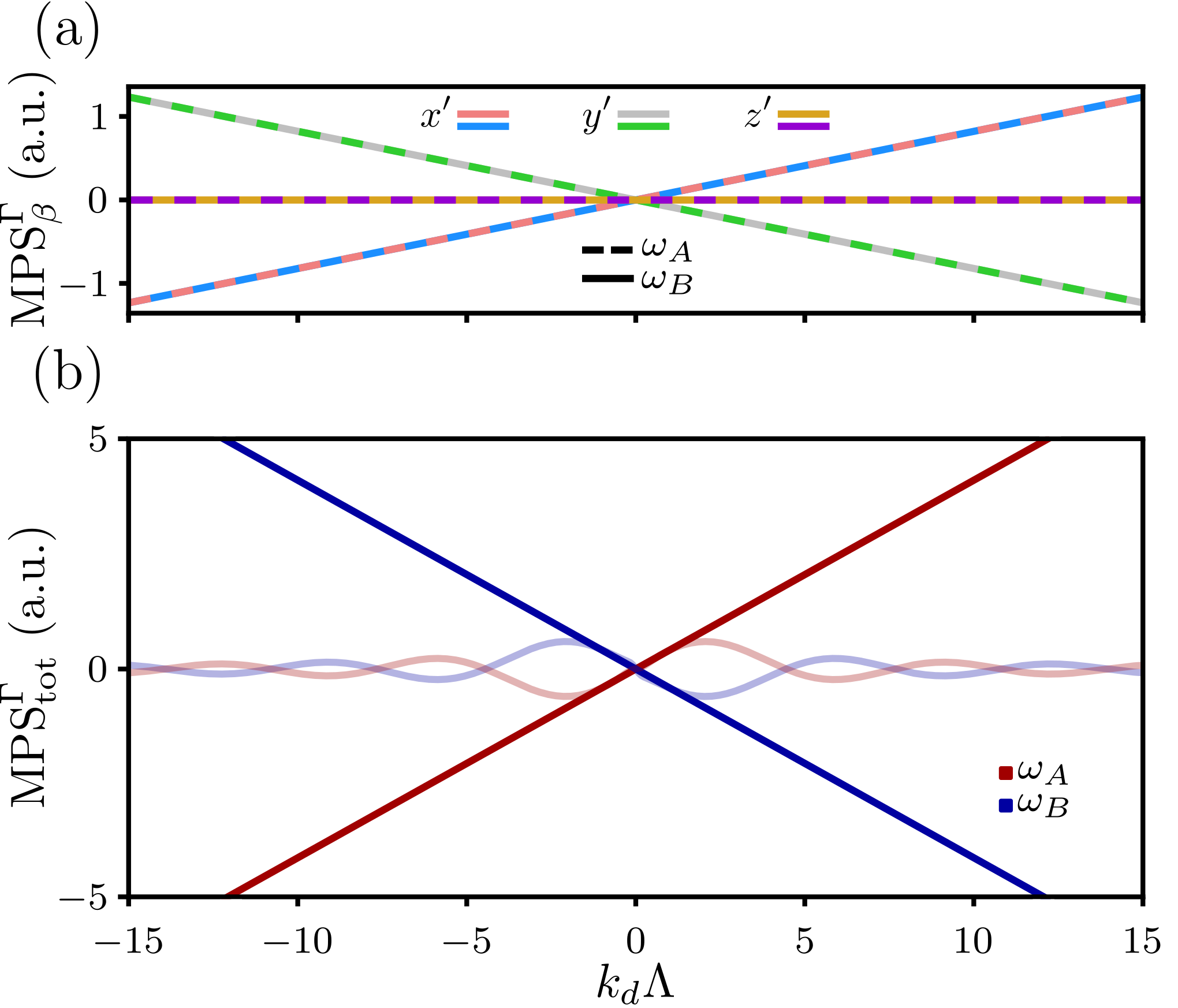}
\caption{Momentum pseudoscalar chirality metrics for the BK system. (a) $\text{MPS}^\Gamma_\beta$ and (b) $\text{MPS}^\Gamma_\text{tot}$ chirality metrics for the BK structure eigenmodes as a function of $k_d \Lambda$. Scaled $\langle \Phi_\text{sca}\rangle_\Gamma$ traces from Fig. \ref{fig3}(b) are included in the background of panel (b) for comparison. 
} 
\label{fig_mps}
\end{figure}
The $\text{MPS}^\Gamma_\beta$ axial chirality metrics for directions $\beta = x', y',z'$ are presented in Fig. \ref{fig_mps}(a) for the $\boldsymbol{\xi}_\Gamma$ eigenmodes of the BK system as a function of $k_d \Lambda$, where $i=1,2$ indexes the two nanorods. The primed coordinate system is selected such that $y'$ is oriented along the vertical arrow in \ref{fig1}(a) and the $z'$ direction is oriented along $\hat{\bf z}$. Although these metrics certainly succeed in capturing some aspect of excitational chirality (except for $\beta = z'$), for a given choice of $\beta$ and $\Lambda$, the MPS metrics for the two BK eigenmodes have the same sign, which is qualitatively different from the behavior of the $\langle \Phi_\text{sca}\rangle_\Gamma$ optical chirality metric shown in Fig. \ref{fig3}(b). However, the axial MPS introduced in Ref. \cite{Abraham2024} is neither derived nor unique, and alternative momentum pseudoscalar metrics can be defined \cite{coles2012chirality,sidorova2021quantitative,RevModPhys.71.1745}. In particular, it can be seen from Eq. \eqref{phi_scaig} that when $k_d \Lambda \ll 1$,
\begin{equation}
\begin{split}
\langle \Phi_{\text{sca}} \rangle_\Gamma &= -\frac{c k_\Gamma^6}{3}\text{Re}\big[d^\Gamma_1\Lambda(d^\Gamma_2)^*\big]\\
&=-\frac{c k_\Gamma^6}{3}\Big(\frac{e}{m\omega}\Big)^2 \text{Re} \Big\{ {\bf p}_\text{tot}\cdot {\bf L}_\text{tot}^* \Big\} \\
&=-\frac{c k_\Gamma^6}{3}\Big(\frac{e}{m\omega}\Big)^2 \text{MPS}^\Gamma_\text{tot}, \\
\end{split}
\label{Phi_MPS}
\end{equation}
where $\text{MPS}^\Gamma_\text{tot} = \langle  {\bf p}_\text{tot} \cdot {\bf L}_\text{tot} \rangle_\Gamma=\sum_{ij}\langle{\bf p}_i\cdot{\bf L}_j\rangle_\Gamma$ is a new momentum pseudoscalar expressed in terms of the total linear $ {\bf p}_\text{tot} = \sum_{i} {\bf p}_i$ and angular  $ {\bf L}_\text{tot} = \sum_{j} {\bf L}_j$ momenta of the two BK nanorods with $i,j=1,2$. Notably, the $\text{MPS}^\Gamma_\text{tot}$ metric (i) is still a pseudoscalar, (ii) removes the necessity of selecting any particular axis $\beta$, and (iii) yields a material excitational chirality metric with qualitative behavior identical to that of the optical chirality metric $\langle \Phi_\text{sca}\rangle_\Gamma$ derived from Maxwell's equations for $k_d \Lambda \ll 1$. The $\text{MPS}^\Gamma_\text{tot}$ metric is presented in Fig. \ref{fig_mps}(b) for the $\boldsymbol{\xi}_\Gamma$ eigenmodes of the BK system as a function of $k_d \Lambda$. When $|k_d \Lambda|$ is not much less than unity, the approximation $j_1(k_\Gamma \Lambda) \approx (1/3) k_\Gamma \Lambda$ underlying the simple connection between $\langle \Phi_{\text{sca}}\rangle_\Gamma$ and $\text{MPS}^\Gamma_\text{tot}$ in Eq. \eqref{Phi_MPS} is lost. 

\section{The Born-Kuhn Structure Under Drive}
\label{BKdrive}

\subsection{Optical Power and Circular Dichroism Spectra}
\label{CD_section}

Despite the intrinsic geometric, excitational, and optical chirality of the BK system's structure, eigenmodes, and eigenfields, characterization of optical activity in chiral molecular and nanoscale systems is often achieved through measurement of circular dichroism (CD) \cite{Warning2021,BK,Lourenco-MartinsHugo2021Opai,Zhiyuan2010}. Like any observable, CD requires the use of a probe to examine the intrinsic eigenspectrum of a material and is defined as the differential extinguished power of left- $(+)$ and right-handed $(-)$ circularly polarized light (LCP/RCP), denoted by $\hat{\bm\epsilon}_0=\hat{\bm\epsilon}_\pm=(\hat{\bf x}\pm i\hat{\bf y})/\sqrt{2}$. Specifically,
\begin{equation}
    \text{CD}(\omega) =  \frac{\langle P_\text{ext} \rangle_+ - \langle P_\text{ext} \rangle_-}{(1/2)[\langle P_\text{ext} \rangle_+ + \langle P_\text{ext} \rangle_-]},
    \label{CD_eq}
\end{equation}
where $\langle P_\text{ext} \rangle_\pm$ is the time-averaged extinction power measured under LCP/RCP excitation. However, according to Barron's dynamic chirality definitions, circularly-polarized light is also truly chiral---parity interconverts the two circularly polarized optical enantiomers, while time reversal does not. Thus, the CD signal convolves the excitational (i.e., optical) chirality of the probing LCP and RCP fields with the excitational chirality of the material target's eigenmodes, and must therefore be interpreted carefully when applied to nanoscale systems where the effects of retardation become important.

Optical driving of the BK dimer can be included by adding the harmonic driving field ${\bf E}_0({\bf x},t)={\bf E}_0({\bf x})e^{-i\omega t}$ to the right hand side of Eq. \eqref{amat}, yielding the time-harmonic amplitudes
\begin{equation}
    \begin{aligned}
d_1&=\alpha_1(\omega)[{\bf E}_0({\bf r}_1)\cdot\hat{\bf{n}}_1+\hat{\bf{n}}_1 \cdot \tensor{\bf{G}}(\omega) \cdot \hat{\bf{n}}_2d_2]\\
d_2&=\alpha_2(\omega)[{\bf E}_0({\bf r}_2)\cdot\hat{\bf{n}}_2+\hat{\bf{n}}_2 \cdot \tensor{\bf{G}}(\omega) \cdot \hat{\bf{n}}_1d_1],
 \end{aligned}
 \label{eom}
\end{equation}
where ${\bf E}_0({\bf x})=\hat{\bm\epsilon}_0E_0 e^{ikz}.$ These coupled equations of motion govern the steady-state dynamics of the driven BK dipoles and are equivalent to a coupled pair of damped and driven harmonic oscillators. From the resulting dipole dynamics together with Poynting’s theorem $-\dot u=\nabla\cdot{\bf S}+{\bf E}\cdot{\bf J}$, the total time-averaged extinction, absorption, and scattering powers within some volume $V$ can be expressed in terms of the incident and scattered optical fields \cite{Jackson2003,Bohren1998}. Specifically, the absorption and scattering powers are $\langle P_{\textrm{abs}}\rangle\equiv-\int d{\bf x}\nabla\cdot\langle{\bf S}\rangle=\int d{\bf x}\langle{\bf E}\cdot{\bf J}\rangle$ and $\langle P_{\textrm{sca}}\rangle\equiv\int d{\bf x}\nabla\cdot\langle{\bf S}_{\textrm{sca}}\rangle= -\int d{\bf x}\langle{\bf E}_\text{s}\cdot{\bf J}\rangle,$ respectively, since the time average $\langle\dot{u}\rangle$ is zero. Here, ${\bf S}=(c/4\pi){\bf E}\times{\bf B}$ and ${\bf S}_{\textrm{sca}}=(c/4\pi){\bf E}_{\textrm{s}}\times{\bf B}_{\textrm{s}}$ are the total and scattered energy fluxes, where $\mathbf{E}=\mathbf{E}_0+\mathbf{E}_\text{s}$ is the total field composed of both incident $\mathbf{E}_0$ and scattered $\mathbf{E}_\text{s}$ components. Definition of the components of the magnetic field $\bf B$ follow in the same manner. Together, these time-averaged powers satisfy the energy conservation equation $\langle P_\text{ext}\rangle=\langle P_\text{abs}\rangle+\langle P_\text{sca}\rangle = \int d{\bf x} \langle{\bf E}_0 \cdot {\bf J}\rangle$ within volume $V$.

\begin{figure*}
    \centering
\includegraphics[width=0.5\textwidth]{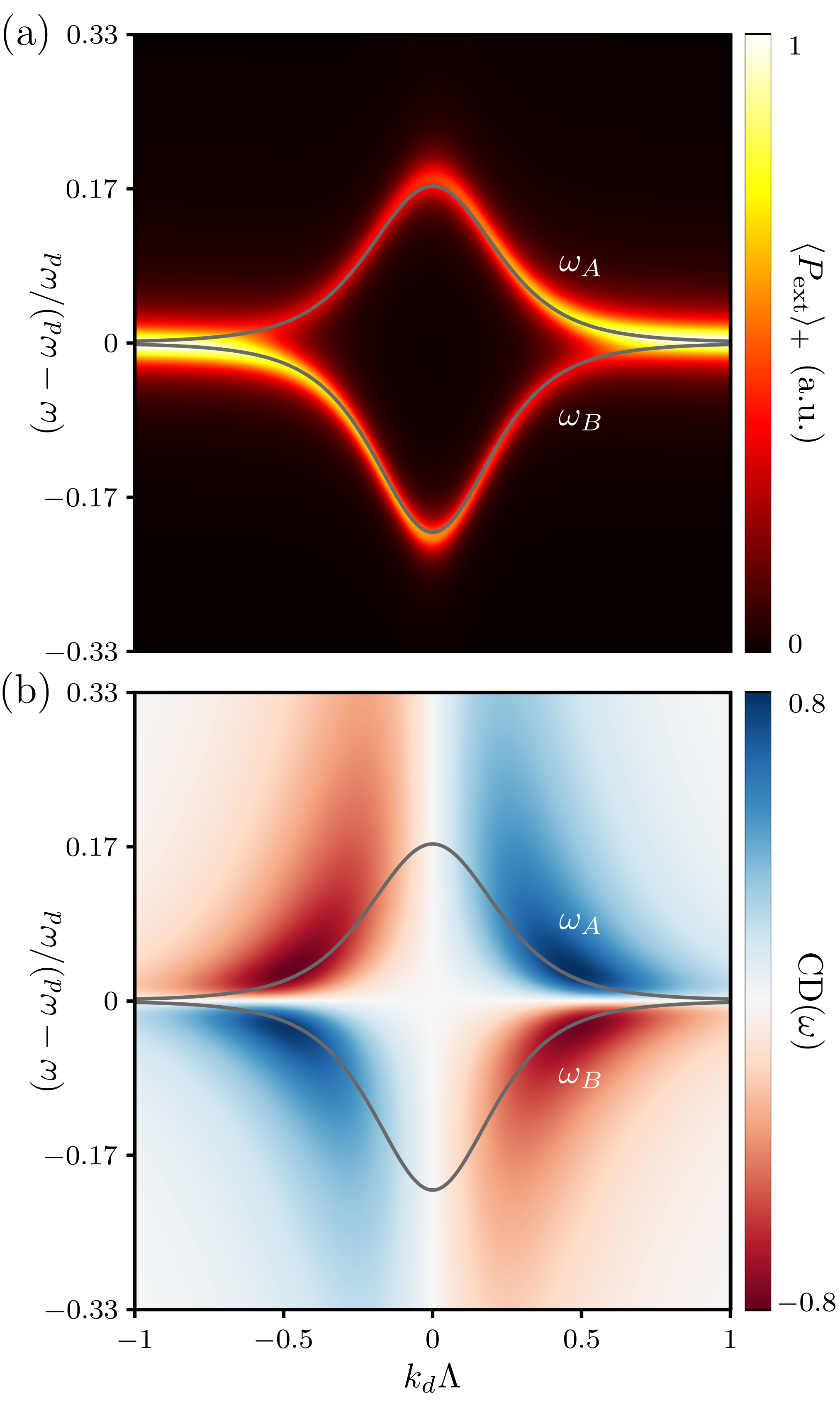}
\caption{Extinction power and circular dichroism spectra of the BK system. (a) Extinction power under LCP ($+$) plane wave drive (Eq. \eqref{powerpm}) and (b) circular dichroism (Eq. \eqref{CD_BK}) versus spectral detuning and inter-rod separation distance within the near-field. The $\omega_\Gamma$ eigenfrequencies are indicated by the gray traces.} 
\label{f33}
\end{figure*}
From these relationships, an exact expression for the optical CD spectrum can be obtained for the BK system. Specifically, by combining the time-averaged extinction powers under $\pm$-circularly polarized drive, 
\begin{equation}
\begin{split}
\langle P_\text{ext} \rangle_\pm &=\frac{\omega}{2}\sum_i\textrm{Im}[(\hat{\bm\epsilon}_\pm E_0({\bf r}_i))^*\cdot{\bf d}_i]\\
&=\frac{\omega}{4} \left|E_0\right|^2  \text{Im} \left[ \frac{\alpha_1 +\alpha_2}{1 - g^2 \alpha_1 \alpha_2}  \pm   \frac{2g\alpha_1\alpha_2 }{1 - g^2 \alpha_1 \alpha_2}\sin(k\Lambda) \right]
\end{split}
\label{powerpm}
\end{equation}
the CD spectrum acquires the compact form
\begin{equation}
    \textrm{CD}(\omega) = \frac{\text{Im} \left[ \frac{4 g \alpha_1 \alpha_2}{1 - g^2 \alpha_1 \alpha_2} \right]}
    {\text{Im} \left[ \frac{\alpha_1 + \alpha_2}{1 - g^2 \alpha_1 \alpha_2} \right]}
    \sin( k \Lambda).
    \label{CD_BK}
\end{equation}
This widely used reporter of optical activity measures the difference in extinction between LCP and RCP light, and is commonly applied to investigate the chirality of molecular targets. Fig. \ref{f33} plots the inter-rod separation and spectral detuning dependence of (a) the power extinguished under LCP ($+$) plane wave drive, and (b) the circular dichroism at inter-rod separation distances within the near-field. It is evident that $\langle P_\text{ext} \rangle_+$ in panel (a) closely tracks the $\omega_\Gamma$ eigenfrequencies (gray traces), while the CD spectrum (panel (b)) exhibits a doubly-odd behavior within the near field and falls to zero rapidly in the intermediate-field and beyond. We note, however, that while CD is widely used as a reporter of optical actvity, it is typically not derived from fundamental governing principles, whereas the optical chirality metrics introduced above are fundamentally established from Maxwell's equations. In the next section, we identify the relationship between CD and the extinction of optical chirality.

\subsection{Chiral Extinction, Absorption, and Scattering Cross Sections}
\label{chiralCS}
In analogy to the extinction, absorption, and scattering powers derived from Poynting's theorem for the conservation of electromagnetic energy in the presence of optical driving, the extinction, absorption, and scattering of optical chirality by a driven system can likewise be derived from the continuity of chirality expression in Eq. \eqref{chi_conservation}. Specifically, the total time-averaged extinguished, absorbed, and scattered optical chiralities within a volume $V$ can be expressed as (see Appendix \ref{chi_ext_sca_abs})
\begin{equation}
\begin{aligned}
    \langle \Phi_{\text{ext}} \rangle
    &= \frac{\omega}{8\pi}\int r^2d\Omega\,\hat{\bf n}\cdot\textrm{Im}[{\bf E}_0\times{\bf E}_{\textrm{s}}^*+{\bf B}_0\times{\bf B}_{\textrm{s}}^*]\\
    \langle \Phi_{\text{abs}} \rangle &= \frac{\omega}{16\pi}\int r^2d\Omega\,\hat{\bf n}\cdot\textrm{Im}[{\bf E}\times{\bf E}^*+{\bf B}\times{\bf B}^*]\\
    \langle \Phi_{\text{sca}} \rangle &= -\frac{\omega}{16\pi}\int r^2d\Omega\,\hat{\bf n}\cdot\textrm{Im}[{\bf E}_{\textrm{s}}\times{\bf E}_{\textrm{s}}^*+{\bf B}_{\textrm{s}}\times{\bf B}_{\textrm{s}}^*],
\end{aligned}
\label{Phi1}
\end{equation}
or equivalently
\begin{equation}
\begin{aligned}
    \langle \Phi_{\text{ext}} \rangle
    &= \frac{1}{2} \int d\mathbf{x} \, \text{Re} \left[ \mathbf{E}_{0} \cdot (\nabla \times \mathbf{J}^*) \right]\\
    \langle \Phi_{\text{abs}} \rangle &= \frac{1}{2} \int d\mathbf{x} \, \text{Re} \left[ \mathbf{E} \cdot (\nabla \times \mathbf{J}^*) \right]\\
    \langle \Phi_{\text{sca}} \rangle &= -\frac{1}{2} \int d\mathbf{x} \, \text{Re} \left[ \mathbf{E}_{\text{s}} \cdot (\nabla \times \mathbf{J}^*) \right],
\end{aligned}
\label{Phi2}
\end{equation}
Here, $\langle \Phi_{\text{abs}}\rangle\equiv-\int d{\bf x}\nabla\cdot\langle\boldsymbol{\varphi}\rangle=-\int d{\bf x}\langle\mathcal{J}\rangle$ and $\langle \Phi_{\text{sca}}\rangle\equiv\int d{\bf x}\nabla\cdot\langle\boldsymbol{\varphi}_{\textrm{sca}}\rangle=\int d{\bf x}\langle\mathcal{J}_{\textrm{sca}}\rangle$, where the total optical chirality flux and source are $\langle\bm{\varphi}\rangle =- (\omega/16\pi)\textrm{Im}[{\bf E}\times{\bf E}^*+{\bf B}\times{\bf B}^*]$ and $\langle{\cal J}\rangle=-(1/4)\textrm{Re}[{\bf E}\cdot\nabla\times{\bf J}^*+{\bf J}\cdot\nabla\times{\bf E}^*]$, respectively, and where the scattered optical chirality flux and source were defined previously in Sec. \ref{SEC_EigSca_OptChi}. In analogy to power, these chirality metrics satisfy the rule $\langle \Phi_\text{ext}\rangle = \langle \Phi_\text{abs}\rangle+\langle \Phi_\text{sca}\rangle$, indicating that the rate at which optical chirality is extinguished from the incident fields within some volume $V$ is equal to the total rate of optical chirality production via scattered fields generated by driven sources less the net rate at which optical chirality exits the volume through its boundary surface $S$.

Given the steady-state BK model dynamics specified in Eq. \eqref{eom}, analytic forms for the observables in Eqs. \eqref{Phi1}-\eqref{Phi2} can be obtained. For a $\hat{\bf z}$-propagating harmonic driving field of arbitrary polarization ${\hat{\boldsymbol{\epsilon}}_0}$, an explicit expression for the optical chirality extinction $\langle \Phi_{\text{ext}} \rangle_{\hat{\boldsymbol{\epsilon}}_0}$ is derived in Appendix \ref{BK_ext}. In the case of 45$^\circ$ polarization, i.e., $\hat{\bm\epsilon}_{0}=\hat{\bm\epsilon}_{\pi/4}$, the ratio of the time-averaged optical chirality extinction to the $\pm$-circularly polarized incident chiral flux \footnote{$\hat{\bm\epsilon}_{0}=\hat{\bm\epsilon}_{\pi/4}=(1/\sqrt{2})[\hat{\bf x}+\hat{\bf y}]=({1}/{2})[(1-i)\hat{\bm\epsilon}_{+}+(1+i)\hat{\bm\epsilon}_{-}]$ incident field polarization is chosen to provide equal forcing to each of the two nanorods composing the BK structure.}
\begin{widetext}
\begin{equation}
   {\igmas}_{\textrm{ext}}^{\pi/4}(\omega)\equiv\frac{ \langle \Phi_{\text{ext}} \rangle_{\pi/4}}{|\langle\bm{\varphi}_0\rangle|}
= \frac{2\pi\omega}{c}\text{Im} \left[ i  \bigg( \frac{\alpha_1 -\alpha_2}{1 - g^2 \alpha_1 \alpha_2}\bigg)   +   \frac{2g\alpha_1\alpha_2 }{1 - g^2 \alpha_1 \alpha_2}\sin(k\Lambda) \right],
\label{chiext}
\end{equation}
\end{widetext}
defines the {\it chiral extinction cross section} ${\igmas}_{\textrm{ext}}^{\pi/4}(\omega)$ in analogy to the way that the optical extinction cross section ${\sigma}_{\textrm{ext}}(\omega)=\langle P_{\text{ext}} \rangle/|\langle{\bf S}_0\rangle|$ is formed from ratio of the extinction power $\langle P_{\text{ext}} \rangle$ to the incident energy flux $\langle{\bf S}_0\rangle=(c/8\pi)\textrm{Re}[{\bf E}_0\times{\bf B}^*_0]$ after time-averaging. Up to (frequency-dependent) normalization, ${\igmas}_{\textrm{ext}}^{\pi/4}(\omega)$ is proportional to CD in Eq. \eqref{CD_BK} provided the two rods are identical, i.e., $\alpha_1(\omega)=\alpha_2(\omega),$ where $\langle\bm{\varphi}_0\rangle =- (\omega/16\pi)\textrm{Im}[{\bf E}_0\times{\bf E}^*_0+{\bf B}_0\times{\bf B}^*_0]=\pm(\omega/8\pi)|E_0|^2\hat{\bf{z}} = \pm(\omega/c)\langle{\bf{S}}_0\rangle$ is the incident chiral flux of $\pm$-circularly polarized light \cite{Bliokh2011}. Similar expressions for chiral scattering and absorption cross sections can be found by analogy. For example, under the same incident field polarization conditions, the {\it chiral scattering cross section} is
\begin{equation}
\begin{aligned}
   {\igmas}_{\textrm{sca}}^{\pi/4}(\omega)\equiv\frac{ \langle \Phi_{\text{sca}} \rangle_{\pi/4}}{|\langle\bm{\varphi}_0\rangle|}= -4\pi \Big(\frac{\omega}{c}\Big)^4 j_1(k\Lambda)\text{Re}\Bigg[\frac{\alpha_1\alpha_2^*e^{i k \Lambda}(1+g\alpha_2e^{-i k \Lambda})(1+g\alpha_1e^{i k \Lambda})^*}{|1 - g^2 \alpha_1 \alpha_2|^2}\Bigg],
   \label{phi_sca_drive}
\end{aligned}
\end{equation}
while the optical chirality absorption cross section is determined by the difference ${\igmas}_{\textrm{abs}}^{\pi/4}(\omega)={\igmas}_{\textrm{ext}}^{\pi/4}(\omega)-{\igmas}_{\textrm{sca}}^{\pi/4}(\omega).$ Extension of these cross sections to arbitrary incident polarization states can be derived from Eq. \eqref{chiext_general}. Nevertheless, Eqs. \eqref{Phi1} and \eqref{Phi2} are valid for arbitrary harmonic driving fields, offering a useful generalization to optical CD, which is not applicable when quantifying differential absorption and scattering rates involving large classes of experimentally relevant situations, e.g., near-field sources.       



\begin{figure*}
    \centering
\includegraphics[width=1\textwidth]{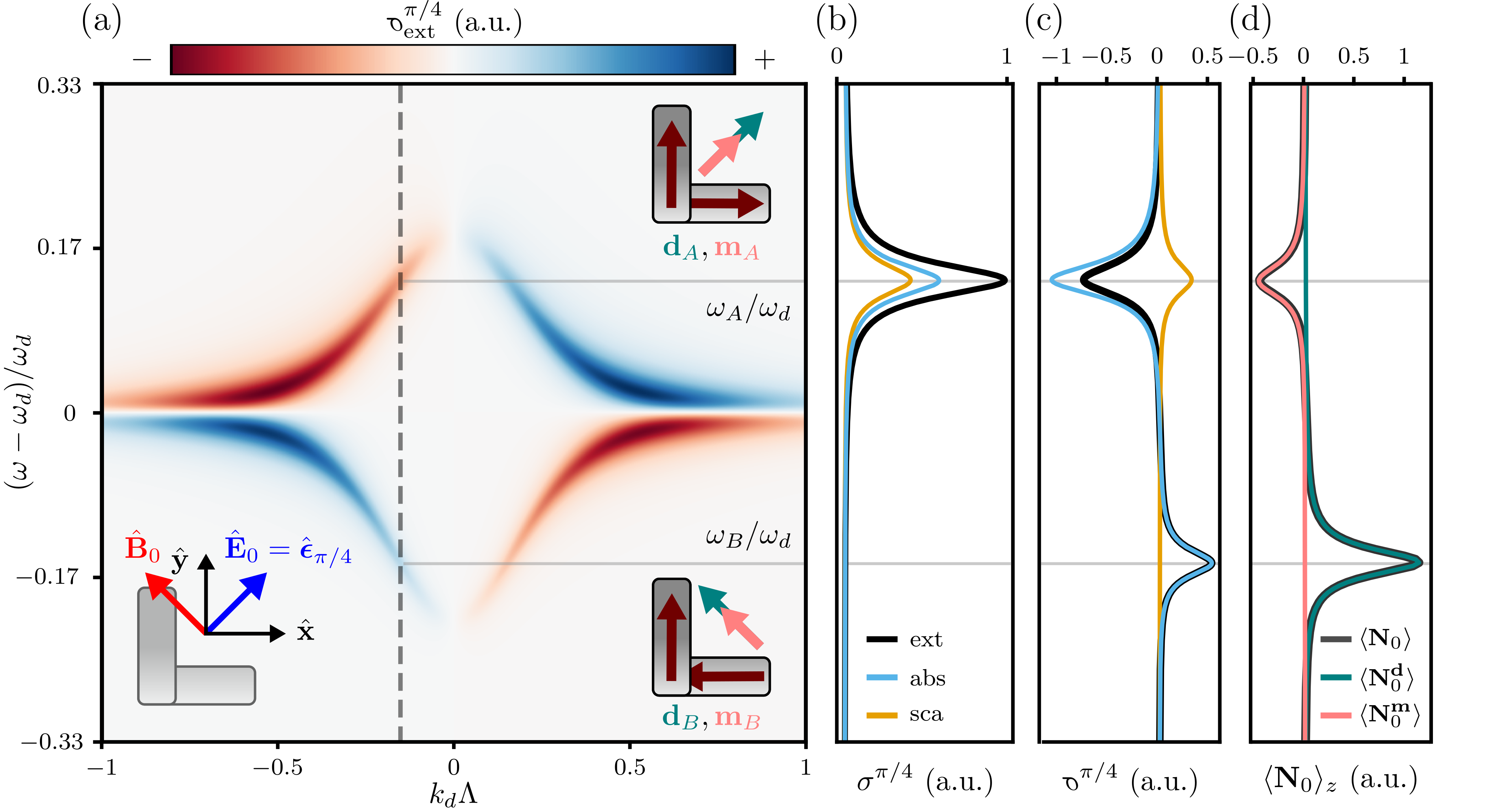}
\caption{Extinction, scattering, and absorption of optical chirality and power by L and D plasmonic BK structures. (a) Chiral extinction cross section ${\igmas}_{\textrm{ext}}^{\pi/4}$ under linearly polarized $\hat{\boldsymbol{\epsilon}}_0 = \hat{\boldsymbol{\epsilon}}_{\pi/4}$ excitation versus detuning $(\omega-\omega_d)/\omega_d$ and interdipole separation distance $k_d\Lambda$. The left inset depicts orientations of the incident plane wave electric ($\hat{\bf E}_0$) and magnetic ($\hat{\bf B}_0$) driving fields. Right insets show the effective co-located electric (${\bf d}_\Gamma$) and magnetic (${\bf m}_\Gamma$) dipoles associated with the $\bm{\xi}_\Gamma$ eigenmodes for the $\Lambda=-10$ nm BK enantiomer indicated by the dashed vertical line. (b) and (c) show extinction, absorption, and scattering spectra of optical power and chirality cross sections, respectively, at $\Lambda=-10$ nm. (d) Total external $\langle {\bf N}_0   \rangle$, electric $\langle {\bf N}_0^{\bf d}\rangle$, and magnetic $\langle {\bf N}_0^{\bf m}\rangle$ torques exerted upon the BK system. $\hat{\boldsymbol{\epsilon}}_0 = \hat{\boldsymbol{\epsilon}}_{\pi/4}$ in all panels.} 
\label{f3}
\end{figure*}
Fig. \ref{f3}(a) shows optical chiral extinction cross section ${\igmas}_{\textrm{ext}}^{\pi/4}$ spectra of the BK system driven by a $\hat{\bf z}$-propagating plane wave polarized along $\hat{\bm\epsilon}_{\pi/4}$ (see lower-left inset) as a function of $k_d\Lambda$. Since ${\igmas}_{\textrm{ext}}^{\pi/4}$ and CD have identical numerators, differing only in (frequency-dependent) normalization, ${\igmas}_{\textrm{ext}}^{\pi/4}$ also exhibits doubly odd behavior in both the detuning $(\omega-\omega_d)/\omega_d$ and the dimensionless nanorod-nanorod separation $k_d\Lambda$ within the near to intermediate field ($0\leq |k_d\Lambda|\leq1$), with ${\igmas}_{\textrm{ext}}^{\pi/4}\to0$ beyond. This behavior of optical chirality extinction implies that CD similarly grows increasingly insensitive to both the geometric chirality and excitational eigenmode chirality of the BK structure when the two nanorods are separated by distances beyond the realm of validity of the quasistatic limit, i.e., beyond the near-field. 


Normalized extinction, absorption, and scattering components of the optical power $\sigma^{\pi/4}$ and the optical chirality ${\igmas}^{\pi/4}$ cross section spectra are presented in Fig. \ref{f3}(b) and Fig. \ref{f3}(c), respectively, for the $\Lambda=-10$ nm BK enantiomer, well within the near-field coupling regime. The insets on the right side of Fig. \ref{f3}(a) depict the orientations of the effective (co-located) electric (${\bf d}_\Gamma$) and magnetic (${\bf m}_\Gamma$) dipole moments associated with the $\bm{\xi}_\Gamma$ eigenmodes of the $\Lambda=-10$ nm BK enantiomer, which is indicated by the dashed vertical line. In the present case, effective multipole moments are obtained through a vector spherical harmonic expansion of the ${\bf E}_\Gamma$ eigenfields as described in Appendix \ref{Multipoles}. The optical power spectra in Fig. \ref{f3}(b) exhibit pronounced responses at the $\omega_A$ mode energy due to strong overlap of ${\bf d}_A$ with ${\bf E}_{0}$, while responses at $\omega_B$ are significantly reduced due to near orthogonality between ${\bf E}_0$ and the ${\bf d}_B$ mode polarization (see insets in panel (a)). On the other hand, ${\bf m}_A\cdot\hat{\bf B}_0=0$ at $\omega_A$, and ${\bf m}_B\cdot\hat{\bf B}_0\neq0$ at $\omega_B$.

Meanwhile, optical chirality is extinguished at both $\bm{\xi}_\Gamma$ mode energies, which can be rationalized by considering the optical torque applied to the BK system by the incident electromagnetic field. The connection between optical torque and optical chirality extinction can be made explicit by leveraging the extension of the optical theorem to optical chirality \cite{Nieto-Vesperinas_2015}. Specifically, under plane wave drive of arbitrary polarization ${\hat{\boldsymbol{\epsilon}}_0}$, Appendix \ref{optical_thm} shows that the optical chirality extinction defined in Eqs. \eqref{Phi1} or \eqref{Phi2} can be expressed as
\begin{equation}
    \langle \Phi_{\text{ext}} \rangle_{\hat{\boldsymbol{\epsilon}}_0} = \frac{\omega}{\hbar}\mathbf{P}_0\cdot\langle{\bf{N}}_0\rangle_{\hat{\boldsymbol{\epsilon}}_0},
    \label{CE}
\end{equation}
where $\langle{\bf{N}}_0\rangle_{\hat{\boldsymbol{\epsilon}}_0}$ is the time-averaged optical torque applied by the incident electromagnetic field with linear momentum $\mathbf{P}_0=\hbar {\bf k}_0=\hbar k\hat{\bf z}$. When ${\bf E}_0$ is a single plane wave with no orbital angular momentum, $\langle {\bf N}_0\rangle_{\hat{\boldsymbol{\epsilon}}_0}$ is the torque associated with changing optical spin angular momentum, providing a simple physical interpretation of Eq. \eqref{CE} \cite{Nieto-Vesperinas_15}.  The total time-averaged applied optical torque $\langle {\bf N}_0\rangle_{\hat{\boldsymbol{\epsilon}}_0}$ experienced by the $\Lambda=-10$ nm BK structure is presented in Fig. \ref{f3}(d) along with its decomposition into contributions arising from the effective BK electric and magnetic dipoles, i.e., 
\begin{equation}
         \langle {\bf N}_0   \rangle _{\hat{\boldsymbol{\epsilon}}_0}= \frac{1}{2} \textrm{Re} (\mathbf{d} \times \mathbf{E}_0^*) + \frac{1}{2} \textrm{Re} (\mathbf{m} \times \mathbf{B}_0^*).
\label{torque_mech}
\end{equation}
Notably, at $\omega_A,$ the $\bm{\xi}_A$ eigenmode experiences an external magnetic torque of $\langle{\bf N}_0^A\rangle_{\hat{\boldsymbol{\epsilon}}_0}= \langle{\bf N}^\mathbf{m}_0 \rangle_{\hat{\boldsymbol{\epsilon}}_0}= ({1}/{2}) \textrm{Re} (\mathbf{m} \times \mathbf{B}_0^*),$ while at $\omega_B,$ the $\bm{\xi}_B$ eigenmode experiences an opposing external electric torque $\langle{\bf N}_0^B\rangle_{\hat{\boldsymbol{\epsilon}}_0}= \langle{\bf N}^\mathbf{d}_0 \rangle_{\hat{\boldsymbol{\epsilon}}_0}= ({1}/{2}) \textrm{Re} (\mathbf{d} \times \mathbf{E}_0^*).$

Although $\langle \Phi_\text{ext} \rangle$ characterizes the rate at which optical chirality is extinguished from applied electromagnetic fields, it does not track the ultimate fate of transferred angular momentum.  Upon excitation, the BK system may dissipate optical chirality through a combination of scattering and/or absorption processes, subject to the constraint $\langle \Phi_\text{ext}\rangle = \langle \Phi_\text{abs}\rangle+\langle \Phi_\text{sca}\rangle$. The rate of dissipation into the radiative channel is characterized by the generalization of the optical chirality scattering (Eq. \eqref{Phi_sca_plus_minus_3}) to include drive. In practice, this amounts to the replacement of eigenfields in Eq. \eqref{Phi_sca_plus_minus_3} by their driven counterparts, which in general will be a weighted superposition of eigenmode fields. Specifically, 
\begin{equation}
\begin{aligned}
\langle \Phi_{\text{sca}}\rangle_{\hat{\boldsymbol{\epsilon}}_0} &\equiv \int r^2d\Omega\,\hat{\bf n}\cdot\langle {\bm\varphi}_{\textrm{sca}}\rangle_{\hat{\boldsymbol{\epsilon}}_0}\\
&=\frac{3\omega}{8\pi}\int d\Omega\,{\bf k}\cdot\textrm{Im}\Big[\frac{k^3}{3}{\bf d}\times{\bf d}^*+\frac{k^3}{3}{\bf m}\times{\bf m}^*\Big]+\frac{\omega k^4}{8\pi}\int d\Omega\,\textrm{Im}\Big[\mathbf{d}^\perp \cdot \mathbf{m}^{\perp *}\Big]\\
&=\frac{3 k^2}{32\pi m} \gamma_\text{rad}(\omega) \int d\Omega \,\hat{\bf n}\cdot\text{Re} \bigg\{ \hat{\bf n} \Big(\mathbf{p}^\perp \cdot \mathbf{L}^{\perp *}\Big) \bigg\},
\end{aligned}
\label{Phi_sca_drive}
\end{equation}
demonstrating that the radiated intrinsic torque $\langle{\bf N}_{\textrm{int}}\rangle=\text{Im}[({k^3}/{3}){\bf d}\times{\bf d}^*+({k^3}/{3}){\bf m}\times{\bf m}^*]$ makes no contribution to the angle-averaged scattered optical chirality, leaving only the canonical spin angular momentum to be dissipated back into the far-field as discussed in Section \ref{SEC_EigSca_OptChi}. 

Electromagnetic reciprocity explains why the $\bm{\xi}_A$ eigenmode at $\omega_A$, which couples directly to the incident $E_0 \hat{\boldsymbol{\epsilon}}_{\pi/4}$ plane wave drive, exhibits appreciable optical chiral scattering in Fig. \ref{f3}(c), whereas the $\bm{\xi}_B$ eigenmode at $\omega_B$ does not. The difference ${\igmas}^{\pi/4}_\text{abs} = {\igmas}^{\pi/4}_\text{ext} - {\igmas}^{\pi/4}_\text{sca} $ characterizes the rate at which (canonical spin) angular momentum is transferred to nonradiative degrees of freedom of the BK structure, including the angular momentum associated with the charge oscillations underlying the BK eigenmodes. We note, e.g., that in molecular systems that are free to rotate, this transfer of angular momentum would also induce molecular mechanical rotational motion, which is not a dynamical degree of freedom in the present model.

\section{Conservation of optical chirality in light-matter interactions: The BK system as an optical wave plate}
\label{QWP}
\begin{figure*}
    \centering
\includegraphics[width=1\textwidth]{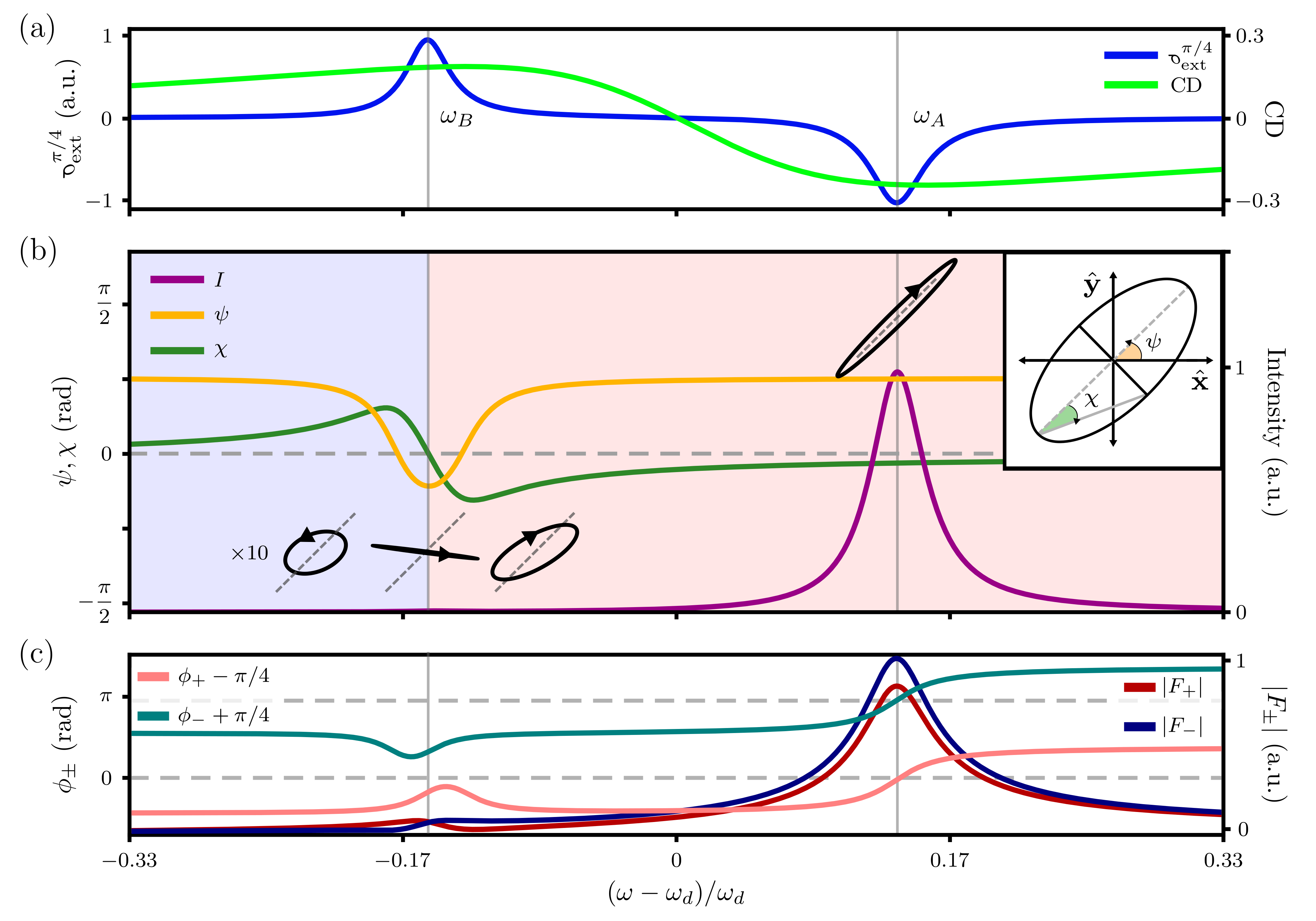}
\caption{Rotation and elliptic polarization of light scattered from the BK system. (a) Spectral comparison of chiral extinction ${\igmas}_{\textrm{ext}}^{\pi/4}$ under 45$^\circ$-polarized plane wave excitation versus CD. (b) Spectra of Stokes parameters (see inset) corresponding to the polarization ellipses of the forward scattered electric field show changes in both ellipticity and/or polarization angle at the $\omega_B$ and $\omega_A$ eigenmode frequencies. Blue ($+$) and red ($-$) regions specify the $\pm$-handedness of the forward scattered electric field. (c) Amplitudes ($F_\pm$) and phases ($\phi_\pm$) of the left- and right-circularly polarized components of the forward scattering amplitude contributing to the chiral extinction cross section. $\hat{\boldsymbol{\epsilon}}_0 = \hat{\boldsymbol{\epsilon}}_{\pi/4}$ in all panels (indicated by dashed gray lines within the polarization ellipses).}
\label{fig4}
\end{figure*}
Having examined the excitational chirality of the BK model system in the absence and presence of external driving, we lastly turn to a driven example where the BK dimer serves as an optical element to convert incident linearly polarized light into a rotated, elliptically polarized state, conserving optical chirality between field and matter throughout the interaction process \cite{LeeHye-Eun2018Aaps}. Fig. \ref{fig4}(a) compares the optical chirality extinction cross section spectrum ${\igmas}_{\textrm{ext}}^{\pi/4}(\omega)$ under $\hat{\bm\epsilon}_{\pi/4}$-polarized incident light to the $\textrm{CD}(\omega)$ spectrum for the $\Lambda=-10$ nm BK structure. Both spectra can be interpreted from generalization of the optical theorem whereby optical chirality extinction \cite{Nieto-Vesperinas_2015}
\begin{equation}
    \langle \Phi_{\text{ext}} \rangle_{\hat{\boldsymbol{\epsilon}}_0}
     =\frac{c}{2k} \text{Re} \Big[\hat{\bm{\epsilon}}_0 E_0\cdot\Big({\bf k}_0\times\mathbf{F}^*({\bf k}_0,{\bf k}_0)\Big)\Big]
 \label{chiral_optical_thm}
\end{equation}
results from the interference between the forward-scattered far-field ${\bf E}_{\textrm{s}}({\bf x})\sim\mathbf{F}(\mathbf{k}_0,\mathbf{k}_0)e^{ikr}/r$ and the transmitted incident field $\hat{\bm{\epsilon}}_0E_0$, where $\mathbf{k}_0=\hat{\bf z}k=\hat{\bf z}(\omega/c)$ is the incident wave vector and 
\begin{equation}
   \mathbf{F}(\mathbf{k},\mathbf{k}_0) = \frac{1}{4\pi i}\mathbf{k}\times\int r^{\prime 2}d\Omega' e^{-i{\bf k}\cdot{\bf x}'}\Big[\frac{\mathbf{k}\times(\hat{\mathbf{n}}'\times\bf{B}_\text{s}({\bf x}'))}{k}-\hat{\mathbf{n}}'\times \mathbf{E}_\text{s}({\bf x}')\Big],
   \label{kirchoff0}
\end{equation}
is the scattering amplitude; see Appendix \ref{optical_thm}. Like the optical theorem for the extinction of power \cite{Jackson2003}, Eq. \eqref{chiral_optical_thm} expressing the optical theorem for optical chirality is an exact result. For the reasons described in Fig. \ref{f33} above, ${\igmas}_{\textrm{ext}}^{\pi/4}(\omega)$ and $\textrm{CD}(\omega)$ both change sign with frequency in the same manner. Their quantitative difference in appearance lies in the different frequency-dependent factors that multiply $\text{Im} \left[{2g\alpha_1\alpha_2 }/({1 - g^2 \alpha_1 \alpha_2})\right]\sin(k\Lambda)$ between Eqs. \eqref{CD_eq} and \eqref{chiext}, where $\alpha_1=\alpha_2.$

Fig. \ref{fig4}(b) illustrates the spectral dependence of the forward scattered electric field polarization state characterized in terms of the Stokes intensity, orientation angle, and ellipticity 
\begin{equation}
\begin{split}
I&=|\hat{\bm{\epsilon}}^*_+\cdot{\bf F}|^2+|\hat{\bm{\epsilon}}^*_-\cdot{\bf F}|^2,\\
\psi &= \frac{1}{2}\tan^{-1}\frac{\textrm{Im}\{(\hat{\bm{\epsilon}}^*_+\cdot{\bf F})^*(\hat{\bm{\epsilon}}^*_-\cdot{\bf F})\}}{\text{Re}\{(\hat{\bm{\epsilon}}^*_+\cdot{\bf F})^*(\hat{\bm{\epsilon}}^*_-\cdot{\bf F})\}},\\
\chi &= \frac{1}{2}\sin^{-1}\frac{|\hat{\bm{\epsilon}}^*_+\cdot{\bf F}|^2-|\hat{\bm{\epsilon}}^*_-\cdot{\bf F}|^2}{|\hat{\bm{\epsilon}}^*_+\cdot{\bf F}|^2+|\hat{\bm{\epsilon}}^*_-\cdot{\bf F}|^2},
\end{split}
\end{equation}
of its forward scattering amplitude ${\bf F }\equiv \mathbf{F}(\mathbf{k}_0,\mathbf{k}_0)$ polarization ellipse, where $\hat{\bm{\epsilon}}_-\times\hat{\bm{\epsilon}}_+=i\hat{\bf z}$ \cite{Jackson2003}. Spectral regions of negative (positive) helicity (black arrows) are indicated by the blue (red) background colors. At all frequencies, it is evident that the incident $\hat{\bm\epsilon}_{\pi/4}$-polarized light excites polarized plasmonic excitations in the BK structure that produce forward scattered electric fields that are rotated and/or elliptically polarized from the incident linear polarization state (dashed 45$^\circ$ gray lines), defined by $\psi_0=\pi/4$ and $\chi_0=0$. In the spectral vicinity of $\omega_B$, the field polarization changes helicity and ellipticity while also rotating significantly from the applied optical polarization direction. In contrast, at frequencies near $\omega_A$, the field polarization becomes elliptically polarized but does not rotate appreciably from the incident optical polarization axis.


The relationship between the forward scattering amplitude and the chiral extinction can be understood via the right- and left- circular polarization content $F_-=\hat{\bm{\epsilon}}^*_-\cdot\mathbf{F}(\mathbf{k}_0,\mathbf{k}_0)$ and $F_+=\hat{\bm{\epsilon}}^*_+\cdot\mathbf{F}(\mathbf{k}_0,\mathbf{k}_0)$ and phase parameters $\phi_-= \text{arg}(F_-)$ and $\phi_+= \text{arg}(F_+)$ characterizing the forward-scattered fields \cite{kramer2017analytic}. For $\hat{\bm\epsilon}_{\pi/4}$-polarized light, Eq. \eqref{chiral_optical_thm} can be written explicitly as
\begin{equation}
\langle\Phi_\text{ext}\rangle_{{\pi}/{4}}(\omega) = \frac{\sqrt{2}cE_0}{4}\Big[|F_-|\cos\Big(\phi_-+\frac{\pi}{4}\Big)+|F_+|\cos\Big(\phi_+-\frac{\pi}{4}\Big)\Big],
    \label{CP_ext}
\end{equation}
where we have taken the driving field amplitude $E_0$ to be real-valued without any loss of generality and the phase shifts in the cosine arguments are the $\pm$-circularly polarized phases $\phi_-^0=\pi/4$ and $\phi_+^0=-\pi/4$ of the incident light field. When expressed in this form, Eq. \eqref{CP_ext} simplifies considerably at the $\omega_B$ and $\omega_A$ frequencies, offering physical insights to interpret the chiral extinction spectrum that would be otherwise difficult to glean from Eq. \eqref{chiral_optical_thm}.



At $\omega_B$, it is apparent from Fig. \ref{fig4}(c) that $|F_+(\omega_B)| = |F_-(\omega_B)|\equiv|F|$, and $\phi_++\phi_- \approx\phi_+-\phi_- \approx  0$.  Under these conditions, Eq. \eqref{CP_ext} reduces to 
\begin{equation}
\langle\Phi_\text{ext}\rangle_{{\pi}/{4}}(\omega_B) \approx \frac{cE_0}{\sqrt{2}}|F|\cos\Big(\psi+\frac{\pi}{4}\Big),
\label{phiw-}
\end{equation} 
which makes explicit the dependence of the chiral extinction upon the scattered field polarization angle $\psi=(\phi_--\phi+)/2$ relative to the initial polarization angle $\psi_0=(\phi^0_--\phi^0_+)/2=\pi/4.$ Although the ellipticity $\chi=\chi_0$, $\psi\neq\psi_0$ at $\omega_B$. As a result, the scattered field is strongly rotated from $\pi/4,$ but remains linearly polarized. However, at frequencies that are slightly red/blue detuned from $\omega_B$, $\psi\approx\psi_0$ and $\chi\neq\chi_0$, causing the scattered field to become elliptically polarized but otherwise tracking the 45$^\circ$-polarization angle of the incident field.


The physical behavior of the scattered field polarization content at $\omega_A$ is quite different from $\omega_B$. Specifically, from
Fig. \ref{fig4}(c) it is evident that $\phi_+-\pi/4 = 0$ and $\phi_-+\pi/4 = \pi$. Thus Eq. \eqref{CP_ext} reduces to
\begin{equation}
\begin{split}
\langle\Phi_\text{ext}\rangle_{{\pi}/{4}}(\omega_A) &\approx \frac{cE_0}{2\sqrt{2}}[|F_+|-|F_-|]\\
&=\frac{cE_0}{2\sqrt{2}}\frac{I_0\sin(2\chi)}{|F_+|+|F_-|}.
\end{split}
\label{phiw+}
\end{equation}
When expressed in this form it is evident that chiral extinction tracks the ellipticity $\chi,$ which at $\omega_A$ is nonzero, while the rotation angle $\psi=\psi_0$ remains fixed at that of incident field. Thus, the forward scattered field becomes elliptically polarized but is otherwise unrotated. Additionally, the scaling of Eq. \eqref{phiw+} with Stokes intensity $I_0$ provides an enhancement of the forward scattered field at $\omega_B$ relative to $\omega_A.$

The different responses at $\omega_B$ relative to $\omega_A$ exposed in Eqs. \eqref{phiw-} and \eqref{phiw+} provide important physical insight to the chiral extinction spectrum presented in Fig. \ref{fig4}(a) and its differences from CD (Eq. \eqref{CD_BK}), despite their similarity in appearance and in mathematical form. Notably, when taken in reference to the incident field, the chiral extinction resonance feature at $\omega_B$ reports upon the rotation in scattered field polarization, while the negative resonance feature at $\omega_A$ results from changes in ellipticity of the scattered field. In contrast, the qualitatively similar positive and negative resonance features in CD reflect differences in extinction of left- versus right-handed circularly polarized light at both $\omega_B$ and $\omega_A$.

Analysis of the phase and amplitude of the BK scattered field also provides further insight to the associated extinction power presented previously in Fig. \ref{f3}(b). In this basis the optical theorem $\langle P_{\text{ext}} \rangle_{\hat{\bm{\epsilon}}_0}=(c/2k)\text{Im} [ (\hat{\bm{\epsilon}}_0E_0)^*\cdot\mathbf{F}({\bf k}_0,{\bf k}_0)]$ reduces to 
\begin{equation}
\langle P_\text{ext}\rangle_{{\pi}/{4}}(\omega) = \frac{cE_0}{2\sqrt{2}k}\Big[|F_-|\sin\Big(\phi_--\frac{\pi}{4}\Big)+|F_+|\sin\Big(\phi_++\frac{\pi}{4}\Big)\Big]
    \label{o_ext}
\end{equation}
when $\hat{\bm{\epsilon}}_0=\hat{\bm\epsilon}_{\pi/4}$. Again, based on Fig. \ref{fig4}(c), $|F_+(\omega_B)| = |F_-(\omega_B)|\equiv|F|$ and $\phi_++\phi_- \approx\phi_+-\phi_- \approx 0$ at $\omega_B$. As a result, we find that
\begin{equation}
\begin{split}
\langle P_\text{ext}\rangle_{\pi/4} (\omega_B)&\approx \frac{cE_0}{\sqrt{2}k}|F|\Big[\sin\Big(\frac{\phi_-+\phi_+}{2}\Big)\cos\Big(\frac{\phi_+-\phi_-}{2}+\frac{\pi}{4}\Big)\Big]\\
&=0,
\end{split}
\end{equation}
destroying the extinction power at $\omega_B$ due to cancellation of the phases (i.e., $\phi_-\approx-\phi_+$) of the forward scattered field. More specifically, the incident phase angle $\psi_0\approx\pi/4$ is rotated to $\psi\approx0,$ but otherwise no power is removed from the incident beam. Oppositely, at $\omega_A$, $\phi_+-\pi/4 = 0$ implies $\phi_++\pi/4 = \pi/2$ and $\phi_-+\pi/4 = \pi$ implies $\phi_--\pi/4 = \pi/2$. Thus, from Eq. \eqref{o_ext},
\begin{equation}
    \langle P_\text{ext}\rangle_{\pi/4}(\omega_A)  = \frac{cE_0}{2\sqrt{2}k} \Big[|F_+|+|F_-|\Big],
\end{equation}
which is nonzero and carries no phase information, precluding the option for any phase cancellation to modulate the power signal. It is important to note that despite these stark differences in physical behavior of the extinguished power at $\omega_A$ and $\omega_B,$ the chiral extinction spectrum in Fig. \ref{fig4}(a) is finite at both $\omega_A$ and $\omega_B$ eigenfrequencies and is a faithful reporter of optical rotational and ellipticity changes. As a consequence of optical chirality conservation in light-matter interactions, we see that the BK system acts as a wave plate, converting incident achiral linearly polarized light into rotated and elliptically polarized chiral light which evolves as a function of driving field.

\section{Conclusion}
\label{conc}
Intense research efforts are currently underway to leverage chiral light and matter degrees of freedom to realize new material functionalities and enable advanced technologies, e.g., optoelectronic \cite{PhysRevResearch.3.L042032} and photonic devices exploiting chiral quantum optics \cite{lodahl2017chiral} and chirality induced spin selectivity \cite{naaman2019chiral, aiello2022chirality, eckvahl2023direct}. Supporting the ongoing development of materials for chiral light control \cite{crassous2023materials}, and their incorporation into optical cavities \cite{ostovar2015through, aiello2022chirality, baranov2023toward, chen20242d}, this work examines the flow of optical chirality between coupled light and matter degrees of freedom in the presence and absence of external electromagnetic drive. Paired with an explicit analysis of the BK system as its geometry is deformed continuously from one handedness to the other, we highlight the distinction between the structural chirality of static objects and the dynamic chirality of excitations associated with a given Hamiltonian. In particular, the parity and time-reversal symmetries of the intrinsic excitational eigenmodes of a material are shown to determine those of their associated electromagnetic eigenfields as dictated by Maxwell's equations.


Beginning from the continuity equation for optical chirality in the absence of external driving, a unique pseudoscalar quantity $\langle \Phi_\text{sca} \rangle_\Gamma$ is derived, which we refer to as the radiated eigenfield chirality and interpret as a chirality metric quantifying the rate at which canonical optical spin angular momentum is radiated away from a localized collection of currents associated with matter eigenmodes indexed by $\Gamma$. For small inter-rod distances $k_d \Lambda$, the sign of the eigenfield chirality of the BK eigenmodes is odd in the separation parameter $\Lambda$, tracking the handedness of the underlying structure. When $|k_d \Lambda|\ll 1$ is not satisfied, however, the handedness of the radiated eigenfield chirality alternates and is thus distinct from the static chirality of the underlying structure. Here we initially assess the sourcing BK eigenmode chirality using the axial momentum pseudoscalar chirality metric $\text{MPS}^\Gamma_\beta=\sum_i\langle p_\beta^i L_\beta^i \rangle_\Gamma$ proposed recently in Ref. \cite{Abraham2024} within the context of molecular vibrational degrees of freedom. However, straightforward connection of $\text{MPS}^\Gamma_\beta$ to $\langle \Phi_\text{sca} \rangle_\Gamma$, or other relevant physical quantities, remains unclear. Instead, we show that in the case of the BK structure with $|k_d \Lambda|\ll 1$, the radiated optical chirality $\langle \Phi_\text{sca} \rangle_\Gamma$ becomes proportional to a related, but distinct, time-averaged pseudoscalar quantity $\text{MPS}^\Gamma_\text{tot} = \langle  {\bf p}_\text{tot} \cdot {\bf L}_\text{tot} \rangle_\Gamma=\sum_{ij}\langle  {\bf p}_{i} \cdot {\bf L}_{j} \rangle_\Gamma$. This excitational chirality measure is derived directly from Maxwell's equations and does not require ad hoc specification of an axial symmetry axis $\beta$, offering potential utility in characterizing the chirality of other collective excitations, e.g.,  vibrations in molecular \cite{Abraham2024, feng2025chiral,sidorova2021quantitative} and atomic crystal \cite{chen2021propagating, Ueda2023,ishito2023truly} systems, as well as plasmonic excitations supported by individual chiral nanostructures \cite{Lee2020,LeeHye-Eun2018Aaps, fan2012chiral,hentschel2017chiral,zhang2019unraveling,Jo_Ryu_Huh_Kim_Seo_Lee_Kwon_Lee_Nam_Kim_2024} and/or structured nanocavities \cite{cerdan2023chiral,acsphotonicsJuarez, baranov2023toward}.

In the presence of external electromagnetic driving, general expressions are presented to describe the time-averaged extinguished, absorbed, and radiated optical chirality within a given volume $V$ in terms of the incident, scattered, and total electric and magnetic fields, or electric fields and sourcing current densities. In the case of plane wave driving, we introduce optical chirality extinction, absorption, and scattering cross sections for arbitrary target systems, as well as specific analytic forms for the BK structure. Together these cross sections satisfy the sum rule ${\igmas}_{\textrm{ext}} = {\igmas}_{\textrm{abs}}+{\igmas}_{\textrm{sca}}$, arising from the conservation of optical chirality. While the functional form of optical chirality extinction is similar to CD, the former encodes information on both optical polarization and ellipticity changes, making it a more general metric of optical activity. Chirality extinction is shown to correspond to a removal of spin angular momentum from the incident field, inducing an optical torque on the structure. Optical chirality extinguished from the incident field is either radiated away or non-radiatively absorbed, as encoded in $\langle \Phi_\text{sca} \rangle$ and $\langle \Phi_\text{abs} \rangle$, respectively. Finally, by invoking the optical theorem for optical chirality, we analyze the chiral extinction cross section of the BK structure and its relation to the polarization state and ellipticity of forward scattered light as a function of the optical driving frequency. This work additionally lays the groundwork for further theoretical analysis and elucidation of recently developed experimental probes of chiral material excitations, including terahertz circular dischroism spectroscopy \cite{choi2022chiral, crassous2023materials} and application of structured optical field probes \cite{ShenYijie2019Ov3y}, as well as near-field probes offering spatial resolutions far below the optical diffraction limit, e.g., chiral extensions of THz s-SNOM nanoscopy \cite{hillenbrand2025visible}, or inelastic scattering of structured electron waves \cite{zanfrognini2019orbital, Lourenco-MartinsHugo2021Opai, BourgeoisMarcR.2022PEEG, bourgeois2023optical, nixon2024inelastic, tavabi2024symmetry, Bourgeois2025}.

\begin{acknowledgments}
This research was supported by the National Science Foundation under award Nos. NSF CHE-2403938 and NSF CHE-2516408.
\end{acknowledgments}
\appendix
\label{app}
\section{Excitational Symmetries of the Electromagnetic Eigenfields}
\label{SEC_Geo_Exc_Sym}
Like their sourcing $\bm{\xi}_\Gamma$ eigenmode electric dipole moments, the ${\bf E}_\Gamma$ eigenfields oscillate harmonically in time at the same eigenfrequencies $\omega_{\Gamma}$ and inherit symmetries from the same material Hamiltonian. To assess the excitational (i.e., optical) chirality of ${\bf E}_\Gamma$, consideration must again be made to the actions of parity and time-reversal. For example, in the case of the L enantiomer,
\begin{equation}
\begin{split}
\hat{\cal P}{\bf E}^{\textrm{L}}_{\Gamma}({\bf x})&=\int d{\bf x}'\Big[\hat{\cal P}\tensor{\bf{G}}({\bf x},{\bf x}',\omega)\hat{\cal P}^{-1}\Big]\cdot\hat{\cal P}\Big[{\bf d}_1^{\textrm{L}\Gamma}\delta({\bf x}'-{\bf r}_1)+{\bf d}_2^{\textrm{L}\Gamma}\delta({\bf x}'-{\bf r}_2)\Big]\\
&=\int d{\bf x}'\tensor{\bf{G}}({\bf x},{\bf x}',\omega)\cdot\Big[\Big(\hat{\cal R}{{\bf d}_1^{\textrm{D}\Gamma}}\Big)\delta(\hat{\cal P}^{-1}[{\bf x}'-{\bf r}_1])+\Big(\hat{\cal R}{{\bf d}_2^{\textrm{D}\Gamma}}\Big)\delta(\hat{\cal P}^{-1}[{\bf x}'-{\bf r}_2])\Big]\\
&=\int d{\bf x}'\tensor{\bf{G}}({\bf x},{\bf x}',\omega)\cdot\hat{\cal R}\Big[{\bf d}_1^{\textrm{D}\Gamma}\delta(-[{\bf x}'-{\bf r}_1])+{\bf d}_2^{\textrm{D}\Gamma}\delta(-[{\bf x}'-{\bf r}_2])\Big]\\
&=\int d{\bf x}'\Big[\hat{\cal R}\tensor{\bf{G}}({\bf x},{\bf x}',\omega)\hat{\cal R}^{-1}\Big]\cdot\hat{\cal R}\Big[{\bf d}_1^{\textrm{D}\Gamma}\delta({\bf x}'-{\bf r}_1)+{\bf d}_2^{\textrm{D}\Gamma}\delta({\bf x}'-{\bf r}_2)\Big]\\
&=\hat{\cal R}\int d{\bf x}'\tensor{\bf{G}}({\bf x},{\bf x}',\omega)\cdot\Big[{\bf d}_1^{\textrm{D}\Gamma}\delta({\bf x}'-{\bf r}_1)+{\bf d}_2^{\textrm{D}\Gamma}\delta({\bf x}'-{\bf r}_2)\Big]\\
&=\hat{\cal R}{\bf E}^{\textrm{D}}_{\Gamma}({\bf x})
\end{split}
\label{E_P}
\end{equation}
while
\begin{equation}
\begin{split}
\hat{\cal T}{\bf E}^{\textrm{L}}_{\Gamma}({\bf x})
&=\int d{\bf x}'\Big[\hat{\cal T}\tensor{\bf{G}}({\bf x},{\bf x}',\omega)\hat{\cal T}^{-1}\Big]\cdot\hat{\cal T}\Big[{\bf d}_1^{\textrm{L}\Gamma}\delta({\bf x}'-{\bf r}_1)+{\bf d}_2^{\textrm{L}\Gamma}\delta({\bf x}'-{\bf r}_2)\Big]\\
&=\int d{\bf x}'\tensor{\bf{G}}({\bf x},{\bf x}',\omega)\cdot\Big[\Big(e^{i\phi}{{\bf d}_1^{\textrm{L}\Gamma}}\Big)\delta({\bf x}'-{\bf r}_1)+\Big(e^{i\phi}{{\bf d}_2^{\textrm{L}\Gamma}}\Big)\delta({\bf x}'-{\bf r}_2)\Big]\\
&=\int d{\bf x}'\tensor{\bf{G}}({\bf x},{\bf x}',\omega)\cdot\Big[{{\bf d}_1^{\textrm{L}\Gamma}}\delta({\bf x}'-{\bf r}_1)+{{\bf d}_2^{\textrm{L}\Gamma}}\delta({\bf x}'-{\bf r}_2)\Big]e^{i\phi}\\
&={\bf E}^{\textrm{L}}_{\Gamma}({\bf x})e^{i\phi},
\end{split}
\label{E_TR}
\end{equation}
by invoking the symmetries of the dipole relay tensor $\tensor{\bf{G}}({\bf x},{\bf x}',\omega)$ under parity ($\hat{\cal P}$), proper rotation ($\hat{\cal R}$), and time-reversal ($\hat{\cal T}$), the invariance of the integration volume and measure under inversion and rotation, together with the parity and time symmetries of the $\bm{\xi}_\Gamma^{\textrm{L}}$ modes displayed in Fig. \ref{fig1}b. Here, $\phi$ is a constant phase angle equal to $\pi$ in the quasistatic limit corresponding to the polarization profiles displayed in Fig. \ref{fig1}(b). Similar analysis can be made for the parity and time symmetries of the magnetic eigenfields.

Thus, parity interconverts eigenfield enantiomers up to proper rotation while time-reversal does not, establishing the true chirality of the eigenfields sourced by truly chiral current densities. Identical conclusions are reached by repeating the above analysis in Eqs. \eqref{E_P} and \eqref{E_TR} beginning from the D enantiomer.

\section{Multipole Expansion of Electromagnetic Fields}
\label{Multipoles}
The scattered electromagnetic fields produced by any localized sources of charge and current can be expressed in the multipole basis, in terms of vector spherical harmonics $\mathbf{X}_{\ell m} ({\bf x})= ({-i}/{\sqrt{\ell(\ell+1)}} )\,\mathbf{x} \times \nabla Y_{\ell m}(\theta,\phi)$ combined with spherical Hankel functions $h_{\ell}^{(1)}(kr)$ satisfying the boundary conditions for outward propagating waves \cite{Jackson2003,Bohren1998}. Specifically,
\begin{equation}
    \begin{aligned}
        \mathbf{E}({\bf x}) &= \sum_{\ell m}\left[\frac{i}{k}a_{\ell m}{\nabla\times}\Big( h_{\ell}^{(1)}(kr)\mathbf{X}_{\ell m}({\bf x}) \Big)+b_{\ell m}h_{\ell}^{(1)}(kr)\mathbf {X}_{\ell m}({\bf x}) \right]\\
        \mathbf{B}({\bf x}) &= \sum_{\ell m}\left[a_{\ell m}h_{\ell}^{(1)}(kr)\mathbf {X}_{\ell m}({\bf x}) -\frac{i}{k}b_{\ell m}{\nabla\times} \Big(h_{\ell}^{(1)}(kr)\mathbf {X }_{\ell m}({\bf x}) \Big)\right],
        \label{expansion}
    \end{aligned}
\end{equation}
where $a_{\ell m}$ and $b_{\ell m}$ are the electric and magnetic multipole coefficients, respectively. The vector spherical harmonics are eigenfunctions of the total angular momentum operators $\hat{\bf{L}}^2$ and $\hat{L}_z$ with eigenvalues $\hbar\ell(\ell+1)$ and $\hbar m$, respectively.

Within the multipole basis, the time-averaged optical chiral density $\langle C\rangle = ({k}/{8\pi})  \text{Im}\left[\mathbf{E} \cdot \mathbf{B}^*\right]$ becomes
\begin{equation}
\begin{aligned}
    \langle C({\bf x},t) \rangle &= \sum_{\ell,m}\frac{k}{8\pi}\text{Im}\bigg[\frac{i^2}{k^2}a_{\ell m}\mathbf{\nabla\times}\Big(h_\ell^{(1)}(kr)\mathbf {X} _{\ell m}({\bf x}) \Big)\cdot b^*_{\ell m}\nabla\times\Big(h_{\ell}^*(kr)\mathbf {X} _{\ell m}({\bf x}) ^*\Big)\\
    &+a^*_{\ell m} h_{\ell}^*(kr)\mathbf {X} _{\ell m}({\bf x}) ^*\cdot b_{\ell m} h_{\ell}(kr)\mathbf {X} _{\ell m}({\bf x}) \bigg].
\end{aligned}
\label{chi_density_exp} 
\end{equation}
Utilizing the relationship between optical chiral density and optical chirality scattering in Eq.\eqref{chi_sca_and_density}, the latter becomes 
\begin{equation}
    \langle \Phi_{\text{sca}}\rangle  
=\frac{c}{2}\int r^2d\Omega\, \sum_{\ell,m}\frac{k}{4\pi}|h_{\ell}(kr)|^2\text{Im}\left[a_{\ell m}^*b_{\ell m} \right]\mathbf {X } _{\ell m}({\bf x}) ^*\cdot \mathbf {X } _{\ell m}({\bf x}).
\label{D3}
\end{equation}
Here the angular integration is performed only in the second term of Eq. \eqref{chi_density_exp} since the two terms contribute equally to the optical chirality density \cite{Jackson2003}. Given the asymptotic form of the spherical Hankel functions $h_\ell(kr)\sim(-i)^{\ell+1}{e^{ikr}}/{kr}$ and the fact that the vector spherical harmonics $\mathbf {X }_{\ell m}({\bf x})$ are normalized to unity, Eq. \eqref{D3} becomes
\begin{equation}
    \langle \Phi_\text{sca}\rangle =  \sum_{\ell m}\frac{c}{8\pi k}\text{Im}\left[a_{\ell m}^*b_{\ell m}\right].
    \label{chiscamult}
\end{equation}
Eq. \ref{chiscamult} is the chiral density radiated into the far field by a chiral scatterer of arbitrary multipole composition \cite{olmos_trigo_CD,nieto2017chiral}. We note from Eq. \ref{chiscamult} that any arbitrary collection of multipoles is chiral only if it contains both 
electric and magnetic multipole coefficients $a_{\ell m}$ and $b_{\ell m}$ with the same $\ell$ and $m$ values and nonzero relative phase.

\section{Derivation of Chiral Extinction, Absorption, and Scattering}
\label{chi_ext_sca_abs}

The extinction, absorption, and scattering of optical chirality can be defined from the appropriate volume integration of the total chiral flux $\langle{\bm\varphi}\rangle$. Specifically, volume integration of the time-averaged optical chirality continuity equation (Eq. \eqref{chi_conservation}) yields
\begin{equation}
        \int d\mathbf{x}\,\bigg\langle \frac{dC}{dt} \bigg\rangle  =-\int d{\bf x}\nabla\cdot\langle\boldsymbol{\varphi}\rangle+\int d{\bf x}\langle\mathcal{J}\rangle =0
\end{equation}
since $\langle dC/dt \rangle=0$ in the case of harmonic time dependence. Optical chirality absorption is defined from the total inward flux of optical chirality, and can be expressed as
\begin{equation}
\begin{aligned}
\langle\Phi_{\text{abs}}\rangle \equiv -\int d{\bf x}\,\nabla\cdot\langle\boldsymbol{\varphi}\rangle =  -\int r^2 d\Omega\,\hat{\mathbf{n}}\cdot\langle\boldsymbol{\varphi}\rangle,
\end{aligned}
\end{equation}
so that
\begin{equation}
\begin{aligned}
    \langle\Phi_{\text{abs}}\rangle &= -\frac{\omega}{16\pi} \int r^2 d\Omega\,  \hat{\mathbf{n}} \cdot\textrm{Im}\Big[\big({\bf E}_0+{\bf E}_{\text{s}})\times\big({\bf E}_0+{\bf E}_{\text{s}}\big)^* +\big({\bf B}_0+{\bf B}_{\text{s}}\big)\times\big({\bf B}_0+{\bf B}_{\text{s}}\big)^*\Big]\\
&= -\frac{\omega}{16\pi} \int r^2 d\Omega\,  \hat{\mathbf{n}} \cdot \text{Im} \Big[ \mathbf{E}_\text{s}^* \times \mathbf{E}_\text{s} + \mathbf{B}_\text{s}^* \times \mathbf{B}_\text{s}  + 2i\, \text{Im}(\mathbf{E}_\text{s}^* \times \mathbf{E}_0 + \mathbf{B}_\text{s}^* \times \mathbf{B}_0) \Big]\\
 &= -\langle \Phi_{\text{sca}} \rangle - \frac{\omega}{8\pi} \int r^2 d\Omega\,  \hat{\mathbf{n}} \cdot \text{Im} \left[ \mathbf{E}_\text{s}^* \times \mathbf{E}_0 + \mathbf{B}_\text{s}^* \times \mathbf{B}_0 \right] \\
&= -\langle \Phi_{\text{sca}} \rangle + \langle \Phi_{\text{ext}} \rangle,
\label{abs2}
\end{aligned}
\end{equation}
where ${\bf E}={\bf E}_0+{\bf E}_{\text{s}}$ and ${\bf B}={\bf B}_0+{\bf B}_{\text{s}}$ are the total electric and magnetic fields, each composed of incident and scattered components. The last term in Eq. \eqref{abs2} is the optical chirality extinction, analogous to the extinction power. Collectively, the optical chirality extinction, absorption, and scattering are thus given by surface integrals over the fields
\begin{equation}
    \begin{aligned}
    \langle \Phi_{\text{ext}} \rangle 
&= -\frac{\omega}{8\pi} \int r^2 d\Omega\,  \hat{\mathbf{n}} \cdot \text{Im} \left[ \mathbf{E}_\text{s}^* \times \mathbf{E}_0 + \mathbf{B}_\text{s}^* \times \mathbf{B}_0 \right]\\
\langle \Phi_{\text{abs}} \rangle 
&= -\frac{\omega}{16\pi} \int r^2 d\Omega\,  \hat{\mathbf{n}} \cdot \text{Im} \left[ \mathbf{E}^* \times \mathbf{E} + \mathbf{B}^* \times \mathbf{B} \right] \\
\langle \Phi_{\text{sca}} \rangle 
&= \frac{\omega}{16\pi} \int r^2 d\Omega\,  \hat{\mathbf{n}} \cdot \text{Im} \left[ \mathbf{E}_\text{s}^* \times \mathbf{E}_\text{s} + \mathbf{B}_\text{s}^* \times \mathbf{B}_\text{s} \right].
\label{final_ext}
\end{aligned}
\end{equation}
Eqs. \eqref{final_ext} may also be written in terms of the incident fields $\mathbf{E}_0, \mathbf{B}_0$ and the induced current $\mathbf{J}$ via Maxwell's equations \cite{Jackson2003} as
\begin{equation}
\begin{aligned}
\langle \Phi_{\text{ext}} \rangle &= -\frac{\omega}{8\pi} \int d\mathbf{x} \, \nabla \cdot \text{Im} \, [\mathbf{E}_\text{s}^* \times \mathbf{E}_0 + \mathbf{B}_\text{s}^* \times \mathbf{B}_0] \\
&= -\frac{\omega}{8\pi} \int d\mathbf{x} \, \text{Im} \, [\mathbf{E}_0 \cdot (\nabla \times \mathbf{E}_\text{s}^*) - \mathbf{E}_\text{s}^* \cdot (\nabla \times \mathbf{E}_0) + \mathbf{B}_0 \cdot (\nabla \times \mathbf{B}_\text{s}^*) - \mathbf{B}_\text{s}^* \cdot (\nabla \times \mathbf{B}_0)] \\
&= -\frac{\omega}{8\pi} \int d\mathbf{x} \, \text{Im}\, \left[ \mathbf{E}_0 \cdot (-ik\mathbf{B}_\text{s}^*) - \mathbf{E}_\text{s}^* \cdot (ik\mathbf{B}_0) + \mathbf{B}_0 \cdot \left( \frac{4\pi}{c}\mathbf{J}^* + ik\mathbf{E}_\text{s}^* \right) - \mathbf{B}_\text{s}^* \cdot (-ik\mathbf{E}_0) \right] \\
&= -\frac{k}{2} \int d\mathbf{x} \, \text{Im} \, [\mathbf{B}_0 \cdot \mathbf{J}^*]\\
&= \frac{1}{2} \int d\mathbf{x} \, \text{Re} \, [(\nabla \times \mathbf{E}_0) \cdot \mathbf{J}^*]\\
&= \frac{1}{2} \int d\mathbf{x} \, \text{Re} \, [\mathbf{E}_0 \cdot (\nabla \times \mathbf{J}^*)]
\end{aligned}
\end{equation}
and
\begin{equation}
\begin{aligned}
\langle \Phi_{\text{sca}} \rangle 
&= \frac{k}{2} \int d\mathbf{x} \, \text{Im} \left[ \mathbf{B}_{\text{s}} \cdot \mathbf{J}^* \right] 
= -\frac{1}{2} \int d\mathbf{x} \, \text{Re} \left[ (\nabla \times \mathbf{E}_{\text{s}}) \cdot \mathbf{J}^* \right] 
= -\frac{1}{2} \int d\mathbf{x} \, \text{Re} \left[ \mathbf{E}_{\text{s}} \cdot (\nabla \times \mathbf{J}^*) \right]\\
\langle \Phi_{\text{abs}} \rangle 
&= -\frac{k}{2} \int d\mathbf{x} \, \text{Im} \left[ \mathbf{B} \cdot \mathbf{J}^* \right] 
= \frac{1}{2} \int d\mathbf{x} \, \text{Re} \left[ \left( \nabla \times \mathbf{E} \right) \cdot \mathbf{J}^* \right] 
= \frac{1}{2} \int d\mathbf{x} \, \text{Re} \left[ \mathbf{E} \cdot \left( \nabla \times \mathbf{J}^* \right) \right].
\end{aligned}
\end{equation}

Note that for the case considered in Sec. \ref{SEC_EigSca_OptChi}, where there is no incident field but matter degrees of freedom are generating fields such that $\mathbf{E}=\mathbf{E}_\Gamma,$ $\langle \Phi_\text{ext} \rangle_\Gamma = 0$ and $\langle \Phi_\text{sca} \rangle_\Gamma = -\langle \Phi_\text{abs} \rangle_\Gamma = \int d \mathbf{x} \langle \mathcal{J}_\text{sca} \rangle$.

\section{Extinction of Optical Chirality in the BK Structure}
\label{BK_ext}
Explicitly defining the polarization vector $\hat{\bm{\epsilon}}_0$ of in-plane polarized incident fields $\mathbf{E}_0 = 
\hat{\bm{\epsilon}}_0E_0e^{ikz}$ and $\mathbf{B}_0 = ({1}/{k})\mathbf{k}_0\times\mathbf{E}_0$ as $\hat{\bm{\epsilon}}_0 = (A\hat{\mathbf{x}}+B\hat{\mathbf{y}})$ so that
\begin{equation}
\mathbf{E}_0(\mathbf{x}) = (A\hat{\mathbf{x}}+B\hat{\mathbf{y}})E_0e^{i kz},
\end{equation}
the coupled dipole equations in Eq. \ref{eom} can be written as
\begin{equation}
    \begin{aligned}
        \mathbf{d}_1 &= \frac{\alpha_1 A+g\alpha_1\alpha_2 e^{i k (z_2-z_1)}B}{1 - g^2 \alpha_1 \alpha_2} \hat{\mathbf{x}}E_0e^{i k z_1}\\
        \mathbf{d}_2 &= \frac{\alpha_2 B+g\alpha_1\alpha_2 e^{i k (z_1-z_2)}A}{1 - g^2 \alpha_1 \alpha_2} \hat{\mathbf{y}}E_0e^{i k z_2},
    \end{aligned}
\end{equation}
where $A$ and $B$ are arbitrary complex-valued amplitudes satisfying $|A|^2+|B|^2=1.$ By inserting the steady-state current $\mathbf{J}({\bf x}) = -i\omega \sum_i \mathbf{d}_i \delta(\mathbf{x} - \mathbf{r}_i)$ sourced by the coupled BK dipoles ${\mathbf{d}}_1=\hat{\mathbf{n}}_1d_1$ and ${\mathbf{d}}_2=\hat{\mathbf{n}}_2d_2$ into Eq. $\ref{Phi2}$, the extinction of optical chirality becomes
\begin{equation}
\begin{split}
\langle \Phi_{\text{ext}} \rangle _{\hat{\boldsymbol{\epsilon}}_0}&= \frac{\omega}{2} \int d\mathbf{x} \, \text{Im} \Big[ (\nabla \times \mathbf{E}_0(\mathbf{x}))^* \cdot \mathbf{d}_1 \delta(\mathbf{x} - \mathbf{r}_1) +  (\nabla \times \mathbf{E}_0(\mathbf{x}))^* \cdot \mathbf{d}_2 \delta(\mathbf{x} - \mathbf{r}_2) \Big]\\
&= \frac{\omega k}{2} \left|E_0\right|^2  \text{Im} \Bigg[ i\bigg( 
    \frac{\alpha_1  AB^* + g\alpha_1\alpha_2  e^{-i k \Lambda}|B|^2}{1 - g^2 \alpha_1 \alpha_2} -   \frac{\alpha_2 A^*B + g\alpha_1\alpha_2  e^{i k \Lambda}|A|^2}{1 - g^2 \alpha_1 \alpha_2} \bigg) \Bigg],
\end{split}
\label{chiext_general}
\end{equation}
recalling that $\hat{\mathbf{n}}_1=\hat{\mathbf{x}}$ and $\hat{\mathbf{n}}_2=\hat{\mathbf{y}}.$ This result holds for any arbitrary polarization $\hat{\bm{\epsilon}}_0 = A\hat{\mathbf{x}}+B\hat{\mathbf{y}}$.

Using Eq. \ref{chiext_general}, particular forms for optical chirality extinction can be derived for specific incident polarization states. For example, under $\pm 45^\circ$-polarization with $A = {1}/{\sqrt{2}}$ and $B = \pm{1}/{\sqrt{2}}$,
\begin{equation}
    \langle\Phi_\text{ext}\rangle_{\pm\pi/4}= \frac{\omega k}{4} \left|E_0\right|^2  \text{Im} \Bigg[ \pm i  \Bigg( 
    \frac{\alpha_1-\alpha_2}{1 - g^2 \alpha_1 \alpha_2}\bigg) +   \frac{ 2g\alpha_1\alpha_2 }{1 - g^2 \alpha_1 \alpha_2} \sin{(k\Lambda)} \Bigg],
    \label{linpoleext}
\end{equation}
while for circular polarization with $A ={1}/{\sqrt{2}}$ and $B = \pm{i}/{\sqrt{2}}$,
\begin{equation}
\begin{split}
    \langle\Phi_\text{ext}\rangle_\pm&= \frac{\omega k}{4} \left|E_0\right|^2  \text{Im} \Bigg[\pm  \Bigg( 
    \frac{\alpha_1+\alpha_2}{1 - g^2 \alpha_1 \alpha_2}\bigg) +   \frac{ 2g\alpha_1\alpha_2 }{1 - g^2 \alpha_1 \alpha_2} \sin{(k
    \Lambda)} \Bigg]\\
    &=\pm\frac{\omega k}{4} \left|E_0\right|^2  \text{Im} \Bigg[\Bigg( 
    \frac{\alpha_1+\alpha_2}{1 - g^2 \alpha_1 \alpha_2}\bigg) \pm  \frac{ 2g\alpha_1\alpha_2 }{1 - g^2 \alpha_1 \alpha_2} \sin{(k
    \Lambda)} \Bigg]\\
    &=\pm k \langle P_\text{ext}\rangle_\pm
    \end{split}
    \label{DC}
\end{equation}
where $\Lambda=z_1-z_2$ and where the extinction power from Eq. \eqref{powerpm} was used in the last equality. Based on Eq. \eqref{DC} together with Eq. \eqref {CD_eq}, CD may be related to chiral extinction as
\begin{equation}
\begin{split}
    \text{CD}(\omega) &=  \frac{\langle P_\text{ext} \rangle_+ - \langle P_\text{ext} \rangle_-}{(1/2)[\langle P_\text{ext} \rangle_+ + \langle P_\text{ext} \rangle_-]}\\
    &=\frac{\langle \Phi_\text{ext} \rangle_+ + \langle \Phi_\text{ext} \rangle_-}{(1/2)[\langle \Phi_\text{ext} \rangle_+ - \langle \Phi_\text{ext} \rangle_-]}.
    \end{split}
    \label{CD_eq2}
\end{equation}

\section{Optical Theorem for Optical Chirality}
\label{optical_thm}

In the far-field, the scattered electric and magnetic fields of localized sources under plane wave optical drive are well-described by a product of a vector scattering amplitude $\mathbf{F}(\mathbf{k},\mathbf{k}_0)$ and a scalar spherical wave component such that \cite{Jackson2003}
\begin{equation}
\begin{split}
    \mathbf{E}_\text{s} &= \mathbf{F}(\mathbf{k},\mathbf{k}_0)\frac{e^{ikr}}{r}\\
    \mathbf{B}_\text{s} &= \hat{\bf n}\times\mathbf{E}_\text{s},
    \end{split}
    \label{scatteringamplitude}
\end{equation} 
where $\mathbf{k}_0$ specifies the incident wave vector, and $\mathbf{k}=k_0 \hat{\mathbf{n}}$. The vector scattering amplitude $\mathbf{F}(\mathbf{k},\mathbf{k}_0)$ can be expressed by the Kirchoff integral relation  as \cite{Jackson2003} 
\begin{equation}
   \mathbf{F}(\mathbf{k},\mathbf{k}_0) = \frac{1}{4\pi i}\mathbf{k}\times\int r^{\prime 2}d\Omega' e^{-i{\bf k}\cdot{\bf x}'}\Big[\frac{\mathbf{k}\times(\hat{\mathbf{n}}'\times\bf{B}_\text{s}({\bf x}'))}{k}-\hat{\mathbf{n}}'\times \mathbf{E}_\text{s}({\bf x}')\Big],
   \label{kirchoff}
\end{equation}
with ${\bf x}'=r'\hat{\bf n}'$, where the ${\bf k}_0$ dependence is implicit within the scattered fields. With the incident fields $\mathbf{E}_0 = 
\hat{\bm{\epsilon}}_0E_0e^{ikz}$ and $\mathbf{B}_0 = ({1}/{k})\mathbf{k}_0\times\mathbf{E}_0$, the optical chirality extinction in Eq. \ref{Phi1} can be written as
\begin{equation}
\begin{split}
    \langle \Phi_{\text{ext}} \rangle_{\hat{\boldsymbol{\epsilon}}_0}
    &= \frac{\omega}{8\pi}\int r'^2d\Omega'\,\hat{\mathbf n}'\cdot\textrm{Im}[{\bf E}_0\times{\bf E}_{\textrm{s}}^*+{\bf B}_0\times{\bf B}_{\textrm{s}}^*] \\
     &= \frac{\omega}{8\pi} \int r'^2d\Omega' \,  \hat{\mathbf{n}}'\cdot \text{Im} \Bigg[E_0e^{ikz'}\hat{\bm{\epsilon}}_0\times\mathbf{E}_\text{s}^*  + E_0e^{ikz}\frac{\mathbf{k}_0\times\hat{\bm{\epsilon}}_0}{k}\times \mathbf{B}_\text{s}^*\Bigg]\\
     &=\frac{\omega}{8\pi} \int r'^2d\Omega' \,  \text{Im} \Bigg[\hat{\bm{\epsilon}}_0 E_0e^{ikz'}\cdot\Big[\frac{\mathbf{k}_0\times\big(\hat{\mathbf{n}}'\times \mathbf{B}_\text{s}^*\big)}{k}-\hat{\mathbf{n}}'\times\mathbf{E}_\text{s}^*\Big]\Bigg].
\end{split}
\end{equation}
Thus, using the vector identity $-k^2{\bf V}={\bf k}_0\times({\bf k}_0\times{\bf V})$, where ${\bf V}={\mathbf{k}_0\times\big(\hat{\mathbf{n}}'\times \mathbf{B}_\text{s}^*\big)}/{k}-\hat{\mathbf{n}}'\times\mathbf{E}_\text{s}^*$,
\begin{equation}
\begin{split}
    \langle \Phi_{\text{ext}} \rangle_{\hat{\boldsymbol{\epsilon}}_0}
     &=\frac{\omega}{8\pi k^2} \text{Im} \Bigg[\hat{\bm{\epsilon}}_0 E_0\cdot{\bf k}_0\times \frac{-4\pi i}{-4\pi i}\Bigg({\bf k}_0\times\int r'^2d\Omega'e^{ikz'}\Big[ \hat{\mathbf{n}}'\times\mathbf{E}_\text{s}^*-\frac{\mathbf{k}_0\times\big(\hat{\mathbf{n}}'\times \mathbf{B}_\text{s}^*\big)}{k}\Big]\Bigg)\Bigg]\\
     &=\frac{\omega}{2 k^2} \text{Re} \Bigg[\hat{\bm{\epsilon}}_0 E_0\cdot{\bf k}_0\times \frac{1}{-4\pi i}\Bigg({\bf k}_0\times\int r'^2d\Omega'e^{ikz'}\Big[\frac{\mathbf{k}_0\times\big(\hat{\mathbf{n}}'\times \mathbf{B}_\text{s}^*\big)}{k}-\hat{\mathbf{n}}'\times\mathbf{E}_\text{s}^*\Big]\Bigg)\Bigg]\\
     &=\frac{c}{2 k} \text{Re} \Big[\hat{\bm{\epsilon}}_0 E_0\cdot\Big({\bf k}_0\times\mathbf{F}^*(\mathbf{k}_0,\mathbf{k}_0)\Big)\Big].
\end{split}
\label{COT}
\end{equation}
Eq. \eqref{COT} is the optical chiral analogue to the optical theorem $\langle P_\text{ext}\rangle _{\hat{\boldsymbol{\epsilon}}_0}= (c/{2k})\text{Im}\Big[(\hat{\bm{\epsilon}}_0E_0)^*\cdot\mathbf{F}(\mathbf{k}_0,\mathbf{k}_0)\Big]$, known as the optical theorem for optical chirality \cite{Nieto-Vesperinas_2015}, with the forward-directed vector scattering amplitude $\mathbf{F}(\mathbf{k}_0,\mathbf{k}_0)$ defined in Eq. \eqref{kirchoff}.

The optical theorem for optical chirality can be exploited to directly connect the extinction of optical chirality $\langle\Phi_\text{ext}\rangle$ to the time-averaged optical torque exerted on the BK system by the incident field.  By representing the BK fields in terms of their effective co-located electric $\mathbf{d}$ and magnetic $\mathbf{m}$ dipole components through vector sperical harmonic expansion, the vector scattering amplitude in the forward direction $\mathbf{F}(\mathbf{k}_0,\mathbf{k}_0)=\mathbf{F}_{\bf d}(\mathbf{k}_0,\mathbf{k}_0)+\mathbf{F}_{\bf m}(\mathbf{k}_0,\mathbf{k}_0),$ where $\mathbf{F}_{\bf d}(\mathbf{k}_0,\mathbf{k}_0) = k^2\mathbf{d}$ and $\mathbf{F}_{\bf m}(\mathbf{k}_0,\mathbf{k}_0)  = -k(\mathbf{k}_0\times\mathbf{m})$. Eq. \eqref{COT} can then be written as
\begin{equation}
\begin{aligned}
    \langle \Phi_{\text{ext}} \rangle_{\hat{\boldsymbol{\epsilon}}_0} &= \frac{c}{2k}\text{Re}\bigg[E_0\hat{\bm{\epsilon}}_0\cdot\Big(\mathbf{k}_0\times\big[k^2\mathbf{d}^*-k\mathbf{k}_0\times\mathbf{m}^*\big]\Big)\bigg]\\
    &= \frac{kc}{2}\text{Re}\bigg[\mathbf{k}_0\cdot\big( \mathbf{d}^*\times \mathbf{E}_0\big)+\mathbf{k}_0\cdot\big(\mathbf{m}^*\times\mathbf{B} _0\big)\bigg]\\
    &= \frac{\omega}{\hbar}\mathbf{P}_0\cdot\text{Re}\bigg[\frac{1}{2}\big( \mathbf{d}^*\times \mathbf{E}_0\big)+\frac{1}{2}\big(\mathbf{m}^*\times\mathbf{B} _0\big)\bigg]\\
    &= \frac{\omega}{\hbar}\mathbf{P}_0\cdot\langle{\bf{N}}_0\rangle_{\hat{\boldsymbol{\epsilon}}_0},
\end{aligned}
\end{equation}
where the optical momentum of the incident field is ${\bf P}_0=\hbar{\bf k}_0$ and $\langle{\bf N}_0\rangle_{\hat{\boldsymbol{\epsilon}}_0}$ is the time-averaged optical torque exerted on the BK system from the incident field only \cite{Nieto-Vesperinas_15}.

\bibliography{references}

\begin{thebibliography}{84}%
\makeatletter
\providecommand \@ifxundefined [1]{%
 \@ifx{#1\undefined}
}%
\providecommand \@ifnum [1]{%
 \ifnum #1\expandafter \@firstoftwo
 \else \expandafter \@secondoftwo
 \fi
}%
\providecommand \@ifx [1]{%
 \ifx #1\expandafter \@firstoftwo
 \else \expandafter \@secondoftwo
 \fi
}%
\providecommand \natexlab [1]{#1}%
\providecommand \enquote  [1]{``#1''}%
\providecommand \bibnamefont  [1]{#1}%
\providecommand \bibfnamefont [1]{#1}%
\providecommand \citenamefont [1]{#1}%
\providecommand \href@noop [0]{\@secondoftwo}%
\providecommand \href [0]{\begingroup \@sanitize@url \@href}%
\providecommand \@href[1]{\@@startlink{#1}\@@href}%
\providecommand \@@href[1]{\endgroup#1\@@endlink}%
\providecommand \@sanitize@url [0]{\catcode `\\12\catcode `\$12\catcode
  `\&12\catcode `\#12\catcode `\^12\catcode `\_12\catcode `\%12\relax}%
\providecommand \@@startlink[1]{}%
\providecommand \@@endlink[0]{}%
\providecommand \url  [0]{\begingroup\@sanitize@url \@url }%
\providecommand \@url [1]{\endgroup\@href {#1}{\urlprefix }}%
\providecommand \urlprefix  [0]{URL }%
\providecommand \Eprint [0]{\href }%
\providecommand \doibase [0]{https://doi.org/}%
\providecommand \selectlanguage [0]{\@gobble}%
\providecommand \bibinfo  [0]{\@secondoftwo}%
\providecommand \bibfield  [0]{\@secondoftwo}%
\providecommand \translation [1]{[#1]}%
\providecommand \BibitemOpen [0]{}%
\providecommand \bibitemStop [0]{}%
\providecommand \bibitemNoStop [0]{.\EOS\space}%
\providecommand \EOS [0]{\spacefactor3000\relax}%
\providecommand \BibitemShut  [1]{\csname bibitem#1\endcsname}%
\let\auto@bib@innerbib\@empty
\bibitem [{\citenamefont {Blackmond}(2010)}]{blackmond2010origin}%
  \BibitemOpen
  \bibfield  {author} {\bibinfo {author} {\bibfnamefont {D.~G.}\ \bibnamefont
  {Blackmond}},\ }\bibfield  {title} {\bibinfo {title} {The origin of
  biological homochirality},\ }\href@noop {} {\bibfield  {journal} {\bibinfo
  {journal} {Cold Spring Harb. Perspect. Biol.}\ }\textbf {\bibinfo {volume}
  {2}},\ \bibinfo {pages} {a002147} (\bibinfo {year} {2010})}\BibitemShut
  {NoStop}%
\bibitem [{\citenamefont {Frank}(1953)}]{frank_spontaneous_1953}%
  \BibitemOpen
  \bibfield  {author} {\bibinfo {author} {\bibfnamefont {F.~C.}\ \bibnamefont
  {Frank}},\ }\bibfield  {title} {\bibinfo {title} {On spontaneous asymmetric
  synthesis},\ }\href@noop {} {\bibfield  {journal} {\bibinfo  {journal}
  {Biochim. Biophys. Acta}\ }\textbf {\bibinfo {volume} {11}},\ \bibinfo
  {pages} {459} (\bibinfo {year} {1953})}\BibitemShut {NoStop}%
\bibitem [{\citenamefont {Liu}\ \emph {et~al.}(2015)\citenamefont {Liu},
  \citenamefont {Han}, \citenamefont {Liu},\ and\ \citenamefont
  {Stoltz}}]{liu2015catalytic}%
  \BibitemOpen
  \bibfield  {author} {\bibinfo {author} {\bibfnamefont {Y.}~\bibnamefont
  {Liu}}, \bibinfo {author} {\bibfnamefont {S.-J.}\ \bibnamefont {Han}},
  \bibinfo {author} {\bibfnamefont {W.-B.}\ \bibnamefont {Liu}},\ and\ \bibinfo
  {author} {\bibfnamefont {B.~M.}\ \bibnamefont {Stoltz}},\ }\bibfield  {title}
  {\bibinfo {title} {Catalytic enantioselective construction of quaternary
  stereocenters: assembly of key building blocks for the synthesis of
  biologically active molecules},\ }\href@noop {} {\bibfield  {journal}
  {\bibinfo  {journal} {Acc. Chem. Res.}\ }\textbf {\bibinfo {volume} {48}},\
  \bibinfo {pages} {740} (\bibinfo {year} {2015})}\BibitemShut {NoStop}%
\bibitem [{\citenamefont {Xu}\ \emph {et~al.}(2022)\citenamefont {Xu},
  \citenamefont {Wang}, \citenamefont {Wang}, \citenamefont {Sun},
  \citenamefont {Choi}, \citenamefont {Kim}, \citenamefont {Hao}, \citenamefont
  {Li}, \citenamefont {Qu}, \citenamefont {Lu}, \citenamefont {Wu},
  \citenamefont {Colombari}, \citenamefont {Gomes}, \citenamefont {Blanco},
  \citenamefont {de~Moura}, \citenamefont {Guo}, \citenamefont {Kuang},
  \citenamefont {Kotov},\ and\ \citenamefont {Xu}}]{xu2022enantiomer}%
  \BibitemOpen
  \bibfield  {author} {\bibinfo {author} {\bibfnamefont {L.}~\bibnamefont
  {Xu}}, \bibinfo {author} {\bibfnamefont {X.}~\bibnamefont {Wang}}, \bibinfo
  {author} {\bibfnamefont {W.}~\bibnamefont {Wang}}, \bibinfo {author}
  {\bibfnamefont {M.}~\bibnamefont {Sun}}, \bibinfo {author} {\bibfnamefont
  {W.~J.}\ \bibnamefont {Choi}}, \bibinfo {author} {\bibfnamefont {J.-Y.}\
  \bibnamefont {Kim}}, \bibinfo {author} {\bibfnamefont {C.}~\bibnamefont
  {Hao}}, \bibinfo {author} {\bibfnamefont {S.}~\bibnamefont {Li}}, \bibinfo
  {author} {\bibfnamefont {A.}~\bibnamefont {Qu}}, \bibinfo {author}
  {\bibfnamefont {M.}~\bibnamefont {Lu}}, \bibinfo {author} {\bibfnamefont
  {X.}~\bibnamefont {Wu}}, \bibinfo {author} {\bibfnamefont {F.~M.}\
  \bibnamefont {Colombari}}, \bibinfo {author} {\bibfnamefont {W.~R.}\
  \bibnamefont {Gomes}}, \bibinfo {author} {\bibfnamefont {A.~L.}\ \bibnamefont
  {Blanco}}, \bibinfo {author} {\bibfnamefont {A.~F.}\ \bibnamefont
  {de~Moura}}, \bibinfo {author} {\bibfnamefont {X.}~\bibnamefont {Guo}},
  \bibinfo {author} {\bibfnamefont {H.}~\bibnamefont {Kuang}}, \bibinfo
  {author} {\bibfnamefont {N.~A.}\ \bibnamefont {Kotov}},\ and\ \bibinfo
  {author} {\bibfnamefont {C.}~\bibnamefont {Xu}},\ }\bibfield  {title}
  {\bibinfo {title} {Enantiomer-dependent immunological response to chiral
  nanoparticles},\ }\href@noop {} {\bibfield  {journal} {\bibinfo  {journal}
  {Nature}\ }\textbf {\bibinfo {volume} {601}},\ \bibinfo {pages} {366}
  (\bibinfo {year} {2022})}\BibitemShut {NoStop}%
\bibitem [{\citenamefont {Warning}\ \emph {et~al.}(2021)\citenamefont
  {Warning}, \citenamefont {Miandashti}, \citenamefont {McCarthy},
  \citenamefont {Zhang}, \citenamefont {Landes},\ and\ \citenamefont
  {Link}}]{Warning2021}%
  \BibitemOpen
  \bibfield  {author} {\bibinfo {author} {\bibfnamefont {L.~A.}\ \bibnamefont
  {Warning}}, \bibinfo {author} {\bibfnamefont {A.~R.}\ \bibnamefont
  {Miandashti}}, \bibinfo {author} {\bibfnamefont {L.~A.}\ \bibnamefont
  {McCarthy}}, \bibinfo {author} {\bibfnamefont {Q.}~\bibnamefont {Zhang}},
  \bibinfo {author} {\bibfnamefont {C.~F.}\ \bibnamefont {Landes}},\ and\
  \bibinfo {author} {\bibfnamefont {S.}~\bibnamefont {Link}},\ }\bibfield
  {title} {\bibinfo {title} {Nanophotonic approaches for chirality sensing},\
  }\href@noop {} {\bibfield  {journal} {\bibinfo  {journal} {ACS Nano}\
  }\textbf {\bibinfo {volume} {15}},\ \bibinfo {pages} {15538} (\bibinfo {year}
  {2021})}\BibitemShut {NoStop}%
\bibitem [{\citenamefont {Almousa}\ \emph {et~al.}(2024)\citenamefont
  {Almousa}, \citenamefont {Weiss},\ and\ \citenamefont
  {Muljarov}}]{Almousa2024}%
  \BibitemOpen
  \bibfield  {author} {\bibinfo {author} {\bibfnamefont {S.~F.}\ \bibnamefont
  {Almousa}}, \bibinfo {author} {\bibfnamefont {T.}~\bibnamefont {Weiss}},\
  and\ \bibinfo {author} {\bibfnamefont {E.~A.}\ \bibnamefont {Muljarov}},\
  }\bibfield  {title} {\bibinfo {title} {Employing quasidegenerate optical
  modes for chiral sensing},\ }\href@noop {} {\bibfield  {journal} {\bibinfo
  {journal} {Phys. Rev. B}\ }\textbf {\bibinfo {volume} {109}},\ \bibinfo
  {pages} {L041410} (\bibinfo {year} {2024})}\BibitemShut {NoStop}%
\bibitem [{\citenamefont {Mun}\ \emph {et~al.}(2020)\citenamefont {Mun},
  \citenamefont {Kim}, \citenamefont {Yang}, \citenamefont {Badloe},
  \citenamefont {Ni}, \citenamefont {Chen}, \citenamefont {Qiu},\ and\
  \citenamefont {Rho}}]{mun_electromagnetic_2020}%
  \BibitemOpen
  \bibfield  {author} {\bibinfo {author} {\bibfnamefont {J.}~\bibnamefont
  {Mun}}, \bibinfo {author} {\bibfnamefont {M.}~\bibnamefont {Kim}}, \bibinfo
  {author} {\bibfnamefont {Y.}~\bibnamefont {Yang}}, \bibinfo {author}
  {\bibfnamefont {T.}~\bibnamefont {Badloe}}, \bibinfo {author} {\bibfnamefont
  {J.}~\bibnamefont {Ni}}, \bibinfo {author} {\bibfnamefont {Y.}~\bibnamefont
  {Chen}}, \bibinfo {author} {\bibfnamefont {C.-W.}\ \bibnamefont {Qiu}},\ and\
  \bibinfo {author} {\bibfnamefont {J.}~\bibnamefont {Rho}},\ }\bibfield
  {title} {\bibinfo {title} {Electromagnetic chirality: from fundamentals to
  nontraditional chiroptical phenomena},\ }\href@noop {} {\bibfield  {journal}
  {\bibinfo  {journal} {Light Sci. Appl.}\ }\textbf {\bibinfo {volume} {9}},\
  \bibinfo {pages} {139} (\bibinfo {year} {2020})}\BibitemShut {NoStop}%
\bibitem [{\citenamefont {Lee}\ \emph {et~al.}(2020)\citenamefont {Lee},
  \citenamefont {Kim}, \citenamefont {Im}, \citenamefont {Balamurugan},\ and\
  \citenamefont {Nam}}]{Lee2020}%
  \BibitemOpen
  \bibfield  {author} {\bibinfo {author} {\bibfnamefont {Y.~Y.}\ \bibnamefont
  {Lee}}, \bibinfo {author} {\bibfnamefont {R.~M.}\ \bibnamefont {Kim}},
  \bibinfo {author} {\bibfnamefont {S.~W.}\ \bibnamefont {Im}}, \bibinfo
  {author} {\bibfnamefont {M.}~\bibnamefont {Balamurugan}},\ and\ \bibinfo
  {author} {\bibfnamefont {K.~T.}\ \bibnamefont {Nam}},\ }\bibfield  {title}
  {\bibinfo {title} {Plasmonic metamaterials for chiral sensing applications},\
  }\href@noop {} {\bibfield  {journal} {\bibinfo  {journal} {Nanoscale}\
  }\textbf {\bibinfo {volume} {12}},\ \bibinfo {pages} {58} (\bibinfo {year}
  {2020})}\BibitemShut {NoStop}%
\bibitem [{\citenamefont {Solomon}\ \emph {et~al.}(2020)\citenamefont
  {Solomon}, \citenamefont {Saleh}, \citenamefont {Poulikakos}, \citenamefont
  {Abendroth}, \citenamefont {Tadesse},\ and\ \citenamefont
  {Dionne}}]{Dionne2020}%
  \BibitemOpen
  \bibfield  {author} {\bibinfo {author} {\bibfnamefont {M.~L.}\ \bibnamefont
  {Solomon}}, \bibinfo {author} {\bibfnamefont {A.~A.~E.}\ \bibnamefont
  {Saleh}}, \bibinfo {author} {\bibfnamefont {L.~V.}\ \bibnamefont
  {Poulikakos}}, \bibinfo {author} {\bibfnamefont {J.~M.}\ \bibnamefont
  {Abendroth}}, \bibinfo {author} {\bibfnamefont {L.~F.}\ \bibnamefont
  {Tadesse}},\ and\ \bibinfo {author} {\bibfnamefont {J.~A.}\ \bibnamefont
  {Dionne}},\ }\bibfield  {title} {\bibinfo {title} {Nanophotonic platforms for
  chiral sensing and separation},\ }\href@noop {} {\bibfield  {journal}
  {\bibinfo  {journal} {Acc. Chem. Res.}\ }\textbf {\bibinfo {volume} {53}},\
  \bibinfo {pages} {588} (\bibinfo {year} {2020})}\BibitemShut {NoStop}%
\bibitem [{\citenamefont {Chattopadhyay}\ and\ \citenamefont
  {Biteen}(2024)}]{Biteen2024}%
  \BibitemOpen
  \bibfield  {author} {\bibinfo {author} {\bibfnamefont {S.}~\bibnamefont
  {Chattopadhyay}}\ and\ \bibinfo {author} {\bibfnamefont {J.~S.}\ \bibnamefont
  {Biteen}},\ }\bibfield  {title} {\bibinfo {title} {Achiral plasmonic antennas
  enhance differential absorption to increase preferential detection of chiral
  single molecules},\ }\href@noop {} {\bibfield  {journal} {\bibinfo  {journal}
  {ACS Meas. Sci. Au}\ }\textbf {\bibinfo {volume} {4}},\ \bibinfo {pages}
  {528} (\bibinfo {year} {2024})}\BibitemShut {NoStop}%
\bibitem [{\citenamefont {Ma}\ \emph {et~al.}(2013)\citenamefont {Ma},
  \citenamefont {Kuang}, \citenamefont {Xu}, \citenamefont {Ding},
  \citenamefont {Xu}, \citenamefont {Wang},\ and\ \citenamefont
  {Kotov}}]{MaWei2013}%
  \BibitemOpen
  \bibfield  {author} {\bibinfo {author} {\bibfnamefont {W.}~\bibnamefont
  {Ma}}, \bibinfo {author} {\bibfnamefont {H.}~\bibnamefont {Kuang}}, \bibinfo
  {author} {\bibfnamefont {L.}~\bibnamefont {Xu}}, \bibinfo {author}
  {\bibfnamefont {L.}~\bibnamefont {Ding}}, \bibinfo {author} {\bibfnamefont
  {C.}~\bibnamefont {Xu}}, \bibinfo {author} {\bibfnamefont {L.}~\bibnamefont
  {Wang}},\ and\ \bibinfo {author} {\bibfnamefont {N.~A.}\ \bibnamefont
  {Kotov}},\ }\bibfield  {title} {\bibinfo {title} {Attomolar {DNA} detection
  with chiral nanorod assemblies},\ }\href@noop {} {\bibfield  {journal}
  {\bibinfo  {journal} {Nat. Commun.}\ }\textbf {\bibinfo {volume} {4}},\
  \bibinfo {pages} {2689} (\bibinfo {year} {2013})}\BibitemShut {NoStop}%
\bibitem [{\citenamefont {Zhang}\ \emph {et~al.}(2024)\citenamefont {Zhang},
  \citenamefont {Hu}, \citenamefont {Ma}, \citenamefont {Li}, \citenamefont
  {Wang}, \citenamefont {Li}, \citenamefont {Movsesyan}, \citenamefont {Wang},
  \citenamefont {Govorov}, \citenamefont {Gan},\ and\ \citenamefont
  {Ding}}]{zhang2024quantum}%
  \BibitemOpen
  \bibfield  {author} {\bibinfo {author} {\bibfnamefont {C.}~\bibnamefont
  {Zhang}}, \bibinfo {author} {\bibfnamefont {H.}~\bibnamefont {Hu}}, \bibinfo
  {author} {\bibfnamefont {C.}~\bibnamefont {Ma}}, \bibinfo {author}
  {\bibfnamefont {Y.}~\bibnamefont {Li}}, \bibinfo {author} {\bibfnamefont
  {X.}~\bibnamefont {Wang}}, \bibinfo {author} {\bibfnamefont {D.}~\bibnamefont
  {Li}}, \bibinfo {author} {\bibfnamefont {A.}~\bibnamefont {Movsesyan}},
  \bibinfo {author} {\bibfnamefont {Z.}~\bibnamefont {Wang}}, \bibinfo {author}
  {\bibfnamefont {A.}~\bibnamefont {Govorov}}, \bibinfo {author} {\bibfnamefont
  {Q.}~\bibnamefont {Gan}},\ and\ \bibinfo {author} {\bibfnamefont
  {T.}~\bibnamefont {Ding}},\ }\bibfield  {title} {\bibinfo {title} {Quantum
  plasmonics pushes chiral sensing limit to single molecules: a paradigm for
  chiral biodetections},\ }\href@noop {} {\bibfield  {journal} {\bibinfo
  {journal} {Nat. Commun.}\ }\textbf {\bibinfo {volume} {15}},\ \bibinfo
  {pages} {2} (\bibinfo {year} {2024})}\BibitemShut {NoStop}%
\bibitem [{\citenamefont {Ueda}\ \emph {et~al.}(2023)\citenamefont {Ueda},
  \citenamefont {García-Fernández}, \citenamefont {Agrestini}, \citenamefont
  {Romao}, \citenamefont {van~den Brink}, \citenamefont {Spaldin},
  \citenamefont {Zhou},\ and\ \citenamefont {Staub}}]{Ueda2023}%
  \BibitemOpen
  \bibfield  {author} {\bibinfo {author} {\bibfnamefont {H.}~\bibnamefont
  {Ueda}}, \bibinfo {author} {\bibfnamefont {M.}~\bibnamefont
  {García-Fernández}}, \bibinfo {author} {\bibfnamefont {S.}~\bibnamefont
  {Agrestini}}, \bibinfo {author} {\bibfnamefont {C.~P.}\ \bibnamefont
  {Romao}}, \bibinfo {author} {\bibfnamefont {J.}~\bibnamefont {van~den
  Brink}}, \bibinfo {author} {\bibfnamefont {N.~A.}\ \bibnamefont {Spaldin}},
  \bibinfo {author} {\bibfnamefont {K.-J.}\ \bibnamefont {Zhou}},\ and\
  \bibinfo {author} {\bibfnamefont {U.}~\bibnamefont {Staub}},\ }\bibfield
  {title} {\bibinfo {title} {Chiral phonons in quartz probed by {X}-rays},\
  }\href@noop {} {\bibfield  {journal} {\bibinfo  {journal} {Nature}\ }\textbf
  {\bibinfo {volume} {618}},\ \bibinfo {pages} {946} (\bibinfo {year}
  {2023})}\BibitemShut {NoStop}%
\bibitem [{\citenamefont {Ishito}\ \emph {et~al.}(2023)\citenamefont {Ishito},
  \citenamefont {Mao}, \citenamefont {Kousaka}, \citenamefont {Togawa},
  \citenamefont {Iwasaki}, \citenamefont {Zhang}, \citenamefont {Murakami},
  \citenamefont {Kishine},\ and\ \citenamefont {Satoh}}]{ishito2023truly}%
  \BibitemOpen
  \bibfield  {author} {\bibinfo {author} {\bibfnamefont {K.}~\bibnamefont
  {Ishito}}, \bibinfo {author} {\bibfnamefont {H.}~\bibnamefont {Mao}},
  \bibinfo {author} {\bibfnamefont {Y.}~\bibnamefont {Kousaka}}, \bibinfo
  {author} {\bibfnamefont {Y.}~\bibnamefont {Togawa}}, \bibinfo {author}
  {\bibfnamefont {S.}~\bibnamefont {Iwasaki}}, \bibinfo {author} {\bibfnamefont
  {T.}~\bibnamefont {Zhang}}, \bibinfo {author} {\bibfnamefont
  {S.}~\bibnamefont {Murakami}}, \bibinfo {author} {\bibfnamefont {J.-i.}\
  \bibnamefont {Kishine}},\ and\ \bibinfo {author} {\bibfnamefont
  {T.}~\bibnamefont {Satoh}},\ }\bibfield  {title} {\bibinfo {title} {Truly
  chiral phonons in $\alpha$-{H}g{S}},\ }\href@noop {} {\bibfield  {journal}
  {\bibinfo  {journal} {Nat. Phys.}\ }\textbf {\bibinfo {volume} {19}},\
  \bibinfo {pages} {35} (\bibinfo {year} {2023})}\BibitemShut {NoStop}%
\bibitem [{\citenamefont {Schilthuizen}\ and\ \citenamefont
  {Davison}(2005)}]{schilthuizen_convoluted_2005}%
  \BibitemOpen
  \bibfield  {author} {\bibinfo {author} {\bibfnamefont {M.}~\bibnamefont
  {Schilthuizen}}\ and\ \bibinfo {author} {\bibfnamefont {A.}~\bibnamefont
  {Davison}},\ }\bibfield  {title} {\bibinfo {title} {The convoluted evolution
  of snail chirality},\ }\href@noop {} {\bibfield  {journal} {\bibinfo
  {journal} {Naturwissenschaften}\ }\textbf {\bibinfo {volume} {92}},\ \bibinfo
  {pages} {504} (\bibinfo {year} {2005})}\BibitemShut {NoStop}%
\bibitem [{\citenamefont {Yu}\ \emph {et~al.}(2020)\citenamefont {Yu},
  \citenamefont {Motloch}, \citenamefont {Pen}, \citenamefont {Yu},
  \citenamefont {Wang}, \citenamefont {Mo}, \citenamefont {Yang},\ and\
  \citenamefont {Jing}}]{yu2020probing}%
  \BibitemOpen
  \bibfield  {author} {\bibinfo {author} {\bibfnamefont {H.-R.}\ \bibnamefont
  {Yu}}, \bibinfo {author} {\bibfnamefont {P.}~\bibnamefont {Motloch}},
  \bibinfo {author} {\bibfnamefont {U.-L.}\ \bibnamefont {Pen}}, \bibinfo
  {author} {\bibfnamefont {Y.}~\bibnamefont {Yu}}, \bibinfo {author}
  {\bibfnamefont {H.}~\bibnamefont {Wang}}, \bibinfo {author} {\bibfnamefont
  {H.}~\bibnamefont {Mo}}, \bibinfo {author} {\bibfnamefont {X.}~\bibnamefont
  {Yang}},\ and\ \bibinfo {author} {\bibfnamefont {Y.}~\bibnamefont {Jing}},\
  }\bibfield  {title} {\bibinfo {title} {Probing primordial chirality with
  galaxy spins},\ }\href@noop {} {\bibfield  {journal} {\bibinfo  {journal}
  {Phys. Rev. Lett.}\ }\textbf {\bibinfo {volume} {124}},\ \bibinfo {pages}
  {101302} (\bibinfo {year} {2020})}\BibitemShut {NoStop}%
\bibitem [{\citenamefont {Barron}(1986)}]{Barron_1986}%
  \BibitemOpen
  \bibfield  {author} {\bibinfo {author} {\bibfnamefont {L.}~\bibnamefont
  {Barron}},\ }\bibfield  {title} {\bibinfo {title} {True and false chirality
  and parity violation},\ }\href@noop {} {\bibfield  {journal} {\bibinfo
  {journal} {Chem. Phys. Lett.}\ }\textbf {\bibinfo {volume} {123}},\ \bibinfo
  {pages} {423–} (\bibinfo {year} {1986})}\BibitemShut {NoStop}%
\bibitem [{\citenamefont {Zabrodsky}\ and\ \citenamefont
  {Avnir}(1995)}]{Zabrodsky1995}%
  \BibitemOpen
  \bibfield  {author} {\bibinfo {author} {\bibfnamefont {H.}~\bibnamefont
  {Zabrodsky}}\ and\ \bibinfo {author} {\bibfnamefont {D.}~\bibnamefont
  {Avnir}},\ }\bibfield  {title} {\bibinfo {title} {Continuous symmetry
  measures. 4. {C}hirality},\ }\href@noop {} {\bibfield  {journal} {\bibinfo
  {journal} {J. Am. Chem. Soc.}\ }\textbf {\bibinfo {volume} {117}},\ \bibinfo
  {pages} {462} (\bibinfo {year} {1995})}\BibitemShut {NoStop}%
\bibitem [{\citenamefont {Pinsky}\ \emph {et~al.}(2008)\citenamefont {Pinsky},
  \citenamefont {Casanova}, \citenamefont {Alemany}, \citenamefont {Alvarez},
  \citenamefont {Avnir}, \citenamefont {Dryzun}, \citenamefont {Kizner},\ and\
  \citenamefont {Sterkin}}]{Pinsky2008}%
  \BibitemOpen
  \bibfield  {author} {\bibinfo {author} {\bibfnamefont {M.}~\bibnamefont
  {Pinsky}}, \bibinfo {author} {\bibfnamefont {D.}~\bibnamefont {Casanova}},
  \bibinfo {author} {\bibfnamefont {P.}~\bibnamefont {Alemany}}, \bibinfo
  {author} {\bibfnamefont {S.}~\bibnamefont {Alvarez}}, \bibinfo {author}
  {\bibfnamefont {D.}~\bibnamefont {Avnir}}, \bibinfo {author} {\bibfnamefont
  {C.}~\bibnamefont {Dryzun}}, \bibinfo {author} {\bibfnamefont
  {Z.}~\bibnamefont {Kizner}},\ and\ \bibinfo {author} {\bibfnamefont
  {A.}~\bibnamefont {Sterkin}},\ }\bibfield  {title} {\bibinfo {title}
  {Symmetry operation measures},\ }\href@noop {} {\bibfield  {journal}
  {\bibinfo  {journal} {J. Comput. Chem.}\ }\textbf {\bibinfo {volume} {29}},\
  \bibinfo {pages} {190} (\bibinfo {year} {2008})}\BibitemShut {NoStop}%
\bibitem [{\citenamefont {Dryzun}\ and\ \citenamefont
  {Avnir}(2011)}]{Dryzun2011}%
  \BibitemOpen
  \bibfield  {author} {\bibinfo {author} {\bibfnamefont {C.}~\bibnamefont
  {Dryzun}}\ and\ \bibinfo {author} {\bibfnamefont {D.}~\bibnamefont {Avnir}},\
  }\bibfield  {title} {\bibinfo {title} {Chirality measures for vectors,
  matrices, operators and functions},\ }\href@noop {} {\bibfield  {journal}
  {\bibinfo  {journal} {ChemPhysChem}\ }\textbf {\bibinfo {volume} {12}},\
  \bibinfo {pages} {197} (\bibinfo {year} {2011})}\BibitemShut {NoStop}%
\bibitem [{\citenamefont {Harris}\ \emph {et~al.}(1999)\citenamefont {Harris},
  \citenamefont {Kamien},\ and\ \citenamefont {Lubensky}}]{RevModPhys.71.1745}%
  \BibitemOpen
  \bibfield  {author} {\bibinfo {author} {\bibfnamefont {A.~B.}\ \bibnamefont
  {Harris}}, \bibinfo {author} {\bibfnamefont {R.~D.}\ \bibnamefont {Kamien}},\
  and\ \bibinfo {author} {\bibfnamefont {T.~C.}\ \bibnamefont {Lubensky}},\
  }\bibfield  {title} {\bibinfo {title} {Molecular chirality and chiral
  parameters},\ }\href@noop {} {\bibfield  {journal} {\bibinfo  {journal} {Rev.
  Mod. Phys.}\ }\textbf {\bibinfo {volume} {71}},\ \bibinfo {pages} {1745}
  (\bibinfo {year} {1999})}\BibitemShut {NoStop}%
\bibitem [{\citenamefont {Tang}\ and\ \citenamefont
  {Cohen}(2010)}]{tang2010optical}%
  \BibitemOpen
  \bibfield  {author} {\bibinfo {author} {\bibfnamefont {Y.}~\bibnamefont
  {Tang}}\ and\ \bibinfo {author} {\bibfnamefont {A.~E.}\ \bibnamefont
  {Cohen}},\ }\bibfield  {title} {\bibinfo {title} {Optical chirality and its
  interaction with matter},\ }\href@noop {} {\bibfield  {journal} {\bibinfo
  {journal} {Phys. Rev. Lett.}\ }\textbf {\bibinfo {volume} {104}},\ \bibinfo
  {pages} {163901} (\bibinfo {year} {2010})}\BibitemShut {NoStop}%
\bibitem [{\citenamefont {Lipkin}(1964)}]{Lipkin_1964}%
  \BibitemOpen
  \bibfield  {author} {\bibinfo {author} {\bibfnamefont {D.~M.}\ \bibnamefont
  {Lipkin}},\ }\bibfield  {title} {\bibinfo {title} {Existence of a new
  conservation law in electromagnetic theory},\ }\href@noop {} {\bibfield
  {journal} {\bibinfo  {journal} {J. Math. Phys.}\ }\textbf {\bibinfo {volume}
  {5}},\ \bibinfo {pages} {696–} (\bibinfo {year} {1964})}\BibitemShut
  {NoStop}%
\bibitem [{\citenamefont {Choi}\ \emph {et~al.}(2022)\citenamefont {Choi},
  \citenamefont {Yano}, \citenamefont {Cha}, \citenamefont {Colombari},
  \citenamefont {Kim}, \citenamefont {Wang}, \citenamefont {Lee}, \citenamefont
  {Sun}, \citenamefont {Kruger}, \citenamefont {de~Moura},\ and\ \citenamefont
  {Kotov}}]{choi2022chiral}%
  \BibitemOpen
  \bibfield  {author} {\bibinfo {author} {\bibfnamefont {W.~J.}\ \bibnamefont
  {Choi}}, \bibinfo {author} {\bibfnamefont {K.}~\bibnamefont {Yano}}, \bibinfo
  {author} {\bibfnamefont {M.}~\bibnamefont {Cha}}, \bibinfo {author}
  {\bibfnamefont {F.~M.}\ \bibnamefont {Colombari}}, \bibinfo {author}
  {\bibfnamefont {J.-Y.}\ \bibnamefont {Kim}}, \bibinfo {author} {\bibfnamefont
  {Y.}~\bibnamefont {Wang}}, \bibinfo {author} {\bibfnamefont {S.~H.}\
  \bibnamefont {Lee}}, \bibinfo {author} {\bibfnamefont {K.}~\bibnamefont
  {Sun}}, \bibinfo {author} {\bibfnamefont {J.~M.}\ \bibnamefont {Kruger}},
  \bibinfo {author} {\bibfnamefont {A.~F.}\ \bibnamefont {de~Moura}},\ and\
  \bibinfo {author} {\bibfnamefont {N.~A.}\ \bibnamefont {Kotov}},\ }\bibfield
  {title} {\bibinfo {title} {Chiral phonons in microcrystals and nanofibrils of
  biomolecules},\ }\href@noop {} {\bibfield  {journal} {\bibinfo  {journal}
  {Nature Photonics}\ }\textbf {\bibinfo {volume} {16}},\ \bibinfo {pages}
  {366} (\bibinfo {year} {2022})}\BibitemShut {NoStop}%
\bibitem [{\citenamefont {Abraham}\ and\ \citenamefont
  {Nitzan}(2024)}]{Abraham2024}%
  \BibitemOpen
  \bibfield  {author} {\bibinfo {author} {\bibfnamefont {E.}~\bibnamefont
  {Abraham}}\ and\ \bibinfo {author} {\bibfnamefont {A.}~\bibnamefont
  {Nitzan}},\ }\bibfield  {title} {\bibinfo {title} {Quantifying the chirality
  of vibrational modes in helical molecular chains},\ }\href@noop {} {\bibfield
   {journal} {\bibinfo  {journal} {Phys. Rev. Lett.}\ }\textbf {\bibinfo
  {volume} {133}},\ \bibinfo {pages} {268001} (\bibinfo {year}
  {2024})}\BibitemShut {NoStop}%
\bibitem [{\citenamefont {Fan}\ and\ \citenamefont
  {Govorov}(2012)}]{fan2012chiral}%
  \BibitemOpen
  \bibfield  {author} {\bibinfo {author} {\bibfnamefont {Z.}~\bibnamefont
  {Fan}}\ and\ \bibinfo {author} {\bibfnamefont {A.~O.}\ \bibnamefont
  {Govorov}},\ }\bibfield  {title} {\bibinfo {title} {Chiral nanocrystals:
  {P}lasmonic spectra and circular dichroism},\ }\href@noop {} {\bibfield
  {journal} {\bibinfo  {journal} {Nano Lett.}\ }\textbf {\bibinfo {volume}
  {12}},\ \bibinfo {pages} {3283} (\bibinfo {year} {2012})}\BibitemShut
  {NoStop}%
\bibitem [{\citenamefont {Hentschel}\ \emph {et~al.}(2017)\citenamefont
  {Hentschel}, \citenamefont {Sch{\"a}ferling}, \citenamefont {Duan},
  \citenamefont {Giessen},\ and\ \citenamefont {Liu}}]{hentschel2017chiral}%
  \BibitemOpen
  \bibfield  {author} {\bibinfo {author} {\bibfnamefont {M.}~\bibnamefont
  {Hentschel}}, \bibinfo {author} {\bibfnamefont {M.}~\bibnamefont
  {Sch{\"a}ferling}}, \bibinfo {author} {\bibfnamefont {X.}~\bibnamefont
  {Duan}}, \bibinfo {author} {\bibfnamefont {H.}~\bibnamefont {Giessen}},\ and\
  \bibinfo {author} {\bibfnamefont {N.}~\bibnamefont {Liu}},\ }\bibfield
  {title} {\bibinfo {title} {Chiral plasmonics},\ }\href@noop {} {\bibfield
  {journal} {\bibinfo  {journal} {Sci. Adv.}\ }\textbf {\bibinfo {volume}
  {3}},\ \bibinfo {pages} {e1602735} (\bibinfo {year} {2017})}\BibitemShut
  {NoStop}%
\bibitem [{\citenamefont {Zhang}\ \emph {et~al.}(2019)\citenamefont {Zhang},
  \citenamefont {Hernandez}, \citenamefont {Smith}, \citenamefont {Jebeli},
  \citenamefont {Dai}, \citenamefont {Warning}, \citenamefont {Baiyasi},
  \citenamefont {McCarthy}, \citenamefont {Guo}, \citenamefont {Chen},
  \citenamefont {Dionne}, \citenamefont {Landes},\ and\ \citenamefont
  {Link}}]{zhang2019unraveling}%
  \BibitemOpen
  \bibfield  {author} {\bibinfo {author} {\bibfnamefont {Q.}~\bibnamefont
  {Zhang}}, \bibinfo {author} {\bibfnamefont {T.}~\bibnamefont {Hernandez}},
  \bibinfo {author} {\bibfnamefont {K.~W.}\ \bibnamefont {Smith}}, \bibinfo
  {author} {\bibfnamefont {S.~A.~H.}\ \bibnamefont {Jebeli}}, \bibinfo {author}
  {\bibfnamefont {A.~X.}\ \bibnamefont {Dai}}, \bibinfo {author} {\bibfnamefont
  {L.}~\bibnamefont {Warning}}, \bibinfo {author} {\bibfnamefont
  {R.}~\bibnamefont {Baiyasi}}, \bibinfo {author} {\bibfnamefont {L.~A.}\
  \bibnamefont {McCarthy}}, \bibinfo {author} {\bibfnamefont {H.}~\bibnamefont
  {Guo}}, \bibinfo {author} {\bibfnamefont {D.-H.}\ \bibnamefont {Chen}},
  \bibinfo {author} {\bibfnamefont {J.~A.}\ \bibnamefont {Dionne}}, \bibinfo
  {author} {\bibfnamefont {C.~F.}\ \bibnamefont {Landes}},\ and\ \bibinfo
  {author} {\bibfnamefont {S.}~\bibnamefont {Link}},\ }\bibfield  {title}
  {\bibinfo {title} {Unraveling the origin of chirality from plasmonic
  nanoparticle-protein complexes},\ }\href@noop {} {\bibfield  {journal}
  {\bibinfo  {journal} {Science}\ }\textbf {\bibinfo {volume} {365}},\ \bibinfo
  {pages} {1475} (\bibinfo {year} {2019})}\BibitemShut {NoStop}%
\bibitem [{\citenamefont {Kuzyk}\ \emph {et~al.}(2012)\citenamefont {Kuzyk},
  \citenamefont {Schreiber}, \citenamefont {Fan}, \citenamefont {Pardatscher},
  \citenamefont {Roller}, \citenamefont {Högele}, \citenamefont {Simmel},
  \citenamefont {Govorov},\ and\ \citenamefont {Liedl}}]{KuzykAnton2012Dsoc}%
  \BibitemOpen
  \bibfield  {author} {\bibinfo {author} {\bibfnamefont {A.}~\bibnamefont
  {Kuzyk}}, \bibinfo {author} {\bibfnamefont {R.}~\bibnamefont {Schreiber}},
  \bibinfo {author} {\bibfnamefont {Z.}~\bibnamefont {Fan}}, \bibinfo {author}
  {\bibfnamefont {G.}~\bibnamefont {Pardatscher}}, \bibinfo {author}
  {\bibfnamefont {E.-M.}\ \bibnamefont {Roller}}, \bibinfo {author}
  {\bibfnamefont {A.}~\bibnamefont {Högele}}, \bibinfo {author} {\bibfnamefont
  {F.~C.}\ \bibnamefont {Simmel}}, \bibinfo {author} {\bibfnamefont {A.~O.}\
  \bibnamefont {Govorov}},\ and\ \bibinfo {author} {\bibfnamefont
  {T.}~\bibnamefont {Liedl}},\ }\bibfield  {title} {\bibinfo {title}
  {{DNA}-based self-assembly of chiral plasmonic nanostructures with tailored
  optical response},\ }\href@noop {} {\bibfield  {journal} {\bibinfo  {journal}
  {Nature}\ }\textbf {\bibinfo {volume} {483}},\ \bibinfo {pages} {311}
  (\bibinfo {year} {2012})}\BibitemShut {NoStop}%
\bibitem [{\citenamefont {Cerd{\'a}n}\ \emph {et~al.}(2023)\citenamefont
  {Cerd{\'a}n}, \citenamefont {Zundel},\ and\ \citenamefont
  {Manjavacas}}]{cerdan2023chiral}%
  \BibitemOpen
  \bibfield  {author} {\bibinfo {author} {\bibfnamefont {L.}~\bibnamefont
  {Cerd{\'a}n}}, \bibinfo {author} {\bibfnamefont {L.}~\bibnamefont {Zundel}},\
  and\ \bibinfo {author} {\bibfnamefont {A.}~\bibnamefont {Manjavacas}},\
  }\bibfield  {title} {\bibinfo {title} {Chiral lattice resonances in
  2.5-dimensional periodic arrays with achiral unit cells},\ }\href@noop {}
  {\bibfield  {journal} {\bibinfo  {journal} {ACS photonics}\ }\textbf
  {\bibinfo {volume} {10}},\ \bibinfo {pages} {1925} (\bibinfo {year}
  {2023})}\BibitemShut {NoStop}%
\bibitem [{\citenamefont {Baranov}\ \emph {et~al.}(2023)\citenamefont
  {Baranov}, \citenamefont {Sch{\"a}fer},\ and\ \citenamefont
  {Gorkunov}}]{baranov2023toward}%
  \BibitemOpen
  \bibfield  {author} {\bibinfo {author} {\bibfnamefont {D.~G.}\ \bibnamefont
  {Baranov}}, \bibinfo {author} {\bibfnamefont {C.}~\bibnamefont
  {Sch{\"a}fer}},\ and\ \bibinfo {author} {\bibfnamefont {M.~V.}\ \bibnamefont
  {Gorkunov}},\ }\bibfield  {title} {\bibinfo {title} {Toward molecular chiral
  polaritons},\ }\href@noop {} {\bibfield  {journal} {\bibinfo  {journal} {ACS
  Photonics}\ }\textbf {\bibinfo {volume} {10}},\ \bibinfo {pages} {2440}
  (\bibinfo {year} {2023})}\BibitemShut {NoStop}%
\bibitem [{\citenamefont {Ostovar~Pour}\ \emph {et~al.}(2015)\citenamefont
  {Ostovar~Pour}, \citenamefont {Rocks}, \citenamefont {Faulds}, \citenamefont
  {Graham}, \citenamefont {Parcha{\v{n}}sk{\`y}}, \citenamefont {Bou{\v{r}}},\
  and\ \citenamefont {Blanch}}]{ostovar2015through}%
  \BibitemOpen
  \bibfield  {author} {\bibinfo {author} {\bibfnamefont {S.}~\bibnamefont
  {Ostovar~Pour}}, \bibinfo {author} {\bibfnamefont {L.}~\bibnamefont {Rocks}},
  \bibinfo {author} {\bibfnamefont {K.}~\bibnamefont {Faulds}}, \bibinfo
  {author} {\bibfnamefont {D.}~\bibnamefont {Graham}}, \bibinfo {author}
  {\bibfnamefont {V.}~\bibnamefont {Parcha{\v{n}}sk{\`y}}}, \bibinfo {author}
  {\bibfnamefont {P.}~\bibnamefont {Bou{\v{r}}}},\ and\ \bibinfo {author}
  {\bibfnamefont {E.~W.}\ \bibnamefont {Blanch}},\ }\bibfield  {title}
  {\bibinfo {title} {Through-space transfer of chiral information mediated by a
  plasmonic nanomaterial},\ }\href@noop {} {\bibfield  {journal} {\bibinfo
  {journal} {Nat. Chem.}\ }\textbf {\bibinfo {volume} {7}},\ \bibinfo {pages}
  {591} (\bibinfo {year} {2015})}\BibitemShut {NoStop}%
\bibitem [{\citenamefont {Yin}\ \emph {et~al.}(2013)\citenamefont {Yin},
  \citenamefont {Sch{\"a}ferling}, \citenamefont {Metzger},\ and\ \citenamefont
  {Giessen}}]{BK}%
  \BibitemOpen
  \bibfield  {author} {\bibinfo {author} {\bibfnamefont {X.}~\bibnamefont
  {Yin}}, \bibinfo {author} {\bibfnamefont {M.}~\bibnamefont
  {Sch{\"a}ferling}}, \bibinfo {author} {\bibfnamefont {B.}~\bibnamefont
  {Metzger}},\ and\ \bibinfo {author} {\bibfnamefont {H.}~\bibnamefont
  {Giessen}},\ }\bibfield  {title} {\bibinfo {title} {Interpreting chiral
  nanophotonic spectra: {T}he plasmonic {B}orn–{K}uhn model},\ }\href@noop {}
  {\bibfield  {journal} {\bibinfo  {journal} {Nano Lett.}\ }\textbf {\bibinfo
  {volume} {13}},\ \bibinfo {pages} {6238} (\bibinfo {year}
  {2013})}\BibitemShut {NoStop}%
\bibitem [{\citenamefont {Kuhn}(1930)}]{Kuhn1930}%
  \BibitemOpen
  \bibfield  {author} {\bibinfo {author} {\bibfnamefont {W.}~\bibnamefont
  {Kuhn}},\ }\bibfield  {title} {\bibinfo {title} {The physical significance of
  optical rotatory power},\ }\href@noop {} {\bibfield  {journal} {\bibinfo
  {journal} {Trans. Faraday Soc.}\ }\textbf {\bibinfo {volume} {26}},\ \bibinfo
  {pages} {293} (\bibinfo {year} {1930})}\BibitemShut {NoStop}%
\bibitem [{\citenamefont {Born}(1918)}]{Born1918}%
  \BibitemOpen
  \bibfield  {author} {\bibinfo {author} {\bibfnamefont {M.}~\bibnamefont
  {Born}},\ }\bibfield  {title} {\bibinfo {title} {Elektronentheorie des
  nat{\"u}rlichen optischen drehungsverm{\"o}gens isotroper und anisotroper
  fl{\"u}ssigkeiten},\ }\href@noop {} {\bibfield  {journal} {\bibinfo
  {journal} {Annalen der Physik}\ }\textbf {\bibinfo {volume} {360}},\ \bibinfo
  {pages} {177} (\bibinfo {year} {1918})}\BibitemShut {NoStop}%
\bibitem [{\citenamefont {Louren{\c{c}}o-Martins}\ \emph
  {et~al.}(2021)\citenamefont {Louren{\c{c}}o-Martins}, \citenamefont
  {G{\'e}rard},\ and\ \citenamefont {Kociak}}]{Lourenco-MartinsHugo2021Opai}%
  \BibitemOpen
  \bibfield  {author} {\bibinfo {author} {\bibfnamefont {H.}~\bibnamefont
  {Louren{\c{c}}o-Martins}}, \bibinfo {author} {\bibfnamefont {D.}~\bibnamefont
  {G{\'e}rard}},\ and\ \bibinfo {author} {\bibfnamefont {M.}~\bibnamefont
  {Kociak}},\ }\bibfield  {title} {\bibinfo {title} {Optical polarization
  analogue in free electron beams},\ }\href@noop {} {\bibfield  {journal}
  {\bibinfo  {journal} {Nat. Phys.}\ }\textbf {\bibinfo {volume} {17}},\
  \bibinfo {pages} {598} (\bibinfo {year} {2021})}\BibitemShut {NoStop}%
\bibitem [{\citenamefont {Bourgeois}\ \emph
  {et~al.}(2022{\natexlab{a}})\citenamefont {Bourgeois}, \citenamefont {Nixon},
  \citenamefont {Chalifour}, \citenamefont {Beutler},\ and\ \citenamefont
  {Masiello}}]{BourgeoisMarcR.2022PEEG}%
  \BibitemOpen
  \bibfield  {author} {\bibinfo {author} {\bibfnamefont {M.~R.}\ \bibnamefont
  {Bourgeois}}, \bibinfo {author} {\bibfnamefont {A.~G.}\ \bibnamefont
  {Nixon}}, \bibinfo {author} {\bibfnamefont {M.}~\bibnamefont {Chalifour}},
  \bibinfo {author} {\bibfnamefont {E.~K.}\ \bibnamefont {Beutler}},\ and\
  \bibinfo {author} {\bibfnamefont {D.~J.}\ \bibnamefont {Masiello}},\
  }\bibfield  {title} {\bibinfo {title} {Polarization-resolved electron energy
  gain nanospectroscopy with phase-structured electron beams},\ }\href@noop {}
  {\bibfield  {journal} {\bibinfo  {journal} {Nano Lett.}\ }\textbf {\bibinfo
  {volume} {22}},\ \bibinfo {pages} {7158} (\bibinfo {year}
  {2022}{\natexlab{a}})}\BibitemShut {NoStop}%
\bibitem [{\citenamefont {Sidorova}\ \emph {et~al.}(2021)\citenamefont
  {Sidorova}, \citenamefont {Bystrov}, \citenamefont {Lutsenko}, \citenamefont
  {Shpigun}, \citenamefont {Belova},\ and\ \citenamefont
  {Likhachev}}]{sidorova2021quantitative}%
  \BibitemOpen
  \bibfield  {author} {\bibinfo {author} {\bibfnamefont {A.}~\bibnamefont
  {Sidorova}}, \bibinfo {author} {\bibfnamefont {V.}~\bibnamefont {Bystrov}},
  \bibinfo {author} {\bibfnamefont {A.}~\bibnamefont {Lutsenko}}, \bibinfo
  {author} {\bibfnamefont {D.}~\bibnamefont {Shpigun}}, \bibinfo {author}
  {\bibfnamefont {E.}~\bibnamefont {Belova}},\ and\ \bibinfo {author}
  {\bibfnamefont {I.}~\bibnamefont {Likhachev}},\ }\bibfield  {title} {\bibinfo
  {title} {Quantitative assessment of chirality of protein secondary structures
  and phenylalanine peptide nanotubes},\ }\href@noop {} {\bibfield  {journal}
  {\bibinfo  {journal} {Nanomaterials}\ }\textbf {\bibinfo {volume} {11}},\
  \bibinfo {pages} {3299} (\bibinfo {year} {2021})}\BibitemShut {NoStop}%
\bibitem [{\citenamefont {Baiyasi}\ \emph {et~al.}(2021)\citenamefont
  {Baiyasi}, \citenamefont {Goldwyn}, \citenamefont {McCarthy}, \citenamefont
  {West}, \citenamefont {Hosseini~Jebeli}, \citenamefont {Masiello},
  \citenamefont {Link},\ and\ \citenamefont {Landes}}]{Trichoidal_Dichroism}%
  \BibitemOpen
  \bibfield  {author} {\bibinfo {author} {\bibfnamefont {R.}~\bibnamefont
  {Baiyasi}}, \bibinfo {author} {\bibfnamefont {H.~J.}\ \bibnamefont
  {Goldwyn}}, \bibinfo {author} {\bibfnamefont {L.~A.}\ \bibnamefont
  {McCarthy}}, \bibinfo {author} {\bibfnamefont {C.~A.}\ \bibnamefont {West}},
  \bibinfo {author} {\bibfnamefont {S.~A.}\ \bibnamefont {Hosseini~Jebeli}},
  \bibinfo {author} {\bibfnamefont {D.~J.}\ \bibnamefont {Masiello}}, \bibinfo
  {author} {\bibfnamefont {S.}~\bibnamefont {Link}},\ and\ \bibinfo {author}
  {\bibfnamefont {C.~F.}\ \bibnamefont {Landes}},\ }\bibfield  {title}
  {\bibinfo {title} {Coupled-dipole modeling and experimental characterization
  of geometry-dependent trochoidal dichroism in nanorod trimers},\ }\href@noop
  {} {\bibfield  {journal} {\bibinfo  {journal} {ACS Photonics}\ }\textbf
  {\bibinfo {volume} {8}},\ \bibinfo {pages} {1159} (\bibinfo {year}
  {2021})}\BibitemShut {NoStop}%
\bibitem [{\citenamefont {Cherqui}\ \emph {et~al.}(2014)\citenamefont
  {Cherqui}, \citenamefont {Bigelow}, \citenamefont {Vaschillo}, \citenamefont
  {Goldwyn},\ and\ \citenamefont {Masiello}}]{cherqui2014combined}%
  \BibitemOpen
  \bibfield  {author} {\bibinfo {author} {\bibfnamefont {C.}~\bibnamefont
  {Cherqui}}, \bibinfo {author} {\bibfnamefont {N.~W.}\ \bibnamefont
  {Bigelow}}, \bibinfo {author} {\bibfnamefont {A.}~\bibnamefont {Vaschillo}},
  \bibinfo {author} {\bibfnamefont {H.}~\bibnamefont {Goldwyn}},\ and\ \bibinfo
  {author} {\bibfnamefont {D.~J.}\ \bibnamefont {Masiello}},\ }\bibfield
  {title} {\bibinfo {title} {Combined tight-binding and numerical
  electrodynamics understanding of the {STEM/EELS} magneto-optical responses of
  aromatic plasmon-supporting metal oligomers},\ }\href@noop {} {\bibfield
  {journal} {\bibinfo  {journal} {ACS Photonics}\ }\textbf {\bibinfo {volume}
  {1}},\ \bibinfo {pages} {1013} (\bibinfo {year} {2014})}\BibitemShut
  {NoStop}%
\bibitem [{\citenamefont {Cherqui}\ \emph {et~al.}(2016)\citenamefont
  {Cherqui}, \citenamefont {Thakkar}, \citenamefont {Li}, \citenamefont
  {Camden},\ and\ \citenamefont {Masiello}}]{cherqui2016characterizing}%
  \BibitemOpen
  \bibfield  {author} {\bibinfo {author} {\bibfnamefont {C.}~\bibnamefont
  {Cherqui}}, \bibinfo {author} {\bibfnamefont {N.}~\bibnamefont {Thakkar}},
  \bibinfo {author} {\bibfnamefont {G.}~\bibnamefont {Li}}, \bibinfo {author}
  {\bibfnamefont {J.~P.}\ \bibnamefont {Camden}},\ and\ \bibinfo {author}
  {\bibfnamefont {D.~J.}\ \bibnamefont {Masiello}},\ }\bibfield  {title}
  {\bibinfo {title} {Characterizing localized surface plasmons using electron
  energy-loss spectroscopy},\ }\href@noop {} {\bibfield  {journal} {\bibinfo
  {journal} {Annu. Rev. Phys. Chem.}\ }\textbf {\bibinfo {volume} {67}},\
  \bibinfo {pages} {331} (\bibinfo {year} {2016})}\BibitemShut {NoStop}%
\bibitem [{\citenamefont {Smith}\ \emph {et~al.}(2019)\citenamefont {Smith},
  \citenamefont {Olafsson}, \citenamefont {Hu}, \citenamefont {Quillin},
  \citenamefont {Idrobo}, \citenamefont {Collette}, \citenamefont {Rack},
  \citenamefont {Camden},\ and\ \citenamefont {Masiello}}]{Smith2019}%
  \BibitemOpen
  \bibfield  {author} {\bibinfo {author} {\bibfnamefont {K.~C.}\ \bibnamefont
  {Smith}}, \bibinfo {author} {\bibfnamefont {A.}~\bibnamefont {Olafsson}},
  \bibinfo {author} {\bibfnamefont {X.}~\bibnamefont {Hu}}, \bibinfo {author}
  {\bibfnamefont {S.~C.}\ \bibnamefont {Quillin}}, \bibinfo {author}
  {\bibfnamefont {J.~C.}\ \bibnamefont {Idrobo}}, \bibinfo {author}
  {\bibfnamefont {R.}~\bibnamefont {Collette}}, \bibinfo {author}
  {\bibfnamefont {P.~D.}\ \bibnamefont {Rack}}, \bibinfo {author}
  {\bibfnamefont {J.~P.}\ \bibnamefont {Camden}},\ and\ \bibinfo {author}
  {\bibfnamefont {D.~J.}\ \bibnamefont {Masiello}},\ }\bibfield  {title}
  {\bibinfo {title} {Direct observation of infrared plasmonic {F}ano
  antiresonances by a nanoscale electron probe},\ }\href@noop {} {\bibfield
  {journal} {\bibinfo  {journal} {Phys. Rev. Lett.}\ }\textbf {\bibinfo
  {volume} {123}},\ \bibinfo {pages} {177401} (\bibinfo {year}
  {2019})}\BibitemShut {NoStop}%
\bibitem [{\citenamefont {Devkota}\ \emph {et~al.}(2019)\citenamefont
  {Devkota}, \citenamefont {Brown}, \citenamefont {Beane}, \citenamefont {Yu},\
  and\ \citenamefont {Hartland}}]{devkota_making_2019}%
  \BibitemOpen
  \bibfield  {author} {\bibinfo {author} {\bibfnamefont {T.}~\bibnamefont
  {Devkota}}, \bibinfo {author} {\bibfnamefont {B.~S.}\ \bibnamefont {Brown}},
  \bibinfo {author} {\bibfnamefont {G.}~\bibnamefont {Beane}}, \bibinfo
  {author} {\bibfnamefont {K.}~\bibnamefont {Yu}},\ and\ \bibinfo {author}
  {\bibfnamefont {G.~V.}\ \bibnamefont {Hartland}},\ }\bibfield  {title}
  {\bibinfo {title} {Making waves: {R}adiation damping in metallic
  nanostructures},\ }\href@noop {} {\bibfield  {journal} {\bibinfo  {journal}
  {J. Chem. Phys.}\ }\textbf {\bibinfo {volume} {151}},\ \bibinfo {pages}
  {080901} (\bibinfo {year} {2019})}\BibitemShut {NoStop}%
\bibitem [{\citenamefont {Yang}\ \emph {et~al.}(2022)\citenamefont {Yang},
  \citenamefont {Xie}, \citenamefont {You},\ and\ \citenamefont
  {Ye}}]{YangYanhe2022Rtpr}%
  \BibitemOpen
  \bibfield  {author} {\bibinfo {author} {\bibfnamefont {Y.}~\bibnamefont
  {Yang}}, \bibinfo {author} {\bibfnamefont {H.}~\bibnamefont {Xie}}, \bibinfo
  {author} {\bibfnamefont {J.}~\bibnamefont {You}},\ and\ \bibinfo {author}
  {\bibfnamefont {W.}~\bibnamefont {Ye}},\ }\bibfield  {title} {\bibinfo
  {title} {Revisiting the plasmon radiation {damping} of gold nanorods},\
  }\href@noop {} {\bibfield  {journal} {\bibinfo  {journal} {Phys. Chem. Chem.
  Phys.}\ }\textbf {\bibinfo {volume} {24}},\ \bibinfo {pages} {4131} (\bibinfo
  {year} {2022})}\BibitemShut {NoStop}%
\bibitem [{\citenamefont {Montoni}\ \emph {et~al.}(2018)\citenamefont
  {Montoni}, \citenamefont {Quillin}, \citenamefont {Cherqui},\ and\
  \citenamefont {Masiello}}]{montoni2018tunable}%
  \BibitemOpen
  \bibfield  {author} {\bibinfo {author} {\bibfnamefont {N.~P.}\ \bibnamefont
  {Montoni}}, \bibinfo {author} {\bibfnamefont {S.~C.}\ \bibnamefont
  {Quillin}}, \bibinfo {author} {\bibfnamefont {C.}~\bibnamefont {Cherqui}},\
  and\ \bibinfo {author} {\bibfnamefont {D.~J.}\ \bibnamefont {Masiello}},\
  }\bibfield  {title} {\bibinfo {title} {Tunable spectral ordering of magnetic
  plasmon resonances in noble metal nanoclusters},\ }\href@noop {} {\bibfield
  {journal} {\bibinfo  {journal} {ACS Photonics}\ }\textbf {\bibinfo {volume}
  {5}},\ \bibinfo {pages} {3272} (\bibinfo {year} {2018})}\BibitemShut
  {NoStop}%
\bibitem [{\citenamefont {Bourgeois}\ \emph
  {et~al.}(2022{\natexlab{b}})\citenamefont {Bourgeois}, \citenamefont {Rossi},
  \citenamefont {Khorasani},\ and\ \citenamefont
  {Masiello}}]{bourgeois2022optical}%
  \BibitemOpen
  \bibfield  {author} {\bibinfo {author} {\bibfnamefont {M.~R.}\ \bibnamefont
  {Bourgeois}}, \bibinfo {author} {\bibfnamefont {A.~W.}\ \bibnamefont
  {Rossi}}, \bibinfo {author} {\bibfnamefont {S.}~\bibnamefont {Khorasani}},\
  and\ \bibinfo {author} {\bibfnamefont {D.~J.}\ \bibnamefont {Masiello}},\
  }\bibfield  {title} {\bibinfo {title} {Optical control over thermal
  distributions in topologically trivial and non-trivial plasmon lattices},\
  }\href@noop {} {\bibfield  {journal} {\bibinfo  {journal} {ACS Photonics}\
  }\textbf {\bibinfo {volume} {9}},\ \bibinfo {pages} {3656} (\bibinfo {year}
  {2022}{\natexlab{b}})}\BibitemShut {NoStop}%
\bibitem [{\citenamefont {Novotny}\ and\ \citenamefont
  {Hecht}(2006)}]{Novotny}%
  \BibitemOpen
  \bibfield  {author} {\bibinfo {author} {\bibfnamefont {L.}~\bibnamefont
  {Novotny}}\ and\ \bibinfo {author} {\bibfnamefont {B.}~\bibnamefont
  {Hecht}},\ }\href@noop {} {\emph {\bibinfo {title} {Principles of
  Nano-Optics}}},\ \bibinfo {edition} {1st}\ ed.\ (\bibinfo  {publisher}
  {Cambridge University Press},\ \bibinfo {address} {Cambridge},\ \bibinfo
  {year} {2006})\BibitemShut {NoStop}%
\bibitem [{\citenamefont {Vernon}\ \emph {et~al.}(2024)\citenamefont {Vernon},
  \citenamefont {Golat}, \citenamefont {Rigouzzo}, \citenamefont {Lim},\ and\
  \citenamefont {Rodríguez-Fortuño}}]{vernon_decomposition_2024}%
  \BibitemOpen
  \bibfield  {author} {\bibinfo {author} {\bibfnamefont {A.~J.}\ \bibnamefont
  {Vernon}}, \bibinfo {author} {\bibfnamefont {S.}~\bibnamefont {Golat}},
  \bibinfo {author} {\bibfnamefont {C.}~\bibnamefont {Rigouzzo}}, \bibinfo
  {author} {\bibfnamefont {E.~A.}\ \bibnamefont {Lim}},\ and\ \bibinfo {author}
  {\bibfnamefont {F.~J.}\ \bibnamefont {Rodríguez-Fortuño}},\ }\bibfield
  {title} {\bibinfo {title} {A decomposition of light’s spin angular momentum
  density},\ }\href@noop {} {\bibfield  {journal} {\bibinfo  {journal} {Light
  Sci. Appl.}\ }\textbf {\bibinfo {volume} {13}},\ \bibinfo {pages} {160}
  (\bibinfo {year} {2024})}\BibitemShut {NoStop}%
\bibitem [{\citenamefont {Bliokh}\ and\ \citenamefont
  {Nori}(2015)}]{bliokh_transverse_2015}%
  \BibitemOpen
  \bibfield  {author} {\bibinfo {author} {\bibfnamefont {K.~Y.}\ \bibnamefont
  {Bliokh}}\ and\ \bibinfo {author} {\bibfnamefont {F.}~\bibnamefont {Nori}},\
  }\bibfield  {title} {\bibinfo {title} {Transverse and longitudinal angular
  momenta of light},\ }\href@noop {} {\bibfield  {journal} {\bibinfo  {journal}
  {Physics Reports}\ }\textbf {\bibinfo {volume} {592}},\ \bibinfo {pages} {1}
  (\bibinfo {year} {2015})}\BibitemShut {NoStop}%
\bibitem [{\citenamefont {Allen}\ \emph {et~al.}(2003)\citenamefont {Allen},
  \citenamefont {Barnett},\ and\ \citenamefont {Padgett}}]{allen_optical_2003}%
  \BibitemOpen
  \bibfield  {author} {\bibinfo {author} {\bibfnamefont {L.}~\bibnamefont
  {Allen}}, \bibinfo {author} {\bibfnamefont {S.~M.}\ \bibnamefont {Barnett}},\
  and\ \bibinfo {author} {\bibfnamefont {M.~J.}\ \bibnamefont {Padgett}},\
  }\href@noop {} {\emph {\bibinfo {title} {Optical Angular Momentum}}},\
  \bibinfo {edition} {1st}\ ed.\ (\bibinfo  {publisher} {CRC Press},\ \bibinfo
  {address} {Boca Raton},\ \bibinfo {year} {2003})\BibitemShut {NoStop}%
\bibitem [{\citenamefont {Barnett}\ and\ \citenamefont
  {Allen}(1994)}]{barnett_orbital_1994}%
  \BibitemOpen
  \bibfield  {author} {\bibinfo {author} {\bibfnamefont {S.~M.}\ \bibnamefont
  {Barnett}}\ and\ \bibinfo {author} {\bibfnamefont {L.}~\bibnamefont
  {Allen}},\ }\bibfield  {title} {\bibinfo {title} {Orbital angular momentum
  and nonparaxial light beams},\ }\href@noop {} {\bibfield  {journal} {\bibinfo
   {journal} {Opt. Commun.}\ }\textbf {\bibinfo {volume} {110}},\ \bibinfo
  {pages} {670} (\bibinfo {year} {1994})}\BibitemShut {NoStop}%
\bibitem [{\citenamefont {Barron}(2009)}]{Barron2009-sp}%
  \BibitemOpen
  \bibfield  {author} {\bibinfo {author} {\bibfnamefont {L.~D.}\ \bibnamefont
  {Barron}},\ }\href@noop {} {\emph {\bibinfo {title} {Molecular Light
  Scattering and Optical Activity}}},\ \bibinfo {edition} {2nd}\ ed.\ (\bibinfo
   {publisher} {Cambridge University Press},\ \bibinfo {address} {Cambridge,
  England},\ \bibinfo {year} {2009})\BibitemShut {NoStop}%
\bibitem [{\citenamefont {Poulikakos}\ \emph {et~al.}(2019)\citenamefont
  {Poulikakos}, \citenamefont {Dionne},\ and\ \citenamefont
  {Garc{\'\i}a-Etxarri}}]{poulikakos2019optical}%
  \BibitemOpen
  \bibfield  {author} {\bibinfo {author} {\bibfnamefont {L.~V.}\ \bibnamefont
  {Poulikakos}}, \bibinfo {author} {\bibfnamefont {J.~A.}\ \bibnamefont
  {Dionne}},\ and\ \bibinfo {author} {\bibfnamefont {A.}~\bibnamefont
  {Garc{\'\i}a-Etxarri}},\ }\bibfield  {title} {\bibinfo {title} {Optical
  helicity and optical chirality in free space and in the presence of matter},\
  }\href@noop {} {\bibfield  {journal} {\bibinfo  {journal} {Symmetry}\
  }\textbf {\bibinfo {volume} {11}},\ \bibinfo {pages} {1113} (\bibinfo {year}
  {2019})}\BibitemShut {NoStop}%
\bibitem [{\citenamefont {Bliokh}\ and\ \citenamefont
  {Nori}(2011)}]{Bliokh2011}%
  \BibitemOpen
  \bibfield  {author} {\bibinfo {author} {\bibfnamefont {K.~Y.}\ \bibnamefont
  {Bliokh}}\ and\ \bibinfo {author} {\bibfnamefont {F.}~\bibnamefont {Nori}},\
  }\bibfield  {title} {\bibinfo {title} {Characterizing optical chirality},\
  }\href@noop {} {\bibfield  {journal} {\bibinfo  {journal} {Phys. Rev. A}\
  }\textbf {\bibinfo {volume} {83}},\ \bibinfo {pages} {021803} (\bibinfo
  {year} {2011})}\BibitemShut {NoStop}%
\bibitem [{\citenamefont {Coles}\ and\ \citenamefont
  {Andrews}(2012)}]{coles2012chirality}%
  \BibitemOpen
  \bibfield  {author} {\bibinfo {author} {\bibfnamefont {M.~M.}\ \bibnamefont
  {Coles}}\ and\ \bibinfo {author} {\bibfnamefont {D.~L.}\ \bibnamefont
  {Andrews}},\ }\bibfield  {title} {\bibinfo {title} {Chirality and angular
  momentum in optical radiation},\ }\href@noop {} {\bibfield  {journal}
  {\bibinfo  {journal} {Phys. Rev. A}\ }\textbf {\bibinfo {volume} {85}},\
  \bibinfo {pages} {063810} (\bibinfo {year} {2012})}\BibitemShut {NoStop}%
\bibitem [{\citenamefont {Barnett}(2010)}]{barnett2010rotation}%
  \BibitemOpen
  \bibfield  {author} {\bibinfo {author} {\bibfnamefont {S.~M.}\ \bibnamefont
  {Barnett}},\ }\bibfield  {title} {\bibinfo {title} {Rotation of
  electromagnetic fields and the nature of optical angular momentum},\
  }\href@noop {} {\bibfield  {journal} {\bibinfo  {journal} {J. Mod. Optic.}\
  }\textbf {\bibinfo {volume} {57}},\ \bibinfo {pages} {1339} (\bibinfo {year}
  {2010})}\BibitemShut {NoStop}%
\bibitem [{\citenamefont
  {Nieto-Vesperinas}(2015{\natexlab{a}})}]{Nieto_2015_oam}%
  \BibitemOpen
  \bibfield  {author} {\bibinfo {author} {\bibfnamefont {M.}~\bibnamefont
  {Nieto-Vesperinas}},\ }\bibfield  {title} {\bibinfo {title} {Optical torque:
  {E}lectromagnetic spin and orbital-angular-momentum conservation laws and
  their significance},\ }\href@noop {} {\bibfield  {journal} {\bibinfo
  {journal} {Phys. Rev. A}\ }\textbf {\bibinfo {volume} {92}},\ \bibinfo
  {pages} {043843} (\bibinfo {year} {2015}{\natexlab{a}})}\BibitemShut
  {NoStop}%
\bibitem [{\citenamefont
  {Nieto-Vesperinas}(2015{\natexlab{b}})}]{Nieto-Vesperinas_15}%
  \BibitemOpen
  \bibfield  {author} {\bibinfo {author} {\bibfnamefont {M.}~\bibnamefont
  {Nieto-Vesperinas}},\ }\bibfield  {title} {\bibinfo {title} {Optical torque
  on small bi-isotropic particles},\ }\href@noop {} {\bibfield  {journal}
  {\bibinfo  {journal} {Opt. Lett.}\ }\textbf {\bibinfo {volume} {40}},\
  \bibinfo {pages} {3021} (\bibinfo {year} {2015}{\natexlab{b}})}\BibitemShut
  {NoStop}%
\bibitem [{\citenamefont {Jackson}(1975)}]{Jackson2003}%
  \BibitemOpen
  \bibfield  {author} {\bibinfo {author} {\bibfnamefont {J.~D.}\ \bibnamefont
  {Jackson}},\ }\href@noop {} {\emph {\bibinfo {title} {Classical
  Electrodynamics}}},\ \bibinfo {edition} {2nd}\ ed.\ (\bibinfo  {publisher}
  {John Wiley \& Sons Ltd.},\ \bibinfo {address} {New York},\ \bibinfo {year}
  {1975})\BibitemShut {NoStop}%
\bibitem [{\citenamefont {Fan}\ and\ \citenamefont
  {Govorov}(2010)}]{Zhiyuan2010}%
  \BibitemOpen
  \bibfield  {author} {\bibinfo {author} {\bibfnamefont {Z.}~\bibnamefont
  {Fan}}\ and\ \bibinfo {author} {\bibfnamefont {A.~O.}\ \bibnamefont
  {Govorov}},\ }\bibfield  {title} {\bibinfo {title} {Plasmonic circular
  dichroism of chiral metal nanoparticle assemblies},\ }\href@noop {}
  {\bibfield  {journal} {\bibinfo  {journal} {Nano Lett.}\ }\textbf {\bibinfo
  {volume} {10}},\ \bibinfo {pages} {2580} (\bibinfo {year}
  {2010})}\BibitemShut {NoStop}%
\bibitem [{\citenamefont {Bohren}\ and\ \citenamefont
  {Huffman}(2008)}]{Bohren1998}%
  \BibitemOpen
  \bibfield  {author} {\bibinfo {author} {\bibfnamefont {C.~F.}\ \bibnamefont
  {Bohren}}\ and\ \bibinfo {author} {\bibfnamefont {D.~R.}\ \bibnamefont
  {Huffman}},\ }\href@noop {} {\emph {\bibinfo {title} {Absorption and
  Scattering of Light by Small Particles}}}\ (\bibinfo  {publisher} {John Wiley
  \& Sons},\ \bibinfo {address} {New York},\ \bibinfo {year}
  {2008})\BibitemShut {NoStop}%
\bibitem [{\citenamefont
  {Nieto-Vesperinas}(2015{\natexlab{c}})}]{Nieto-Vesperinas_2015}%
  \BibitemOpen
  \bibfield  {author} {\bibinfo {author} {\bibfnamefont {M.}~\bibnamefont
  {Nieto-Vesperinas}},\ }\bibfield  {title} {\bibinfo {title} {Optical theorem
  for the conservation of electromagnetic helicity: {S}ignificance for
  molecular energy transfer and enantiomeric discrimination by circular
  dichroism},\ }\href@noop {} {\bibfield  {journal} {\bibinfo  {journal} {Phys.
  Rev. A}\ }\textbf {\bibinfo {volume} {92}},\ \bibinfo {pages} {023813}
  (\bibinfo {year} {2015}{\natexlab{c}})}\BibitemShut {NoStop}%
\bibitem [{\citenamefont {Lee}\ \emph {et~al.}(2018)\citenamefont {Lee},
  \citenamefont {Ahn}, \citenamefont {Mun}, \citenamefont {Lee}, \citenamefont
  {Kim}, \citenamefont {Cho}, \citenamefont {Chang}, \citenamefont {Kim},
  \citenamefont {Rho},\ and\ \citenamefont {Nam}}]{LeeHye-Eun2018Aaps}%
  \BibitemOpen
  \bibfield  {author} {\bibinfo {author} {\bibfnamefont {H.-E.}\ \bibnamefont
  {Lee}}, \bibinfo {author} {\bibfnamefont {H.-Y.}\ \bibnamefont {Ahn}},
  \bibinfo {author} {\bibfnamefont {J.}~\bibnamefont {Mun}}, \bibinfo {author}
  {\bibfnamefont {Y.~Y.}\ \bibnamefont {Lee}}, \bibinfo {author} {\bibfnamefont
  {M.}~\bibnamefont {Kim}}, \bibinfo {author} {\bibfnamefont {N.~H.}\
  \bibnamefont {Cho}}, \bibinfo {author} {\bibfnamefont {K.}~\bibnamefont
  {Chang}}, \bibinfo {author} {\bibfnamefont {W.~S.}\ \bibnamefont {Kim}},
  \bibinfo {author} {\bibfnamefont {J.}~\bibnamefont {Rho}},\ and\ \bibinfo
  {author} {\bibfnamefont {K.~T.}\ \bibnamefont {Nam}},\ }\bibfield  {title}
  {\bibinfo {title} {Amino-acid- and peptide-directed synthesis of chiral
  plasmonic gold nanoparticles},\ }\href@noop {} {\bibfield  {journal}
  {\bibinfo  {journal} {Nature}\ }\textbf {\bibinfo {volume} {556}},\ \bibinfo
  {pages} {360} (\bibinfo {year} {2018})}\BibitemShut {NoStop}%
\bibitem [{\citenamefont {Kramer}\ \emph {et~al.}(2017)\citenamefont {Kramer},
  \citenamefont {Sch{\"a}ferling}, \citenamefont {Weiss}, \citenamefont
  {Giessen},\ and\ \citenamefont {Brixner}}]{kramer2017analytic}%
  \BibitemOpen
  \bibfield  {author} {\bibinfo {author} {\bibfnamefont {C.}~\bibnamefont
  {Kramer}}, \bibinfo {author} {\bibfnamefont {M.}~\bibnamefont
  {Sch{\"a}ferling}}, \bibinfo {author} {\bibfnamefont {T.}~\bibnamefont
  {Weiss}}, \bibinfo {author} {\bibfnamefont {H.}~\bibnamefont {Giessen}},\
  and\ \bibinfo {author} {\bibfnamefont {T.}~\bibnamefont {Brixner}},\
  }\bibfield  {title} {\bibinfo {title} {Analytic optimization of near-field
  optical chirality enhancement},\ }\href@noop {} {\bibfield  {journal}
  {\bibinfo  {journal} {ACS Photonics}\ }\textbf {\bibinfo {volume} {4}},\
  \bibinfo {pages} {396} (\bibinfo {year} {2017})}\BibitemShut {NoStop}%
\bibitem [{\citenamefont {Gao}\ \emph {et~al.}(2021)\citenamefont {Gao},
  \citenamefont {Addison}, \citenamefont {Mele},\ and\ \citenamefont
  {Rappe}}]{PhysRevResearch.3.L042032}%
  \BibitemOpen
  \bibfield  {author} {\bibinfo {author} {\bibfnamefont {L.}~\bibnamefont
  {Gao}}, \bibinfo {author} {\bibfnamefont {Z.}~\bibnamefont {Addison}},
  \bibinfo {author} {\bibfnamefont {E.~J.}\ \bibnamefont {Mele}},\ and\
  \bibinfo {author} {\bibfnamefont {A.~M.}\ \bibnamefont {Rappe}},\ }\bibfield
  {title} {\bibinfo {title} {Intrinsic {F}ermi-surface contribution to the bulk
  photovoltaic effect},\ }\href
  {https://doi.org/10.1103/PhysRevResearch.3.L042032} {\bibfield  {journal}
  {\bibinfo  {journal} {Phys. Rev. Res.}\ }\textbf {\bibinfo {volume} {3}},\
  \bibinfo {pages} {L042032} (\bibinfo {year} {2021})}\BibitemShut {NoStop}%
\bibitem [{\citenamefont {Lodahl}\ \emph {et~al.}(2017)\citenamefont {Lodahl},
  \citenamefont {Mahmoodian}, \citenamefont {Stobbe}, \citenamefont
  {Rauschenbeutel}, \citenamefont {Schneeweiss}, \citenamefont {Volz},
  \citenamefont {Pichler},\ and\ \citenamefont {Zoller}}]{lodahl2017chiral}%
  \BibitemOpen
  \bibfield  {author} {\bibinfo {author} {\bibfnamefont {P.}~\bibnamefont
  {Lodahl}}, \bibinfo {author} {\bibfnamefont {S.}~\bibnamefont {Mahmoodian}},
  \bibinfo {author} {\bibfnamefont {S.}~\bibnamefont {Stobbe}}, \bibinfo
  {author} {\bibfnamefont {A.}~\bibnamefont {Rauschenbeutel}}, \bibinfo
  {author} {\bibfnamefont {P.}~\bibnamefont {Schneeweiss}}, \bibinfo {author}
  {\bibfnamefont {J.}~\bibnamefont {Volz}}, \bibinfo {author} {\bibfnamefont
  {H.}~\bibnamefont {Pichler}},\ and\ \bibinfo {author} {\bibfnamefont
  {P.}~\bibnamefont {Zoller}},\ }\bibfield  {title} {\bibinfo {title} {Chiral
  quantum optics},\ }\href@noop {} {\bibfield  {journal} {\bibinfo  {journal}
  {Nature}\ }\textbf {\bibinfo {volume} {541}},\ \bibinfo {pages} {473}
  (\bibinfo {year} {2017})}\BibitemShut {NoStop}%
\bibitem [{\citenamefont {Naaman}\ \emph {et~al.}(2019)\citenamefont {Naaman},
  \citenamefont {Paltiel},\ and\ \citenamefont {Waldeck}}]{naaman2019chiral}%
  \BibitemOpen
  \bibfield  {author} {\bibinfo {author} {\bibfnamefont {R.}~\bibnamefont
  {Naaman}}, \bibinfo {author} {\bibfnamefont {Y.}~\bibnamefont {Paltiel}},\
  and\ \bibinfo {author} {\bibfnamefont {D.~H.}\ \bibnamefont {Waldeck}},\
  }\bibfield  {title} {\bibinfo {title} {Chiral molecules and the electron
  spin},\ }\href@noop {} {\bibfield  {journal} {\bibinfo  {journal} {Nat. Rev.
  Chem.}\ }\textbf {\bibinfo {volume} {3}},\ \bibinfo {pages} {250} (\bibinfo
  {year} {2019})}\BibitemShut {NoStop}%
\bibitem [{\citenamefont {Aiello}\ \emph {et~al.}(2022)\citenamefont {Aiello},
  \citenamefont {Abendroth}, \citenamefont {Abbas}, \citenamefont {Afanasev},
  \citenamefont {Agarwal}, \citenamefont {Banerjee}, \citenamefont {Beratan},
  \citenamefont {Belling}, \citenamefont {Berche}, \citenamefont {Botana},
  \citenamefont {Caram}, \citenamefont {Celardo}, \citenamefont {Cuniberti},
  \citenamefont {Garcia-Etxarri}, \citenamefont {Dianat}, \citenamefont
  {Diez-Perez}, \citenamefont {Guo}, \citenamefont {Gutierrez}, \citenamefont
  {Herrmann}, \citenamefont {Hihath}, \citenamefont {Kale}, \citenamefont
  {Kurian}, \citenamefont {Lai}, \citenamefont {Liu}, \citenamefont {Lopez},
  \citenamefont {Medina}, \citenamefont {Mujica}, \citenamefont {Naaman},
  \citenamefont {Noormandipour}, \citenamefont {Palma}, \citenamefont
  {Paltiel}, \citenamefont {Petuskey}, \citenamefont {Ribeiro-Silva},
  \citenamefont {Saenz}, \citenamefont {Santos}, \citenamefont
  {Solyanik-Gorgone}, \citenamefont {Sorger}, \citenamefont {Stemer},
  \citenamefont {Ugalde}, \citenamefont {Valdes-Curiel}, \citenamefont
  {Varela}, \citenamefont {Waldeck}, \citenamefont {Wasielewski}, \citenamefont
  {Weiss}, \citenamefont {Zacharias},\ and\ \citenamefont
  {Wang}}]{aiello2022chirality}%
  \BibitemOpen
  \bibfield  {author} {\bibinfo {author} {\bibfnamefont {C.~D.}\ \bibnamefont
  {Aiello}}, \bibinfo {author} {\bibfnamefont {J.~M.}\ \bibnamefont
  {Abendroth}}, \bibinfo {author} {\bibfnamefont {M.}~\bibnamefont {Abbas}},
  \bibinfo {author} {\bibfnamefont {A.}~\bibnamefont {Afanasev}}, \bibinfo
  {author} {\bibfnamefont {S.}~\bibnamefont {Agarwal}}, \bibinfo {author}
  {\bibfnamefont {A.~S.}\ \bibnamefont {Banerjee}}, \bibinfo {author}
  {\bibfnamefont {D.~N.}\ \bibnamefont {Beratan}}, \bibinfo {author}
  {\bibfnamefont {J.~N.}\ \bibnamefont {Belling}}, \bibinfo {author}
  {\bibfnamefont {B.}~\bibnamefont {Berche}}, \bibinfo {author} {\bibfnamefont
  {A.}~\bibnamefont {Botana}}, \bibinfo {author} {\bibfnamefont {J.~R.}\
  \bibnamefont {Caram}}, \bibinfo {author} {\bibfnamefont {G.~L.}\ \bibnamefont
  {Celardo}}, \bibinfo {author} {\bibfnamefont {G.}~\bibnamefont {Cuniberti}},
  \bibinfo {author} {\bibfnamefont {A.}~\bibnamefont {Garcia-Etxarri}},
  \bibinfo {author} {\bibfnamefont {A.}~\bibnamefont {Dianat}}, \bibinfo
  {author} {\bibfnamefont {I.}~\bibnamefont {Diez-Perez}}, \bibinfo {author}
  {\bibfnamefont {Y.}~\bibnamefont {Guo}}, \bibinfo {author} {\bibfnamefont
  {R.}~\bibnamefont {Gutierrez}}, \bibinfo {author} {\bibfnamefont
  {C.}~\bibnamefont {Herrmann}}, \bibinfo {author} {\bibfnamefont
  {J.}~\bibnamefont {Hihath}}, \bibinfo {author} {\bibfnamefont
  {S.}~\bibnamefont {Kale}}, \bibinfo {author} {\bibfnamefont {P.}~\bibnamefont
  {Kurian}}, \bibinfo {author} {\bibfnamefont {Y.-C.}\ \bibnamefont {Lai}},
  \bibinfo {author} {\bibfnamefont {T.}~\bibnamefont {Liu}}, \bibinfo {author}
  {\bibfnamefont {A.}~\bibnamefont {Lopez}}, \bibinfo {author} {\bibfnamefont
  {E.}~\bibnamefont {Medina}}, \bibinfo {author} {\bibfnamefont
  {V.}~\bibnamefont {Mujica}}, \bibinfo {author} {\bibfnamefont
  {R.}~\bibnamefont {Naaman}}, \bibinfo {author} {\bibfnamefont
  {M.}~\bibnamefont {Noormandipour}}, \bibinfo {author} {\bibfnamefont {J.~L.}\
  \bibnamefont {Palma}}, \bibinfo {author} {\bibfnamefont {Y.}~\bibnamefont
  {Paltiel}}, \bibinfo {author} {\bibfnamefont {W.}~\bibnamefont {Petuskey}},
  \bibinfo {author} {\bibfnamefont {J.~C.}\ \bibnamefont {Ribeiro-Silva}},
  \bibinfo {author} {\bibfnamefont {J.~J.}\ \bibnamefont {Saenz}}, \bibinfo
  {author} {\bibfnamefont {E.~J.~G.}\ \bibnamefont {Santos}}, \bibinfo {author}
  {\bibfnamefont {M.}~\bibnamefont {Solyanik-Gorgone}}, \bibinfo {author}
  {\bibfnamefont {V.~J.}\ \bibnamefont {Sorger}}, \bibinfo {author}
  {\bibfnamefont {D.~M.}\ \bibnamefont {Stemer}}, \bibinfo {author}
  {\bibfnamefont {J.~M.}\ \bibnamefont {Ugalde}}, \bibinfo {author}
  {\bibfnamefont {A.}~\bibnamefont {Valdes-Curiel}}, \bibinfo {author}
  {\bibfnamefont {S.}~\bibnamefont {Varela}}, \bibinfo {author} {\bibfnamefont
  {D.~H.}\ \bibnamefont {Waldeck}}, \bibinfo {author} {\bibfnamefont {M.~R.}\
  \bibnamefont {Wasielewski}}, \bibinfo {author} {\bibfnamefont {P.~S.}\
  \bibnamefont {Weiss}}, \bibinfo {author} {\bibfnamefont {H.}~\bibnamefont
  {Zacharias}},\ and\ \bibinfo {author} {\bibfnamefont {Q.~H.}\ \bibnamefont
  {Wang}},\ }\bibfield  {title} {\bibinfo {title} {A chirality-based quantum
  leap},\ }\href@noop {} {\bibfield  {journal} {\bibinfo  {journal} {ACS Nano}\
  }\textbf {\bibinfo {volume} {16}},\ \bibinfo {pages} {4989} (\bibinfo {year}
  {2022})}\BibitemShut {NoStop}%
\bibitem [{\citenamefont {Eckvahl}\ \emph {et~al.}(2023)\citenamefont
  {Eckvahl}, \citenamefont {Tcyrulnikov}, \citenamefont {Chiesa}, \citenamefont
  {Bradley}, \citenamefont {Young}, \citenamefont {Carretta}, \citenamefont
  {Krzyaniak},\ and\ \citenamefont {Wasielewski}}]{eckvahl2023direct}%
  \BibitemOpen
  \bibfield  {author} {\bibinfo {author} {\bibfnamefont {H.~J.}\ \bibnamefont
  {Eckvahl}}, \bibinfo {author} {\bibfnamefont {N.~A.}\ \bibnamefont
  {Tcyrulnikov}}, \bibinfo {author} {\bibfnamefont {A.}~\bibnamefont {Chiesa}},
  \bibinfo {author} {\bibfnamefont {J.~M.}\ \bibnamefont {Bradley}}, \bibinfo
  {author} {\bibfnamefont {R.~M.}\ \bibnamefont {Young}}, \bibinfo {author}
  {\bibfnamefont {S.}~\bibnamefont {Carretta}}, \bibinfo {author}
  {\bibfnamefont {M.~D.}\ \bibnamefont {Krzyaniak}},\ and\ \bibinfo {author}
  {\bibfnamefont {M.~R.}\ \bibnamefont {Wasielewski}},\ }\bibfield  {title}
  {\bibinfo {title} {Direct observation of chirality-induced spin selectivity
  in electron donor--acceptor molecules},\ }\href@noop {} {\bibfield  {journal}
  {\bibinfo  {journal} {Science}\ }\textbf {\bibinfo {volume} {382}},\ \bibinfo
  {pages} {197} (\bibinfo {year} {2023})}\BibitemShut {NoStop}%
\bibitem [{\citenamefont {Crassous}\ \emph {et~al.}(2023)\citenamefont
  {Crassous}, \citenamefont {Fuchter}, \citenamefont {Freedman}, \citenamefont
  {Kotov}, \citenamefont {Moon}, \citenamefont {Beard},\ and\ \citenamefont
  {Feldmann}}]{crassous2023materials}%
  \BibitemOpen
  \bibfield  {author} {\bibinfo {author} {\bibfnamefont {J.}~\bibnamefont
  {Crassous}}, \bibinfo {author} {\bibfnamefont {M.~J.}\ \bibnamefont
  {Fuchter}}, \bibinfo {author} {\bibfnamefont {D.~E.}\ \bibnamefont
  {Freedman}}, \bibinfo {author} {\bibfnamefont {N.~A.}\ \bibnamefont {Kotov}},
  \bibinfo {author} {\bibfnamefont {J.}~\bibnamefont {Moon}}, \bibinfo {author}
  {\bibfnamefont {M.~C.}\ \bibnamefont {Beard}},\ and\ \bibinfo {author}
  {\bibfnamefont {S.}~\bibnamefont {Feldmann}},\ }\bibfield  {title} {\bibinfo
  {title} {Materials for chiral light control},\ }\href@noop {} {\bibfield
  {journal} {\bibinfo  {journal} {Nat. Rev. Mater.}\ }\textbf {\bibinfo
  {volume} {8}},\ \bibinfo {pages} {365} (\bibinfo {year} {2023})}\BibitemShut
  {NoStop}%
\bibitem [{\citenamefont {Chen}\ \emph {et~al.}(2024)\citenamefont {Chen},
  \citenamefont {Salij}, \citenamefont {Parrish}, \citenamefont {Rasch},
  \citenamefont {Zinna}, \citenamefont {Brown}, \citenamefont {Pescitelli},
  \citenamefont {Urraci}, \citenamefont {Aronica}, \citenamefont {Dhavamani},
  \citenamefont {Arnold}, \citenamefont {Wasielewski}, \citenamefont {di~Bari},
  \citenamefont {Tempelaar},\ and\ \citenamefont {Goldsmith}}]{chen20242d}%
  \BibitemOpen
  \bibfield  {author} {\bibinfo {author} {\bibfnamefont {T.-L.}\ \bibnamefont
  {Chen}}, \bibinfo {author} {\bibfnamefont {A.}~\bibnamefont {Salij}},
  \bibinfo {author} {\bibfnamefont {K.~A.}\ \bibnamefont {Parrish}}, \bibinfo
  {author} {\bibfnamefont {J.~K.}\ \bibnamefont {Rasch}}, \bibinfo {author}
  {\bibfnamefont {F.}~\bibnamefont {Zinna}}, \bibinfo {author} {\bibfnamefont
  {P.~J.}\ \bibnamefont {Brown}}, \bibinfo {author} {\bibfnamefont
  {G.}~\bibnamefont {Pescitelli}}, \bibinfo {author} {\bibfnamefont
  {F.}~\bibnamefont {Urraci}}, \bibinfo {author} {\bibfnamefont {L.~A.}\
  \bibnamefont {Aronica}}, \bibinfo {author} {\bibfnamefont {A.}~\bibnamefont
  {Dhavamani}}, \bibinfo {author} {\bibfnamefont {M.~S.}\ \bibnamefont
  {Arnold}}, \bibinfo {author} {\bibfnamefont {M.~R.}\ \bibnamefont
  {Wasielewski}}, \bibinfo {author} {\bibfnamefont {L.}~\bibnamefont
  {di~Bari}}, \bibinfo {author} {\bibfnamefont {R.}~\bibnamefont {Tempelaar}},\
  and\ \bibinfo {author} {\bibfnamefont {R.~H.}\ \bibnamefont {Goldsmith}},\
  }\bibfield  {title} {\bibinfo {title} {A 2{D} chiral microcavity based on
  apparent circular dichroism},\ }\href@noop {} {\bibfield  {journal} {\bibinfo
   {journal} {Nat. Commun.}\ }\textbf {\bibinfo {volume} {15}},\ \bibinfo
  {pages} {3072} (\bibinfo {year} {2024})}\BibitemShut {NoStop}%
\bibitem [{\citenamefont {Feng}\ \emph {et~al.}(2025)\citenamefont {Feng},
  \citenamefont {Abraham}, \citenamefont {Subotnik},\ and\ \citenamefont
  {Nitzan}}]{feng2025chiral}%
  \BibitemOpen
  \bibfield  {author} {\bibinfo {author} {\bibfnamefont {J.}~\bibnamefont
  {Feng}}, \bibinfo {author} {\bibfnamefont {E.}~\bibnamefont {Abraham}},
  \bibinfo {author} {\bibfnamefont {J.}~\bibnamefont {Subotnik}},\ and\
  \bibinfo {author} {\bibfnamefont {A.}~\bibnamefont {Nitzan}},\ }\bibfield
  {title} {\bibinfo {title} {Chiral vibrational modes in small molecules},\
  }\href@noop {} {\bibfield  {journal} {\bibinfo  {journal} {arXiv preprint
  arXiv:2503.03048v1}\ } (\bibinfo {year} {2025})}\BibitemShut {NoStop}%
\bibitem [{\citenamefont {Chen}\ \emph {et~al.}(2021)\citenamefont {Chen},
  \citenamefont {Wu}, \citenamefont {Zhu}, \citenamefont {Yang},\ and\
  \citenamefont {Zhang}}]{chen2021propagating}%
  \BibitemOpen
  \bibfield  {author} {\bibinfo {author} {\bibfnamefont {H.}~\bibnamefont
  {Chen}}, \bibinfo {author} {\bibfnamefont {W.}~\bibnamefont {Wu}}, \bibinfo
  {author} {\bibfnamefont {J.}~\bibnamefont {Zhu}}, \bibinfo {author}
  {\bibfnamefont {S.~A.}\ \bibnamefont {Yang}},\ and\ \bibinfo {author}
  {\bibfnamefont {L.}~\bibnamefont {Zhang}},\ }\bibfield  {title} {\bibinfo
  {title} {Propagating chiral phonons in three-dimensional materials},\
  }\href@noop {} {\bibfield  {journal} {\bibinfo  {journal} {Nano Lett.}\
  }\textbf {\bibinfo {volume} {21}},\ \bibinfo {pages} {3060} (\bibinfo {year}
  {2021})}\BibitemShut {NoStop}%
\bibitem [{\citenamefont {Jo}\ \emph {et~al.}(2024)\citenamefont {Jo},
  \citenamefont {Ryu}, \citenamefont {Huh}, \citenamefont {Kim}, \citenamefont
  {Seo}, \citenamefont {Lee}, \citenamefont {Kwon}, \citenamefont {Lee},
  \citenamefont {Nam},\ and\ \citenamefont
  {Kim}}]{Jo_Ryu_Huh_Kim_Seo_Lee_Kwon_Lee_Nam_Kim_2024}%
  \BibitemOpen
  \bibfield  {author} {\bibinfo {author} {\bibfnamefont {J.}~\bibnamefont
  {Jo}}, \bibinfo {author} {\bibfnamefont {J.}~\bibnamefont {Ryu}}, \bibinfo
  {author} {\bibfnamefont {J.-H.}\ \bibnamefont {Huh}}, \bibinfo {author}
  {\bibfnamefont {H.}~\bibnamefont {Kim}}, \bibinfo {author} {\bibfnamefont
  {D.~H.}\ \bibnamefont {Seo}}, \bibinfo {author} {\bibfnamefont
  {J.}~\bibnamefont {Lee}}, \bibinfo {author} {\bibfnamefont {M.}~\bibnamefont
  {Kwon}}, \bibinfo {author} {\bibfnamefont {S.}~\bibnamefont {Lee}}, \bibinfo
  {author} {\bibfnamefont {K.~T.}\ \bibnamefont {Nam}},\ and\ \bibinfo {author}
  {\bibfnamefont {M.}~\bibnamefont {Kim}},\ }\bibfield  {title} {\bibinfo
  {title} {Direct three-dimensional observation of the plasmonic near-fields of
  a nanoparticle with circular dichroism},\ }\href@noop {} {\bibfield
  {journal} {\bibinfo  {journal} {ACS Nano}\ }\textbf {\bibinfo {volume}
  {18}},\ \bibinfo {pages} {32769–} (\bibinfo {year} {2024})}\BibitemShut
  {NoStop}%
\bibitem [{\citenamefont {Juarez}\ \emph {et~al.}(2024)\citenamefont {Juarez},
  \citenamefont {Freire-Fernández}, \citenamefont {Khorasani}, \citenamefont
  {Bourgeois}, \citenamefont {Wang}, \citenamefont {Masiello}, \citenamefont
  {Schatz},\ and\ \citenamefont {Odom}}]{acsphotonicsJuarez}%
  \BibitemOpen
  \bibfield  {author} {\bibinfo {author} {\bibfnamefont {X.~G.}\ \bibnamefont
  {Juarez}}, \bibinfo {author} {\bibfnamefont {F.}~\bibnamefont
  {Freire-Fernández}}, \bibinfo {author} {\bibfnamefont {S.}~\bibnamefont
  {Khorasani}}, \bibinfo {author} {\bibfnamefont {M.~R.}\ \bibnamefont
  {Bourgeois}}, \bibinfo {author} {\bibfnamefont {Y.}~\bibnamefont {Wang}},
  \bibinfo {author} {\bibfnamefont {D.~J.}\ \bibnamefont {Masiello}}, \bibinfo
  {author} {\bibfnamefont {G.~C.}\ \bibnamefont {Schatz}},\ and\ \bibinfo
  {author} {\bibfnamefont {T.~W.}\ \bibnamefont {Odom}},\ }\bibfield  {title}
  {\bibinfo {title} {Chiral optical properties of plasmonic kagome lattices},\
  }\href@noop {} {\bibfield  {journal} {\bibinfo  {journal} {ACS Photonics}\
  }\textbf {\bibinfo {volume} {11}},\ \bibinfo {pages} {673} (\bibinfo {year}
  {2024})}\BibitemShut {NoStop}%
\bibitem [{\citenamefont {Shen}\ \emph {et~al.}(2019)\citenamefont {Shen},
  \citenamefont {Wang}, \citenamefont {Xie}, \citenamefont {Min}, \citenamefont
  {Fu}, \citenamefont {Liu}, \citenamefont {Gong},\ and\ \citenamefont
  {Yuan}}]{ShenYijie2019Ov3y}%
  \BibitemOpen
  \bibfield  {author} {\bibinfo {author} {\bibfnamefont {Y.}~\bibnamefont
  {Shen}}, \bibinfo {author} {\bibfnamefont {X.}~\bibnamefont {Wang}}, \bibinfo
  {author} {\bibfnamefont {Z.}~\bibnamefont {Xie}}, \bibinfo {author}
  {\bibfnamefont {C.}~\bibnamefont {Min}}, \bibinfo {author} {\bibfnamefont
  {X.}~\bibnamefont {Fu}}, \bibinfo {author} {\bibfnamefont {Q.}~\bibnamefont
  {Liu}}, \bibinfo {author} {\bibfnamefont {M.}~\bibnamefont {Gong}},\ and\
  \bibinfo {author} {\bibfnamefont {X.}~\bibnamefont {Yuan}},\ }\bibfield
  {title} {\bibinfo {title} {Optical vortices 30 years on: {OAM} manipulation
  from topological charge to multiple singularities},\ }\href@noop {}
  {\bibfield  {journal} {\bibinfo  {journal} {Light Sci. Appl.}\ }\textbf
  {\bibinfo {volume} {8}},\ \bibinfo {pages} {1} (\bibinfo {year}
  {2019})}\BibitemShut {NoStop}%
\bibitem [{\citenamefont {Hillenbrand}\ \emph {et~al.}(2025)\citenamefont
  {Hillenbrand}, \citenamefont {Abate}, \citenamefont {Liu}, \citenamefont
  {Chen},\ and\ \citenamefont {Basov}}]{hillenbrand2025visible}%
  \BibitemOpen
  \bibfield  {author} {\bibinfo {author} {\bibfnamefont {R.}~\bibnamefont
  {Hillenbrand}}, \bibinfo {author} {\bibfnamefont {Y.}~\bibnamefont {Abate}},
  \bibinfo {author} {\bibfnamefont {M.}~\bibnamefont {Liu}}, \bibinfo {author}
  {\bibfnamefont {X.}~\bibnamefont {Chen}},\ and\ \bibinfo {author}
  {\bibfnamefont {D.~N.}\ \bibnamefont {Basov}},\ }\bibfield  {title} {\bibinfo
  {title} {Visible-to-{TH}z near-field nanoscopy},\ }\href@noop {} {\bibfield
  {journal} {\bibinfo  {journal} {Nat. Rev. Mater.}\ }\textbf {\bibinfo
  {volume} {10}},\ \bibinfo {pages} {285} (\bibinfo {year} {2025})}\BibitemShut
  {NoStop}%
\bibitem [{\citenamefont {Zanfrognini}\ \emph {et~al.}(2019)\citenamefont
  {Zanfrognini}, \citenamefont {Rotunno}, \citenamefont {Frabboni},
  \citenamefont {Sit}, \citenamefont {Karimi}, \citenamefont {Hohenester},\
  and\ \citenamefont {Grillo}}]{zanfrognini2019orbital}%
  \BibitemOpen
  \bibfield  {author} {\bibinfo {author} {\bibfnamefont {M.}~\bibnamefont
  {Zanfrognini}}, \bibinfo {author} {\bibfnamefont {E.}~\bibnamefont
  {Rotunno}}, \bibinfo {author} {\bibfnamefont {S.}~\bibnamefont {Frabboni}},
  \bibinfo {author} {\bibfnamefont {A.}~\bibnamefont {Sit}}, \bibinfo {author}
  {\bibfnamefont {E.}~\bibnamefont {Karimi}}, \bibinfo {author} {\bibfnamefont
  {U.}~\bibnamefont {Hohenester}},\ and\ \bibinfo {author} {\bibfnamefont
  {V.}~\bibnamefont {Grillo}},\ }\bibfield  {title} {\bibinfo {title} {Orbital
  angular momentum and energy loss characterization of plasmonic excitations in
  metallic nanostructures in {TEM}},\ }\href@noop {} {\bibfield  {journal}
  {\bibinfo  {journal} {ACS Photonics}\ }\textbf {\bibinfo {volume} {6}},\
  \bibinfo {pages} {620} (\bibinfo {year} {2019})}\BibitemShut {NoStop}%
\bibitem [{\citenamefont {Bourgeois}\ \emph {et~al.}(2023)\citenamefont
  {Bourgeois}, \citenamefont {Nixon}, \citenamefont {Chalifour},\ and\
  \citenamefont {Masiello}}]{bourgeois2023optical}%
  \BibitemOpen
  \bibfield  {author} {\bibinfo {author} {\bibfnamefont {M.~R.}\ \bibnamefont
  {Bourgeois}}, \bibinfo {author} {\bibfnamefont {A.~G.}\ \bibnamefont
  {Nixon}}, \bibinfo {author} {\bibfnamefont {M.}~\bibnamefont {Chalifour}},\
  and\ \bibinfo {author} {\bibfnamefont {D.~J.}\ \bibnamefont {Masiello}},\
  }\bibfield  {title} {\bibinfo {title} {Optical polarization analogs in
  inelastic free-electron scattering},\ }\href
  {https://doi.org/10.1126/sciadv.adj6038} {\bibfield  {journal} {\bibinfo
  {journal} {Sci. Adv.}\ }\textbf {\bibinfo {volume} {9}},\ \bibinfo {pages}
  {eadj6038} (\bibinfo {year} {2023})}\BibitemShut {NoStop}%
\bibitem [{\citenamefont {Nixon}\ \emph {et~al.}(2024)\citenamefont {Nixon},
  \citenamefont {Chalifour}, \citenamefont {Bourgeois}, \citenamefont
  {Sanchez},\ and\ \citenamefont {Masiello}}]{nixon2024inelastic}%
  \BibitemOpen
  \bibfield  {author} {\bibinfo {author} {\bibfnamefont {A.~G.}\ \bibnamefont
  {Nixon}}, \bibinfo {author} {\bibfnamefont {M.}~\bibnamefont {Chalifour}},
  \bibinfo {author} {\bibfnamefont {M.~R.}\ \bibnamefont {Bourgeois}}, \bibinfo
  {author} {\bibfnamefont {M.}~\bibnamefont {Sanchez}},\ and\ \bibinfo {author}
  {\bibfnamefont {D.~J.}\ \bibnamefont {Masiello}},\ }\bibfield  {title}
  {\bibinfo {title} {Inelastic scattering of transversely structured free
  electrons from nanophotonic targets: Theory and computation},\ }\href@noop {}
  {\bibfield  {journal} {\bibinfo  {journal} {Phys. Rev. A}\ }\textbf {\bibinfo
  {volume} {109}},\ \bibinfo {pages} {043502} (\bibinfo {year}
  {2024})}\BibitemShut {NoStop}%
\bibitem [{\citenamefont {Tavabi}\ \emph {et~al.}(2024)\citenamefont {Tavabi},
  \citenamefont {Rosi}, \citenamefont {Ravelli}, \citenamefont {Gijsbers},
  \citenamefont {Rotunno}, \citenamefont {Guner}, \citenamefont {Zhang},
  \citenamefont {Roncaglia}, \citenamefont {Belsito}, \citenamefont {Pozzi},
  \citenamefont {Denneulin}, \citenamefont {Gazzadi}, \citenamefont {Ghosh},
  \citenamefont {Nijland}, \citenamefont {Frabboni}, \citenamefont {Peters},
  \citenamefont {Karimi}, \citenamefont {Tiemeijer}, \citenamefont
  {Dunin-Borkowski},\ and\ \citenamefont {Grillo}}]{tavabi2024symmetry}%
  \BibitemOpen
  \bibfield  {author} {\bibinfo {author} {\bibfnamefont {A.}~\bibnamefont
  {Tavabi}}, \bibinfo {author} {\bibfnamefont {P.}~\bibnamefont {Rosi}},
  \bibinfo {author} {\bibfnamefont {R.}~\bibnamefont {Ravelli}}, \bibinfo
  {author} {\bibfnamefont {A.}~\bibnamefont {Gijsbers}}, \bibinfo {author}
  {\bibfnamefont {E.}~\bibnamefont {Rotunno}}, \bibinfo {author} {\bibfnamefont
  {T.}~\bibnamefont {Guner}}, \bibinfo {author} {\bibfnamefont
  {Y.}~\bibnamefont {Zhang}}, \bibinfo {author} {\bibfnamefont
  {A.}~\bibnamefont {Roncaglia}}, \bibinfo {author} {\bibfnamefont
  {L.}~\bibnamefont {Belsito}}, \bibinfo {author} {\bibfnamefont
  {G.}~\bibnamefont {Pozzi}}, \bibinfo {author} {\bibfnamefont
  {T.}~\bibnamefont {Denneulin}}, \bibinfo {author} {\bibfnamefont
  {G.}~\bibnamefont {Gazzadi}}, \bibinfo {author} {\bibfnamefont
  {M.}~\bibnamefont {Ghosh}}, \bibinfo {author} {\bibfnamefont
  {R.}~\bibnamefont {Nijland}}, \bibinfo {author} {\bibfnamefont
  {S.}~\bibnamefont {Frabboni}}, \bibinfo {author} {\bibfnamefont
  {P.}~\bibnamefont {Peters}}, \bibinfo {author} {\bibfnamefont
  {E.}~\bibnamefont {Karimi}}, \bibinfo {author} {\bibfnamefont
  {P.}~\bibnamefont {Tiemeijer}}, \bibinfo {author} {\bibfnamefont
  {R.}~\bibnamefont {Dunin-Borkowski}},\ and\ \bibinfo {author} {\bibfnamefont
  {V.}~\bibnamefont {Grillo}},\ }\bibfield  {title} {\bibinfo {title} {Symmetry
  and planar chirality measured with a log-polar transformation in a
  transmission electron microscope},\ }\href@noop {} {\bibfield  {journal}
  {\bibinfo  {journal} {Phys. Rev. Appl.}\ }\textbf {\bibinfo {volume} {22}},\
  \bibinfo {pages} {014083} (\bibinfo {year} {2024})}\BibitemShut {NoStop}%
\bibitem [{\citenamefont {Bourgeois}\ \emph {et~al.}(2025)\citenamefont
  {Bourgeois}, \citenamefont {Rossi},\ and\ \citenamefont
  {Masiello}}]{Bourgeois2025}%
  \BibitemOpen
  \bibfield  {author} {\bibinfo {author} {\bibfnamefont {M.~R.}\ \bibnamefont
  {Bourgeois}}, \bibinfo {author} {\bibfnamefont {A.~W.}\ \bibnamefont
  {Rossi}},\ and\ \bibinfo {author} {\bibfnamefont {D.~J.}\ \bibnamefont
  {Masiello}},\ }\bibfield  {title} {\bibinfo {title} {Strategy for direct
  detection of chiral phonons with phase-structured free electrons},\
  }\href@noop {} {\bibfield  {journal} {\bibinfo  {journal} {Phys. Rev. Lett.}\
  }\textbf {\bibinfo {volume} {134}},\ \bibinfo {pages} {026902} (\bibinfo
  {year} {2025})}\BibitemShut {NoStop}%
\bibitem [{\citenamefont {Olmos-Trigo}\ \emph {et~al.}(2024)\citenamefont
  {Olmos-Trigo}, \citenamefont {Lasa-Alonso}, \citenamefont {G\'omez-Viloria},
  \citenamefont {Molina-Terriza},\ and\ \citenamefont
  {Garc\'{\i}a-Etxarri}}]{olmos_trigo_CD}%
  \BibitemOpen
  \bibfield  {author} {\bibinfo {author} {\bibfnamefont {J.}~\bibnamefont
  {Olmos-Trigo}}, \bibinfo {author} {\bibfnamefont {J.}~\bibnamefont
  {Lasa-Alonso}}, \bibinfo {author} {\bibfnamefont {I.}~\bibnamefont
  {G\'omez-Viloria}}, \bibinfo {author} {\bibfnamefont {G.}~\bibnamefont
  {Molina-Terriza}},\ and\ \bibinfo {author} {\bibfnamefont {A.}~\bibnamefont
  {Garc\'{\i}a-Etxarri}},\ }\bibfield  {title} {\bibinfo {title} {Capturing
  near-field circular dichroism enhancements from far-field measurements},\
  }\href@noop {} {\bibfield  {journal} {\bibinfo  {journal} {Phys. Rev. Res.}\
  }\textbf {\bibinfo {volume} {6}},\ \bibinfo {pages} {013151} (\bibinfo {year}
  {2024})}\BibitemShut {NoStop}%
\bibitem [{\citenamefont {Nieto-Vesperinas}(2017)}]{nieto2017chiral}%
  \BibitemOpen
  \bibfield  {author} {\bibinfo {author} {\bibfnamefont {M.}~\bibnamefont
  {Nieto-Vesperinas}},\ }\bibfield  {title} {\bibinfo {title} {Chiral optical
  fields: a unified formulation of helicity scattered from particles and
  dichroism enhancement},\ }\href@noop {} {\bibfield  {journal} {\bibinfo
  {journal} {Philos. Trans. R. Soc. A}\ }\textbf {\bibinfo {volume} {375}},\
  \bibinfo {pages} {20160314} (\bibinfo {year} {2017})}\BibitemShut {NoStop}%
\end{thebibliography}%
\end{document}